\documentclass[12pt]{article}
\usepackage{amsmath}
\usepackage{amsthm}
\usepackage{graphicx}
\usepackage{enumerate}
\usepackage{natbib}
\usepackage{url} 
\usepackage{subcaption}
\usepackage{float}
\usepackage{caption}
\usepackage{amssymb}
\usepackage{amsfonts}
\usepackage{multirow}
\usepackage{mathtools}
\usepackage{setspace}
\usepackage{bm}
\usepackage{titlesec}
\usepackage{algorithmicx}
\usepackage{adjustbox} 
\usepackage{algorithm}
\usepackage{array}
\usepackage{chngcntr}
\usepackage{dcolumn} 
\usepackage{booktabs} 
\usepackage{caption} 
\usepackage{algpseudocode}
\usepackage{mathrsfs}

\usepackage[colorlinks=true, urlcolor=black, linkcolor=black, citecolor = blue]{hyperref}

\newcommand{\blind}{1}

\begin{document}

\bibliographystyle{apalike} 
\def\spacingset#1{\renewcommand{\baselinestretch}%
{#1}\small\normalsize} \spacingset{1}


\if1\blind
{
  \title{\bf An Efficient Class of Bayesian Generalized Quadratic Nonlinear Dynamic Models with Application to Birth Rate Monitoring}
  \author{Madelyn Clinch\\
    Department of Statistics, Florida State University\\
    and \\
    Jonathan R. Bradley\thanks{
    This work is partially supported by the National Science Foundation (NSF)'s Division of Mathematical Sciences (DMS) through the NSF grant DMS-\(2310756\). \textit{}}\hspace{.2cm} \\
     Department of Statistics, Florida State University}
     \date{}
  \maketitle
} \fi

\if0\blind
{
  \bigskip
  \bigskip
  \bigskip
  \begin{center}
    {\LARGE\bf Exact MCMC-free Bayesian Inference for Data of Any Size with Application to Joint Spatial Analysis of Fine Particulate Matter and Aerosol Optical Thickness}
\end{center}
  \medskip
} \fi

\bigskip
\begin{abstract}
\noindent {Many real-world spatio-temporal processes exhibit nonlinear dynamics that can often be described through stochastic partial differential equations. These models are flexible and scientifically motivated, however, implementing them in a fully Bayesian framework can be computationally challenging. We are motivated by birth rate data, which has important implications for public health and are known to follow nonlinear dynamics. We propose a covariance calibration strategy that specifies the covariance matrix of a linear mixed effects model to be close in Frobenius norm to that of a Generalized Quadratic Nonlinearity (GQN) model. We refer to this as Frobenius norm matching. This allows us to model nonlinear dynamics using an easier to implement linear framework. The calibrated linear model is efficiently implemented using Exact Posterior Regression (EPR), a recently proposed Bayesian model that enables sampling of fixed and random effects directly from the posterior distribution. We provide simulation studies that compare to implementations using MCMC. Finally, we use this approach to analyze Florida county-level birth rate data from 1990-2023. Our results indicate that our non-linear spatio-temporal model outperforms linear dynamic spatio-temporal models for this data, and identifies covariate effects consistent with existing literature, all while avoiding the computational difficulties of MCMC.}
\end{abstract}

\noindent%
\vfill

\newpage
\spacingset{1.9} 
\section{Introduction}\label{s:intro}
Birth rates have been declining in recent years, and in many regions, have fallen below the threshold required for population replacement \citep{aitken2022changing, bhattacharjee2024global, jarzebski2021ageing}. These changes in population dynamics can have adverse economic effects by contributing to labor shortages and increasing the dependency ratio, thereby requiring greater allocation of resources to support the elderly population \citep{mccollough2023impact, kearney2022causes, comolli2021spreading}. Many studies attribute declining birth rates to improved access to education for women and greater availability of reproductive health services \citep{aitken2022changing, liu2020education, cheng2022global}. Understanding and forecasting future birth rate trends is critical for effective public planning, including the allocation of resources for schools, healthcare, and other essential services \citep{bhattacharjee2024global}.

Our motivating application is the study of population dynamics using nonlinear spatio-temporal models, with a focus on modeling birth rates. Birth rates are a key component in population growth models \citep{henderson2019ecological} and have gained increased attention due to the decline in U.S. birth rates in recent years \citep{kearney2022puzzle}. Accurate spatial predictions and forecasting of birth rates is important for informing health policy and planning for resource allocation \citep{vanella2023stochastic, tzitiridou2024predicting, vollset2020fertility}. The reaction-diffusion PDE is commonly used for modeling population dynamics \citep{alessio2023reaction, roques2016modelling} and is a special case of the broader class known as the Generalized Quadratic Nonlinearity (GQN) model \citep{wikle2010general}. Prior studies have also highlighted the presence of nonlinear temporal dynamics in birth processes \citep{tang2002density, neverova2016dynamics, wolpin1984estimable}, further motivating the use of a GQN model in this setting.

While the GQN is a flexible framework that can be used to model complex spatio-temporal dynamics, it does pose computational difficulties. Many real-world spatio-temporal processes exhibit complex interactions that are nonlinear, nonstationary, and non-separable \citep{wikle2010general}, and are often expressed through a stochastic partial differential equation (SPDE), which is used to define or approximate the spatial-temporal covariance structure. SPDEs are known to be difficult to fit using Markov Chain Monte Carlo (MCMC) \citep{inla, cameletti2013spatio, sigrist2015stochastic}, which is a standard Bayesian technique for sampling from a generic posterior distribution. There is often times a large number of parameters to estimate, which can be computationally burdensome. In addition, these models often require the use of a Metropolis-Hastings step as full conditional distributions for use in a Gibbs sampler cannot be derived leading to practical limitations like parameter tuning and convergence checks. When discretized and written as a hierarchical model, many classes of stochastic PDEs can be written with the GQN framework \citep{wikle2010general}, but inference via MCMC can be infeasible for spatio-temporal datasets with large parameter spaces such as the GQN.

To address this issue, we propose an alternative approach that calibrates an easier to implement linear mixed effects model to be close to the GQN, enabling more efficient posterior sampling. Specifically, we define the covariance matrix of the linear mixed model to be close in Frobenius Norm \citep{horn2012matrix} to that of a GQN. We call this approach Frobenius norm matching. Similar covariance calibration strategies have been used in the literature \citep{bonas2024calibrated,  raim2021spatio, bradley2019spatio, mcdermott2016model, bradley2015multivariate}. By aligning the covariance structures in terms of the Frobenius norm, we are able to capture the non-linear dynamics present in the GQN while taking advantage of the computational efficiency of implementing a linear model. {However, Frobenius norm matching alone does not solve the computational challenges posed by the high-dimensional parameter space of the GQN, as it still requires the use of many basis functions making MCMC implementation difficult.}

{After covariance calibration, the linear model can be fitted with an efficient Bayesian sampler called Exact Posterior Regression (EPR) \citep{bradley2024generating}.  EPR is a recently introduced MCMC-free sampler that allows for simulating the fixed and random effects of a spatial generalized linear mixed effects model (GLMM) directly from the posterior distribution by expanding the parameter space through the inclusion of what is referred to as the discrepancy term. The posterior distribution takes the form of a generalized conjugate multivariate distribution which can be sampled from directly. EPR has been shown to significantly reduce the computation time compared to traditional MCMC techniques, while producing similar to better spatial predictions and regression estimation. In previous work, EPR has been applied to univariate spatial models \citep{bradley2024generating}, multivariate spatial models \citep{clinch2024exact, zhou2024multiscale}, and spatio-temporal models for count data \citep{pan2024bayesian}. In this paper, we extend EPR by calibrating it to a nonlinear dynamic spatio-temporal model  using the Frobenius norm matching, resulting in computationally efficient Bayesian inference for our motivating birth rate data.}

In summary, we make four main contributions to the literature on spatio-temporal modeling and population dynamics. First, we assess whether county-level birth rate data exhibits linear or nonlinear dynamics by comparing a linear mixed model calibrated to a GQN with a first order vector auto-regressive (VAR(1)) model. Second, we demonstrate that our approach enables efficient modeling of Florida county-level birth rate data enabling spatial interpolation, temporal forecasting, and estimation of regression coefficients. Third, we develop the Frobenius norm matching strategy that aligns the covariance structure of a linear mixed effects model with that of a complex nonlinear spatio-temporal model. Lastly, we utilize Exact Posterior Regression to fit the calibrated linear model, allowing for posterior inference without the use of MCMC achieving significant computational efficiency. 

The rest of this paper is organized as follows. Section \ref{s:motivate} introduces our motivating data. Section \ref{s:prelim} reviews the VAR(1) and GQN dynamic models. Section \ref{s:method} describes the Frobenius norm matching strategy, discusses a change of support approach for using spatio-temporal basis function at the areal level instead of point reference level, and then finally presents the EPR model. Section \ref{s:simstudy} presents a simulation study comparing our calibrated linear model implemented with EPR compared to various implementations using MCMC. Section \ref{ch2:s:analysis} presents the results from our analysis of the Florida birth rate data. Finally, we conclude with Section \ref{ch2:s:discuss} which provides a discussion and future directions. The R code used for the simulations and analysis are available in Supplementary Appendix \ref{s:code}.
\section{Motivating Data}\label{s:motivate}
{
Numerous studies have provided evidence of nonlinear temporal dynamics in birth rate data \citet{tzitiridou2024predicting, heckman1989forecasting, de2016dynamic, drepper1994nonlinear}. Birth rates are a key component of population dynamics and are often modeled using stochastic PDEs, many of which can be written within the GQN framework. Motivated by the literature, we explore the suitability of GQN models for analyzing county-level birth rate data in Florida over the years 1990-2023.}
{
While most existing literature focuses on temporal modeling, our goal is to fit a spatio-temporal model that captures nonlinear spatio-temporal dynamics. As an initial step, we implemented a univariate GQN time series model in Stan and compared its one-step ahead forecast performance against an autogressive model of order one (AR(1)). Table \ref{single.county} presents forecasting errors for various counties in Florida. }

\begin{table}[H]
\centering
\caption{Forecast error for different covariance structure for various counties in Florida.}
\begin{tabular}{lcccc}
\toprule
Model & Taylor & Collier & Hernando & Bradford \\ 
\midrule
GQN  & 1.707 & 0.001 & 0.112 & 1.695  \\
AR(1) & 1.978 & 0.009 & 0.117 & 1.897  \\
\bottomrule
\end{tabular}\label{single.county}
\end{table}
{
These preliminary results indicate potential nonlinear dynamics in birth rate data, motivating the use of the GQN framework for our spatio-temporal analysis.  We attempted to fit a Generalized Quadratic Nonlinearity model using Hamiltonian Monte Carlo (HMC) implemented via Stan to the entire Florida birth rate data. The implementation failed to complete due to memory constraints arising from the extremely high dimensionality of the parameter space. Specifically, the GQN model includes coefficient matrices with \(n^2\) and \(n^3\) parameters, where \(n\) denotes the number of counties, making posterior sampling of these parameters computationally infeasible. This challenge highlights the need for a scalable alternative that can capture the nonlinear dynamics without requiring substantial dimension reduction or sacrificing model complexity. The results presented later in Section \ref{ch2:s:analysis} demonstrate that the calibrated GQN model outperforms standard linear dynamic models for the full space-time birth rate data. }

\section{Preliminaries}\label{s:prelim}
\subsection{Review: First Order Vector Autoregessive Model}\label{method:var1}
{Our literature review and exploratory analysis in Section \ref{s:motivate} suggests that birth rates have nonlinear dynamics.} However, to our knowledge no one has yet formally modeled county-level birth rate data using a nonlinear spatio-temporal model. Thus, an important goal of our analysis is to formally determine whether or not birth rates evolve nonlinearly. To do this we compare our model with GQN dynamics to a standard linear dynamic model. In particular, let \(U_t(\mathbf{s}_i)\) be a spatio-temporal process observed at location \(\mathbf{s}_i\) and time \(t\) with spatial domain defined by \(n\) locations \(D_s \equiv \{\mathbf{s}_1, \dots, \mathbf{s}_n\}\) and temporal domain defined by \(T\) time points \(D_t \equiv \{1, \dots, T\}\). A commonly used linear dynamic model is the first order vector autoregressive process, which assumes the current value \(U_t(\mathbf{s}_i)\) depends on a linear combination of all the values at the previous time \(\{U_{t-1}(\mathbf{s}_1), \dots U_{t-1}(\mathbf{s}_n)\}\). The VAR(1) model can be defined as the following expression 
\begin{align}
        U_{t}(\mathbf{s}_i) = \sum_{j = 1}^n a_{ij}U_{t-1}(\mathbf{s}_j) + \eta_t(\mathbf{s}_i)
    \label{eq:var1}
\end{align}
where the coefficients \(a_{ij}\) can be thought of as weights. These weights make up the matrix \(\mathbf{A} = (a_{ij})_{1 \leq i,j \leq n}\) which is typically referred to as the propagator or transition matrix. The term \(\eta_t(\mathbf{s}_i)\) is assumed to be the mean zero Gaussian distributed noise with spatial covariance.  
\subsection{Review: General Quadratic Nonlinearity}\label{s:gqn}
A flexible class of nonlinear dynamic spatio-temporal models is given by the Generalized Quadratic Nonlinearity (GQN) model originally introduced in \citet{wikle2010general}. The GQN framework is well suited to model nonlinear dynamic spatio-temporal processes as it can capture non-separable and non-stationary spatio-temporal processes. Specifically, the GQN is defined as the following general expression,
\begin{align}
    U_{t}(\mathbf{s}_i) = \sum_{j = 1}^n a_{ij}U_{t-1}(\mathbf{s}_j) + \sum_{k = 1}^n \sum_{l = 1}^n b_{i, kl}U_{t-1}(\mathbf{s}_k)g(U_{t-1}(\mathbf{s}_l); \boldsymbol{\theta}_g) + \eta_t(\mathbf{s}_i)
    \label{eq:gqn}
\end{align}
where \(U_t(\mathbf{s}_i)\) denotes the latent process at location \(\mathbf{s}_i\) and time t. The first term on the right-hand side of Equation (\ref{eq:gqn}) represents a linear combination of the neighboring values from the previous time point, while the second term is a quadratic interaction which consists of the product between \(U_{t-1}(\mathbf{s}_k)\) and a nonlinear transformation of \(U_{t-1}(\mathbf{s}_l)\) enabling the model to account for complex second-order spatio-temporal dependencies. The function \(g(\cdot)\) depends on parameters \(\boldsymbol{\theta}_g\) and in our application of population dynamics is defined as \(g(\cdot) = \gamma_0(\mathbf{s}_i)\text{exp}(1 - U_{t-1}(\mathbf{s}_i)/\gamma_1(\mathbf{s}_i))\) which corresponds to the discretization of a reaction-diffusion partial differential equation (PDE) \citep{cressie2011statistics}, where \(\gamma_0\) is the growth parameter and \(\gamma_1\) is the carrying capacity. See Supplementary Appendix \ref{appen:pde} for more details on the reaction-diffusion PDE and its discretization. The term \({\eta}_t(\mathbf{s}_i)\) is a Gaussian distributed noise term with spatial dependence.

A challenge with the GQN model is the rapid growth of the parameter space as the number of spatial locations increases. To see this note that the matrix \(\mathbf{A} = (a_{ij})_{1 \leq i,j \leq n}\) has \(n^2\) parameters and the matrix \(\mathbf{B} = (b_{ijk})_{1 \leq i,j,k \leq n}\) has \(n^3\) parameters. Estimating these parameters via computationally expensive Markov Chian Monte Carlo (MCMC) is practically infeasible. In this paper, we adopt a strategy based on Frobenius norm matching, which calibrates a linear mixed effects model to be close to the GQN. This approach enables computationally feasible inference without the use of MCMC.

\subsection{Review: The Generalized Conjugate Multivariate Distribution}\label{rev:gcm}
{As we saw from Section \ref{s:motivate}, implementing MCMC via Stan posed practical challenges and memory limitations, even for our relatively small motivating dataset due to the large parameter space. MCMC algorithms often have poor mixing and slow convergence in scenarios of high-dimensional parameter spaces \citep{septier2015overview, gilks1996strategies, robert2018accelerating}. To by-pass the need for MCMC we consider a new framework called exact posterior regression, whose posterior distribution follows a generalized conjugate multivariate (GCM) distribution. In this section we give a review of the GCM which was developed in \citet{bradley2024generating} and is defined through a transformation of a vector of correlated and non-identically distributed Diacosnis-Ylvisaker (DY) \citep{DY} random variables.} Let  \(\mathbf{w} = (w_i: i = 1 \dots, n)^{\prime}\) denote the vector of DY random variables and define the \(n\)-dimensional GCM random vector \(\mathbf{h} = \boldsymbol{\mu} + \mathbf{V}\mathbf{D}(\boldsymbol{\theta})\mathbf{w}\). Marginalizing over the parameter \(\boldsymbol{\theta}\), the probability density function is given by
\begin{align*}
    \text{GCM}(\mathbf{h}|\boldsymbol{\mu}, \mathbf{V}, \boldsymbol{\alpha}, \boldsymbol{\kappa}) = \int \frac{f(\boldsymbol{\theta})}{\mathscr{N}(\boldsymbol{\theta})} \text{exp}\left[\boldsymbol{\alpha}^{\prime}\mathbf{D}(\boldsymbol{\theta})^{-1}\mathbf{V}^{-1}(\mathbf{h}-\boldsymbol{\mu}) - \boldsymbol{\kappa}^{\prime} \boldsymbol{\psi}\{\mathbf{D}(\boldsymbol{\theta})^{-1}\mathbf{V}^{-1}(\mathbf{h}- \boldsymbol{\mu}) \} \right] d\boldsymbol{\theta}.
\end{align*}
The term \(\boldsymbol{\mu}\) is an \(n\)-dimensional location parameter, \(\mathbf{V}\) is an \( n \times n\) invertible matrix, \(\boldsymbol{\theta}\) is a generic \(d\)-dimensional parameter vector with probability distribution function \(f(\boldsymbol{\theta})\). The \(n \times n\) matrix valued function \(\mathbf{D}(\boldsymbol{\theta})\) exists for all \(\boldsymbol{\theta}\). The vector \(\boldsymbol{\psi} = (\psi_1(\cdot), \dots, \psi_n(\cdot))^{\prime}\) is the \(n\)-dimensional vector of unit log-partition functions, and \(\mathcal{N}(\boldsymbol{\theta})\) is a normalizing constant proportional to \(\text{det}(\mathbf{D}(\boldsymbol{\theta}))\). The \(i\)-th element \(w_i \sim DY(w_i \vert \alpha_i, \kappa_i)\) independently, where the DY probability density function is $DY(w_{i}\vert \alpha_{i}, \kappa_{i}) \propto \text{exp} \left\{\alpha_{i}w_{i} - \kappa_{i}\psi_{i}(w_{i})\right\}$, allowing for non-identical distributions across elements. The \(n\)-dimensional vectors \(\boldsymbol{\alpha} = (\alpha_1, \dots, \alpha_n)^{\prime}\) and \(\boldsymbol{\kappa} = (\kappa_1, \dots, \kappa_n)^{\prime}\). To sample from the  marginal distribution of \(\mathbf{h}|\boldsymbol{\mu}, \mathbf{V}, \boldsymbol{\alpha}, \boldsymbol{\kappa}\), first sample \(\boldsymbol{\theta}\) from \(f(\boldsymbol{\theta})\) and then compute \(\boldsymbol{\mu} + \mathbf{V}\mathbf{D}(\boldsymbol{\theta})\mathbf{w}\).

\section{Methodology}\label{s:method}
\subsection{Frobenius Norm Matching}\label{ss:fnm}
In this section, we present the details for the covariance calibration strategy we refer to as Frobenius norm matching. By this, we mean that we estimate the covariance parameter for a linear mixed effects model by minimizing the Frobenius norm between the empirical covariance matrix generated by the GQN and the implied covariance structure of the linear model. Specifically, this procedure minimizes the sum of squared element-wise differences between the two covariance matrices.

This approach allows us to calibrate the covariance structure of a linear mixed effects model so that it closely approximates the covariance structure induced by the GQN defined in Section \ref{s:gqn}, while avoiding the computational difficulties that arise when fitting the complex nonlinear model. Let \(\widehat{\boldsymbol{\Sigma}}_{GQN}\) be the empirical covariance matrix computed from the latent process \(\mathbf{u}\) defined in Equation (\ref{eq:gqn}). To compute \(\widehat{\boldsymbol{\Sigma}}_{GQN}\), we use a Monte Carlo approximation based on \(R\) simulated replicates of the GQN model in Equation (\ref{eq:gqn}), where each replicate is simulated by sampling from the prior distributions of the model parameters. Let \( \mathbf{U}_{rep} = 
    (\mathbf{u}_{\text{rep}}^{(1)\prime}, \mathbf{u}_{\text{rep}}^{(2)\prime},\cdots,\mathbf{u}_{\text{rep}}^{(R)\prime}) \in \mathbb{R}^{nT\times R}\) where each \(\mathbf{u}_{rep}^{(r)}\) is an \(nT\)-dimensional vector whose elements are simulated according to Equation (\ref{eq:gqn}) and stacked across all locations and time points in our study. Let the mean across \(R\) replicates be \(\boldsymbol{\mu} = \frac{1}{R} \sum_{r=1}^{R} \mathbf{u}_{\text{rep}}^{(r)} \in \mathbb{R}^{nT}\). Then the space-time covariance matrix is computed with \(\frac{1}{R}(\mathbf{U}_{rep} - \boldsymbol{\mu} \mathbf{1}_R^{\prime})(\mathbf{U}_{rep} - \boldsymbol{\mu} \mathbf{1}_R^{\prime})^{\prime}\), where \(\mathbf{1}_R\) is an \(R\)-dimensional vector of ones.

We approximate \(\widehat{\boldsymbol{\Sigma}}_{GQN}\) using a reduced rank covariance structure of the form \(\mathbf{G}\boldsymbol{\Sigma}_{\eta}\mathbf{G}^{\prime}\) where \(\mathbf{G}\) is an \(n \times r\) matrix of basis functions, and \(\boldsymbol{\Sigma}_{\eta}\) is and \(r \times r\) positive define matrix. The goal is to choose a value of \(\boldsymbol{\Sigma}_{\eta}\) close to \(\widehat{\boldsymbol{\Sigma}}_{GQN}\). To specify \(\text{cov}(\mathbf{G}\boldsymbol{\eta})\) to be close to \(\boldsymbol{\Sigma}_{GQN}\) we minimize the squared Frobenius norm given by \(\vert \vert \widehat{\boldsymbol{\Sigma}}_{GQN} - \mathbf{G}\boldsymbol{\Sigma}_{\eta}\mathbf{G}^{\prime}\vert \vert^2_F\). Where \(\vert\vert \mathbf{P} \vert \vert_F^2 \equiv \text{trace}(\mathbf{P}^{\prime}\mathbf{P})\) denotes the Frobenius norm and \(\mathbf{P}\) is a generic square matrix. Minimizing the Frobenius norm results in the following closed form solution 
    \begin{align}
        \mathbf{K} &= \text{argmin}_{\boldsymbol{\Sigma}_{\eta}}(\vert \vert \mathbf{G}\boldsymbol{\Sigma}_{\eta}\mathbf{G}^{\prime} - \widehat{\boldsymbol{\Sigma}}_{GQN}\vert \vert_F^2) =(\mathbf{G}^{\prime}\mathbf{G})^{-1}\mathbf{G}^{\prime}\widehat{\boldsymbol{\Sigma}}_{GQN}\mathbf{G}(\mathbf{G}^{\prime}\mathbf{G})^{-1}
        \label{eq.fnm}
    \end{align}
The matrix \(\mathbf{K}\) serves as a value for \(\boldsymbol{\Sigma}_{\eta}\) that best aligns the covariance structure of the GQN. See Supplementary Appendix \ref{appen:fnm.proof} for proof of Equation (\ref{eq.fnm}).

\subsection{Change of Support} 
To use our Frobenius norm matching strategy, we first need to construct the spatio-temporal basis function matrix \(\mathbf{G}\). Basis functions are often defined for point-referenced data, however our motivating dataset consists of areal data, with birth rates observed at the county level. To account for the mismatch between the support of the basis functions and the data, we use a Change of Support (COS) approach  \citep{waller2004applied} which allows us to derive the spatio-temporal basis functions at the areal level by integrating the point referenced basis functions over the spatial domain of each county. Consider bisquare basis functions \citep{beaton1974fitting} defined as follows
\begin{align*}
    {g}_{j}(\mathbf{s}, t) = \left[1 - \left(\frac{\left\vert\left\vert \begin{pmatrix}
        \mathbf{s} \\ t 
    \end{pmatrix} -  \begin{pmatrix}
        \mathbf{c}_j \\ p_j 
    \end{pmatrix}\right\vert\right\vert}{\gamma} \right)^2 \right]^2 I\left(\left\vert\left\vert \begin{pmatrix}
        \mathbf{s} \\ t 
    \end{pmatrix} -  \begin{pmatrix}
        \mathbf{c}_j \\ p_j 
    \end{pmatrix}\right\vert\right\vert \leq \gamma \right), \hspace{2mm} j = 1,\dots,r
\end{align*}
where \(\mathbf{s}\) denotes the spatial location, \(t\) denotes time, \(\mathbf{c}_j\) and \(p_j\) are pre-specified spatial and temporal knot points. The knot points are constructed to be evenly spaced across the spatial and temporal domains with \(r = r_s \times r_t\) total basis functions, were \(r_s\) and \(r_t\) denote the number of spatial and temporal knots, respectively. In general, other classes of basis functions can be used.

To obtain the basis functions for areal data, we integrate the point level basis functions over the areal units (i.e. counties). Specifically, for each areal unit \(A \in \{1, \dots, 67\}\) and year \(t \in \{1, \dots, 34\}\), the integrated basis functions are defined as 
\begin{align}
    {g}_j(A, t) = \frac{1}{\vert A \vert}\int_A {g}_j(\mathbf{s}, t) \hspace{1mm} d\mathbf{s}
    \label{int.G}
\end{align}
where \({g}_j(\mathbf{s}, t)\) denotes the \(j\)-th spatio-temporal basis function and \(\vert A \vert\) is the area of the region/county \(A\). Since the integral in Equation (\ref{int.G}) cannot be computed with a closed form solution, we approximate it using Monte Carlo. That is, 
\begin{align*}
    {g}_j(A, t) \approx \frac{1}{Q} \sum_{q =1}^{Q} {g}_j(\mathbf{s}_q, t) 
\end{align*}
where \(\mathbf{s}_1, \dots, \mathbf{s}_Q\) are uniformly sampled within region \(A\) and \(Q\) is chosen to be sufficiently large to ensure an accurate approximation while remaining computationally feasible. Then the \(n \times r\) matrix \(\mathbf{G}\) has \((i,j)\)-th element \(g_j(A_i, t_i)\) where \((A_i, t_i)\) is the \(i\)-th observation.

\subsection{Exact Posterior Regression}\label{ss:epr}
{In our motivating application, we use exact posterior regression to avoid the memory limitations and convergence challenges that come with MCMC, as discussed in previous sections. EPR addresses these challenges by expanding the parameter space of a mixed effects model by introducing an additional additive term. This expanded parameter space has a posterior distribution that is GCM (see Section \ref{rev:gcm} for details) and can be sampled from directly, thereby avoiding MCMC. EPR allows for computationally efficient inference which is desirable for our application to birth rate data due to the high dimensional parameter space of a GQN.}   

We consider an \(n\)-dimensional data vector defined as \(\mathbf{z} \equiv (Z_1(s_1),\) \(\dots Z_1(s_{n_1}), Z_2(s_1),\) \(\dots, Z_2(s_{n_2}),\) \(\dots, Z_T(s_1), \dots, Z_T(s_{n_T}))^{\prime}\), where \(Z_t(\mathbf{s}_i)\) denotes the observation at location \(\mathbf{s}_i\) and time \(t\). There are \(T\) time points, with \(n_t\) spatial locations at each time, resulting in a total of \(n = \sum_{t =1}^Tn_t\) observations. This setup allows for a varying number of observed spatial locations across time points. For our motivating application we assume \(\mathbf{z}\) is Gaussian distributed with mean \(\mathbf{y}\) defined by the discrepancy model
\begin{align*}
    \mathbf{y} = \mathbf{X}\boldsymbol{\beta} + \mathbf{L}\boldsymbol{\eta} + \boldsymbol{\xi} - \boldsymbol{\tau}_{y}
\end{align*}
and variance \(\boldsymbol{\Sigma}_z = \text{diag}(\sigma_1^2, \dots \sigma_n^2)\). The matrix \(\textbf{X} \in \mathbb{R}^{n\times p}\) contains \(p\) known covariates with corresponding regression coefficients \(\boldsymbol{\beta} \in \mathbb{R}^{p}\). The vector \(\boldsymbol{\xi} \in \mathbb{R}^{n}\) is a mean-zero Gaussian random effects. The vector \(\boldsymbol{\tau}_y \in \mathbb{R}^n\) is the discrepancy term which allows for explicitly modeling the non-zero error between the true latent process \(\mathbf{y}_{true}\) and the mixed effects model \(\mathbf{y}^* = \mathbf{X}\boldsymbol{\beta} + \mathbf{L}\boldsymbol{\eta} + \boldsymbol{\xi}\). The random effect \(\mathbf{L}\boldsymbol{\eta}\) captures the spatio-temporal dependence, where \(\mathbf{L}\in \mathbb{R}^{n \times r}\), \(\boldsymbol{\eta} \in \mathbb{R}^r\) and \
\begin{align*}
    \text{Cov}(\mathbf{L}\boldsymbol{\eta}) = \mathbf{L}\mathbf{L}^{\prime} = \sigma_{\eta}^2\mathbf{G}\mathbf{K}\mathbf{G}^{\prime}.
\end{align*}
We have that \(\mathbf{L} = \mathbf{G}\mathbf{K}^{1/2}\) is the matrix square root of positive definite matrix \(\text{cov}(\mathbf{L}\boldsymbol{\eta}) \in \mathbb{R}^{n\times n}\). Recall that \(\mathbf{G}\mathbf{K}\mathbf{G}^{\prime}\) is close to the covariance structure of the GQN. The hierarchical model for EPR is defined as follows
\begin{align}
    f(\mathbf{z} \vert \boldsymbol{\beta}, \boldsymbol{\eta}, \boldsymbol{\xi}, \boldsymbol{\tau}, \boldsymbol{\theta}) &\sim N(\mathbf{z} \vert \textbf{X}\boldsymbol{\beta}+\mathbf{L}\boldsymbol{\eta}+\boldsymbol{\xi}-\boldsymbol{\tau}_{y}, \boldsymbol{\Sigma}_z) \notag \\ f(\boldsymbol{\xi}\vert\boldsymbol{\beta},\boldsymbol{\eta},\boldsymbol{\theta},\mathbf{q}) &\sim N(\boldsymbol{\xi}\vert \boldsymbol{\tau}_{\xi}, \sigma_{\xi}^2\mathbf{I}_n)\notag\\
    \boldsymbol{\beta}|\boldsymbol{\theta},\mathbf{q} &\sim N(\boldsymbol{\beta} \vert \mathbf{D}_{\beta}\boldsymbol{\tau}_{\beta}, \boldsymbol{\Sigma}_{\beta}(\boldsymbol{\theta})) \notag\\
    \boldsymbol{\eta}|\boldsymbol{\theta},\mathbf{q}  &\sim N(\boldsymbol{\eta} \vert \mathbf{D}_{\eta}\boldsymbol{\tau}_{\eta}, \boldsymbol{\Sigma}_{\eta}(\boldsymbol{\theta})) \notag\\
    f(\mathbf{q}) &= 1 \notag\\
    &f(\boldsymbol{\theta}),
    \label{ch3:epr.gqn.hier.mod}
\end{align}
where \(\boldsymbol{\theta} \in \mathbb{R}^d\) is a generic parameter with prior distribution \(f(\boldsymbol{\theta})\). The term \(\mathbf{D}_{\beta}\) is the matrix square root of positive definite matrix \(\boldsymbol{\Sigma}_{\beta}(\boldsymbol{\theta}) = \mathbf{D}_{\beta}(\boldsymbol{\theta})\mathbf{D}_{\beta}(\boldsymbol{\theta})^{\prime}\) and \(\mathbf{D}_{\eta}\) is the matrix square root of positive definite matrix \(\boldsymbol{\Sigma}_{\eta}(\boldsymbol{\theta}) = \mathbf{D}_{\eta}(\boldsymbol{\theta})\mathbf{D}_{\eta}(\boldsymbol{\theta})^{\prime}\). In practice, we set \(\mathbf{D}_{\beta} = \sigma_{\beta}\mathbf{I}_p\), \(\mathbf{D}_{\eta} = \sigma_{\eta}\mathbf{I}_r\), and \(\boldsymbol{\theta} = \{\sigma^2_1, \dots \sigma^2_n, \sigma_{\beta}^2, \sigma_{\eta}^2, \sigma_{\xi}^2\}\). The notation \(N(\mathbf{z}\vert \boldsymbol{\mu}, \boldsymbol{\Sigma})\) is the shorthand notation for the multivariate normal distribution. The discrepancy parameter \(\boldsymbol{\tau} \in \mathbb{R}^{2n + p + r}\) is given by 
\begin{align}
    \boldsymbol{\tau}(\boldsymbol{\theta}, \mathbf{q}) = \begin{pmatrix}
\boldsymbol{\tau}_{y}(\boldsymbol{\theta}, \mathbf{q})\\
\boldsymbol{\tau}_{\beta}(\boldsymbol{\theta}, \mathbf{q}) \\
 \boldsymbol{\tau}_{\eta}(\boldsymbol{\theta}, \mathbf{q}) \\
\boldsymbol{\tau}_{\xi}(\boldsymbol{\theta}, \mathbf{q})
\end{pmatrix} = -\mathbf{D}(\boldsymbol{\theta})^{-1}\mathbf{Q}\mathbf{q} = -\begin{pmatrix}
 \mathbf{I}_{n} & \bm{0}_{n,p} & \bm{0}_{n,r} & \bm{0}_{n,n}\\
 \bm{0}_{p,n} &  \mathbf{D}_{\beta}(\boldsymbol{\theta})^{-1} & \bm{0}_{p,r} & \bm{0}_{p,n}\\
 \bm{0}_{r,n} & \bm{0}_{r,p} & \mathbf{D}_{\eta}(\boldsymbol{\theta})^{-1} & \bm{0}_{r,n} \\
 \bm{0}_{n,n} & \bm{0}_{n,p} &  \bm{0}_{n,r} & \frac{1}{\sigma_{\xi}}\bm{I}_{n} \\
\end{pmatrix}
 \mathbf{Q}\mathbf{q}.
 \label{ch3:tau.eq}
\end{align}
This parameter is a function of \(\boldsymbol{\theta}\) and \(\mathbf{q}\). Typically in the literature the unknown parameter \(\mathbf{q} \equiv \mathbf{0}_{n,1}\), an \(n\)-dimensional vector of zeros. EPR instead assumes an improper  prior for \(\mathbf{q}\). The matrix \(\mathbf{Q} \in \mathbb{R}^{(2n + p +r)\times n}\) are the eigenvectors of the orthogonal complement of the matrix \(\mathbf{H} \in \mathbb{R}^{(2n + p + r) \times (n +p+r)}\) given by 
\begin{align*}
    \mathbf{H} = \begin{pmatrix}
 \mathbf{I}_{n} & \mathbf{X} & \mathbf{L}\\
 \bm{0}_{p,n} &  \mathbf{I}_p & \bm{0}_{p,r} \\
 \bm{0}_{r,n} & \bm{0}_{r,p} & \mathbf{I}_r \\
 \mathbf{I}_{n} & \bm{0}_{n,p} &  \bm{0}_{n,r} \\
\end{pmatrix}.
\end{align*}
When additional components are introduced into a model, the expanded parameter space may lead to concerns about confounding or collinearity among the parameters. The particular specification of \(\boldsymbol{\tau}\) in Equation (\ref{ch3:tau.eq}) circumvents these issues because multiplying \(\mathbf{q}\) by \(\mathbf{D}(\boldsymbol{\theta})^{-1}\mathbf{Q}\) ensures that \(\mathbf{q}\) and \((\boldsymbol{\eta}^{\prime}, \boldsymbol{\beta}^{\prime}, \boldsymbol{\eta}^{\prime})^{\prime}\) are orthogonal.

Multiplying the data, process, and parameter models in (\ref{ch3:epr.gqn.hier.mod}), and marginalizing across \(\boldsymbol{\theta}\) results in the following posterior distribution 
\begin{align}
    (\boldsymbol{\xi}',\boldsymbol{\beta}', \boldsymbol{\eta}', \mathbf{q}')' \vert \mathbf{z} &\sim \text{GCM}(\boldsymbol{\alpha}, \boldsymbol{\kappa}, \mathbf{0}_{2n + p + r, 1}, \mathbf{V}, \pi, \mathbf{D}; \boldsymbol{\psi}),
    \label{ch3:gcm.post}
\end{align}
where \(\boldsymbol{\alpha} = (\mathbf{z}^{\prime}\mathbf{D}_{\sigma}, \bm{0}_{1, n + p + r})^{\prime}\) and \(\boldsymbol{\kappa} = (\frac{1}{2}\mathbf{1}_{1,n}\mathbf{D}_{\sigma}^{\prime}, \frac{1}{2}\mathbf{1}_{1, n + p + r})^{\prime}\), \(\mathbf{D}_{\sigma} = \text{diag}(\frac{1}{\sigma_i^2}, i = 1, \dots, n)\), \(\mathbf{V}^{-1} = (\mathbf{H}, \mathbf{Q})\), and \(\mathbf{\psi}(\cdot) = (\cdot)^2\). See Section \ref{rev:gcm} for a review of the GCM distribution. Independent replicates from this posterior distribution can be sampled directly using the following equation
\begin{align}
    \begin{pmatrix}
        \boldsymbol{\xi}_{rep} \\
        \boldsymbol{\beta}_{rep} \\
        \boldsymbol{\eta}_{rep}
    \end{pmatrix}  &= (\mathbf{H}'\mathbf{H})^{-1}\mathbf{H}' \begin{pmatrix}
        \mathbf{y}_{rep} \\
        \mathbf{w}_{\beta} \\
        \mathbf{w}_{\eta} \\
        \mathbf{w}_{\xi}
    \end{pmatrix}, \label{equation.comp}
\end{align}
\noindent where the subscript ``rep'' denotes a posterior replicate of \(\boldsymbol{\xi}\), \(\boldsymbol{\beta}\), and \(\boldsymbol{\eta}\), the term $\mathbf{y}_{rep}$ consists of independent DY random variables where the shape and rate parameters are \(\boldsymbol{\alpha}\) and \(\boldsymbol{\kappa}\). The terms \(\mathbf{w}_{\beta}\), \(\mathbf{w}_{\eta}\), and \(\mathbf{w}_{\xi}\) are obtained by first sampling $\bm{\theta}^{*}$ from its respective prior distribution and then \(\mathbf{w}_{\xi}\) is sampled from a mean zero normal distribution with covariance $\sigma_{\xi}^2\mathbf{I}_n$, \(\mathbf{w}_{\beta}\) is sampled from a mean zero normal distribution with covariance $\textbf{D}_{\beta}(\bm{\theta}^{*})\textbf{D}_{\beta}(\bm{\theta}^{*})^{\prime}$, and \(\mathbf{w}_{\eta}\) is sampled from a mean zero normal distribution with covariance $\textbf{D}_{\eta}(\bm{\theta}^{*})\textbf{D}_{\eta}(\bm{\theta}^{*})^{\prime}$. See Supplementary Appendix \ref{appen:epr.proof} for proof of Equations (\ref{ch3:gcm.post}) and (\ref{equation.comp}) which
follow from Theorem 3.1 and 3.2 in \citet{bradley2024generating}.

\section{Simulation Study}\label{s:simstudy}
\subsection{Comparison to GQN}\label{ss:simstudy1}
We first simulate data from a GQN model with a relatively small parameter space to allow for direct comparison between the GQN fitted via MCMC and the Frobenius norm matching (FNM) strategy implemented with EPR. We simulate data on a two-dimensional spatial grid of \(n_t = 10 \times 10\) locations over \(T = 3, 7, 11, 15\) time points. The observed data \(\mathbf{Z}_t(\mathbf{s}_i)\) is simulated from a Gaussian distribution with mean \({Y}_t(\mathbf{s}_i) = \mathbf{x}_t(\mathbf{s}_i)'\boldsymbol{\beta} + {U}_t(\mathbf{s}_i)\) and variance \(\sigma_z^2 = 0.3\) where \(\mathbf{x}_t\) is defined to be the \(n\)-dimensional vector of ones for an intercept only model with a true regression coefficient \(\beta = 1\). The latent process \(U_t(\mathbf{s}_i)\) is simulated from Equation (\ref{eq:gqn}) where the coefficients \(a_{ij}\) and \(b_{i,kl}\) are defined as follows: 
\begin{align*}
a_{ij} =
\begin{cases}
\delta, & \text{if } \mathbf{s}_j \in \mathcal{N}(\mathbf{s}_i) \\
0, & \text{otherwise}
\end{cases},\hspace{2mm}
b_{i,kl} =
\begin{cases}
\nu, & \text{if } k = l = i \\
0, & \text{otherwise}
\end{cases}
\end{align*}
where \(\delta = 0.14\), \(\nu = 1\), and \(\mathcal{N}(\mathbf{s}_i)\) is defined to be the set of locations that are directly adjacent to \(\mathbf{s}_i\), specifically the locations directly above, below, to the left, and to the right of \(\mathbf{s}_i\) and also including \(\mathbf{s_i}\) itself. The nonlinear function \(g(\cdot)\) is defined in Section \ref{s:gqn}, with parameters \(\gamma_0 = 0.05\) and \(\gamma_1 = 10\). We assume \(\boldsymbol{\eta}_t\sim \text{MVN}(\mathbf{0}, \boldsymbol{\Sigma}_{\eta})\) and \(\mathbf{U}_{0} \sim \text{MVN}(\mathbf{0}, \boldsymbol{\Sigma}_0)\) where \(\boldsymbol{\Sigma}_{\eta} = \sigma_{\eta}^2 \text{exp}\left(-\frac{\mathbf{D}}{\phi} \right)\), \(\sigma^2_{\eta} = 0.5\), \(\boldsymbol{\Sigma}_0 = \sigma_0^2 \text{exp}\left(-\frac{\mathbf{D}}{\phi} \right)\), \(\sigma^2_0 = 1\), \(\phi=10\), and \(\mathbf{D}\) is the \(n_t \times n_t\) matrix of pairwise Euclidean distances between spatial locations. This model contains a total of \(10\) unknown parameters to sample via MCMC. Again, this simplified model specification was intentionally chosen to allow for direct implementation of the GQN model using MCMC while keeping the computation time manageable.

\begin{table}[H]
\centering
\begin{tabular}{llccccc}
\toprule
 & Approach & Forecast & MSE & MSPE & CRPS & CPU Time \\ 
\midrule
\multicolumn{7}{l}{\textbf{Scenario:} $T=3$} \\ 
& GQN - MCMC &  0.147 & 0.096 & 0.064 & 0.282 & 30.12 hrs \\
& FNM - EPR  &0.139& 0.100 & 0.096 & 0.439 & 9.20 sec \\
\addlinespace
\multicolumn{7}{l}{\textbf{Scenario:} $T=7$} \\
& GQN - MCMC & 0.316 & 0.017 & 0.071 & 0.3267 & 4.28 days \\
& FNM - EPR  &0.270 & 0.003 & 0.105 & 0.394 & 1.40 mins \\
\addlinespace
\multicolumn{7}{l}{\textbf{Scenario:} $T=11$} \\
& GQN - MCMC & 0.494 & 0.003 & 0.067 & 0.405 & 7.46 days \\
& FNM - EPR  &  0.489 & 0.010 & 0.106 & 0.532 & 4.84 mins \\
\addlinespace
\multicolumn{7}{l}{\textbf{Scenario:} $T=15$} \\
& GQN - MCMC &0.857 & 0.010 & 0.068 & 0.549 & 12.40 days \\
& FNM - EPR & 0.741 & 0.012 & 0.109 & 0.589 & 28.66 mins \\
\bottomrule
\end{tabular}
\caption{Results comparing the GQN implemented with MCMC compared to the Frobenius norm matching approach implemented with EPR. Each model is used to Forecast the \(T + 1\) time point.}
\label{fig:mcmc.compare}
\end{table}

\begin{figure}[H]
    \centering
    
    \begin{subfigure}{0.3\textwidth}
        \includegraphics[width=\linewidth]{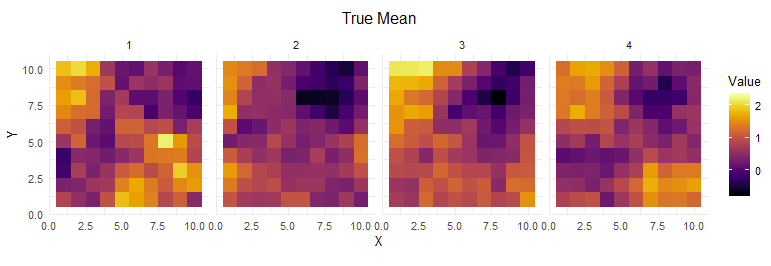}
    \end{subfigure}
    \hfill
    \begin{subfigure}{0.3\textwidth}
        \includegraphics[width=\linewidth]{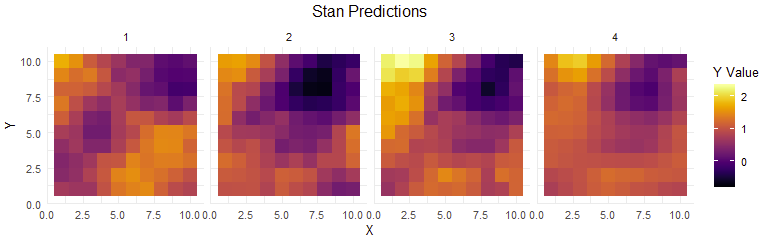}
    \end{subfigure}
    \hfill
    \begin{subfigure}{0.3\textwidth}
        \includegraphics[width=\linewidth]{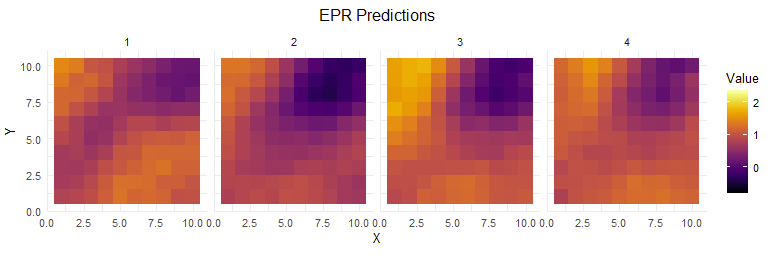}
    \end{subfigure}
    
    \par\bigskip
    \begin{subfigure}{0.3\textwidth}
        \includegraphics[width=\linewidth]{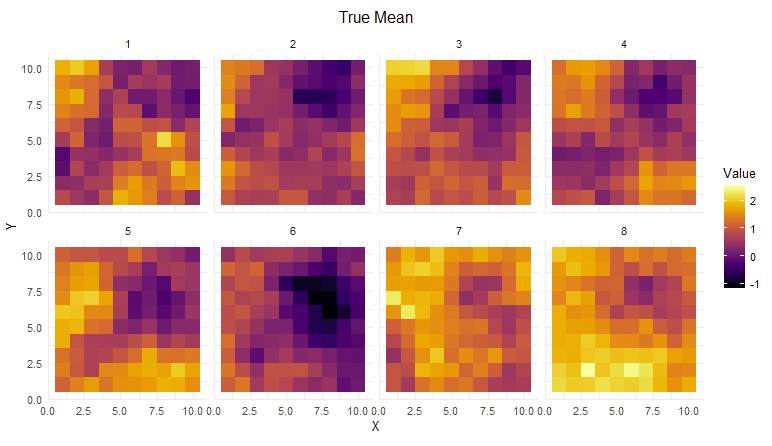}
    \end{subfigure}
    \hfill
    \begin{subfigure}{0.3\textwidth}
        \includegraphics[width=\linewidth]{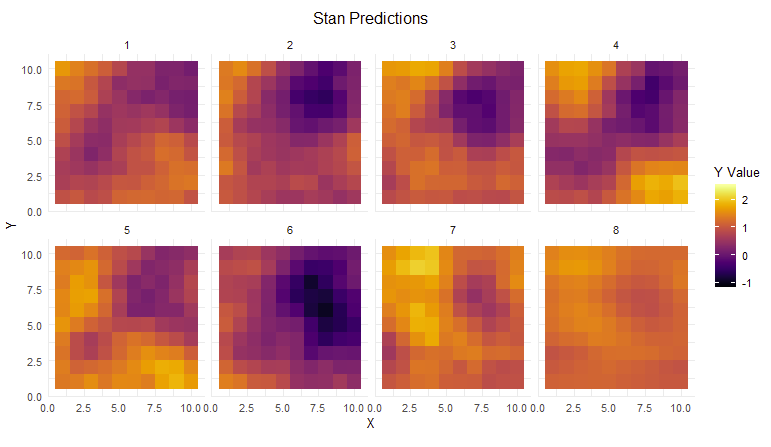}
    \end{subfigure}
    \hfill
    \begin{subfigure}{0.3\textwidth}
        \includegraphics[width=\linewidth]{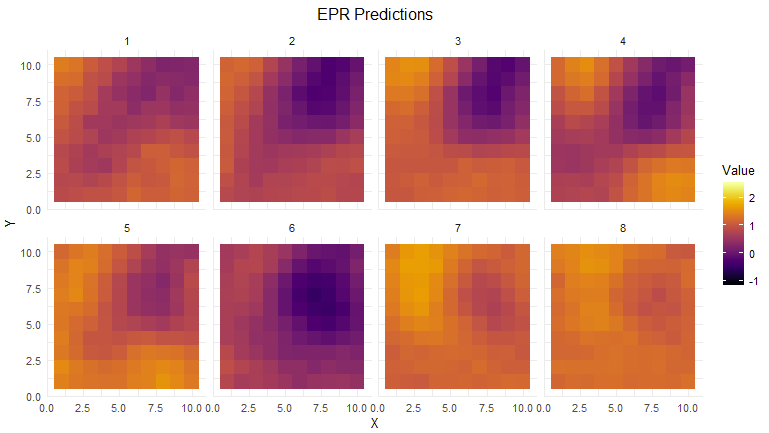}
    \end{subfigure}
    
    \par\bigskip
    \begin{subfigure}{0.3\textwidth}
        \includegraphics[width=\linewidth]{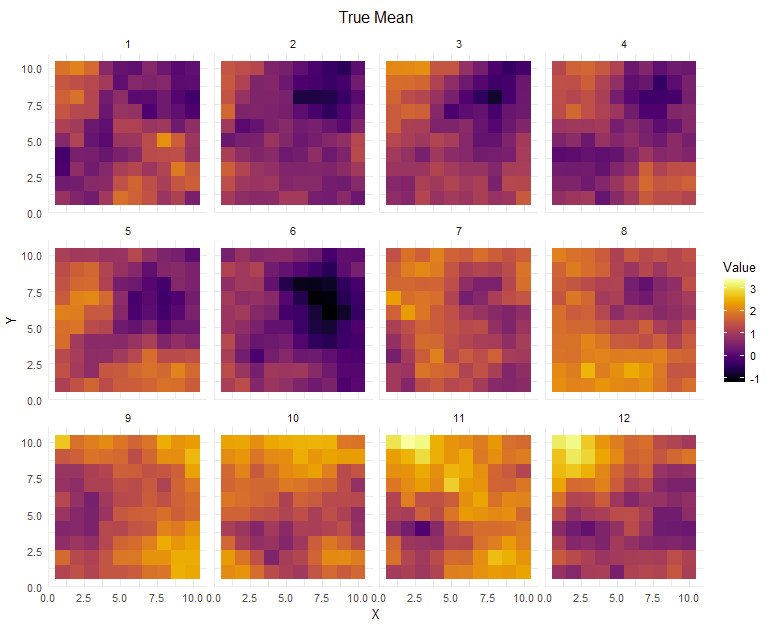}
    \end{subfigure}
    \hfill
    \begin{subfigure}{0.3\textwidth}
        \includegraphics[width=\linewidth]{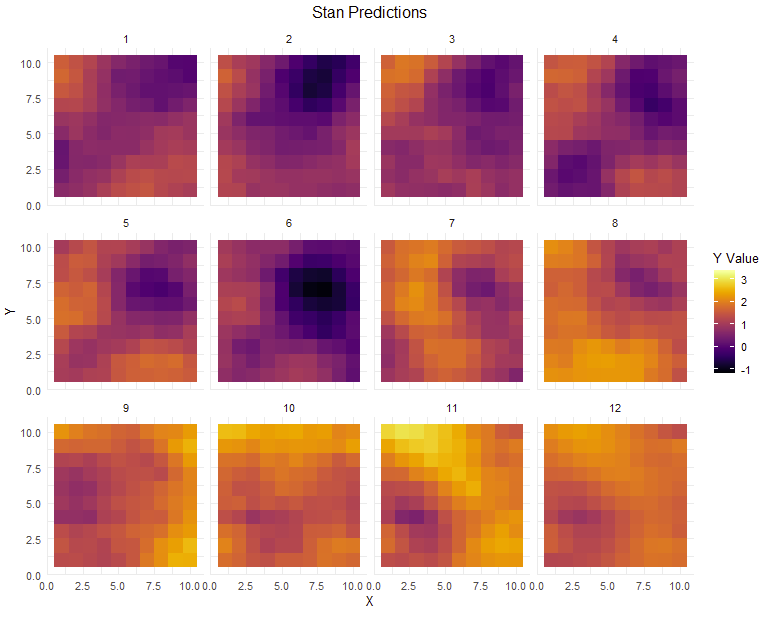}
    \end{subfigure}
    \hfill
    \begin{subfigure}{0.3\textwidth}
        \includegraphics[width=\linewidth]{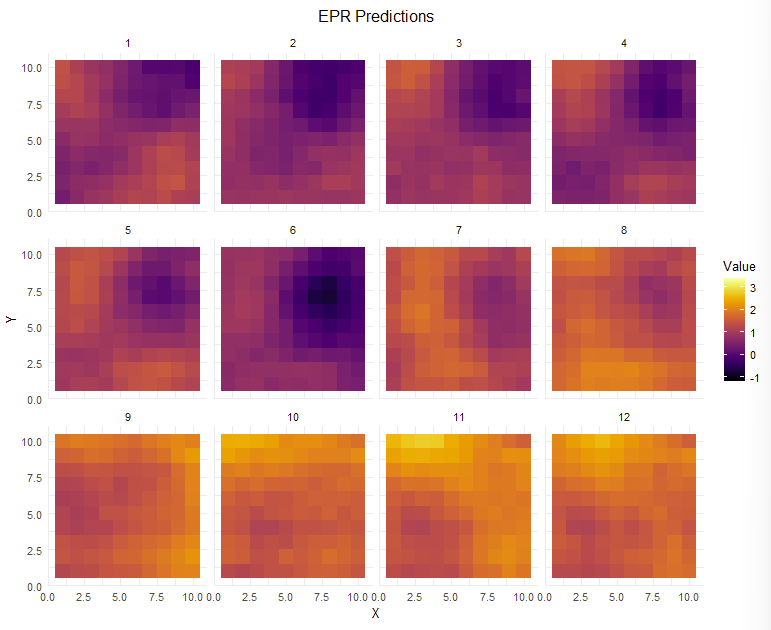}
    \end{subfigure}
    
    \par\bigskip
    \begin{subfigure}{0.3\textwidth}
        \includegraphics[width=\linewidth]{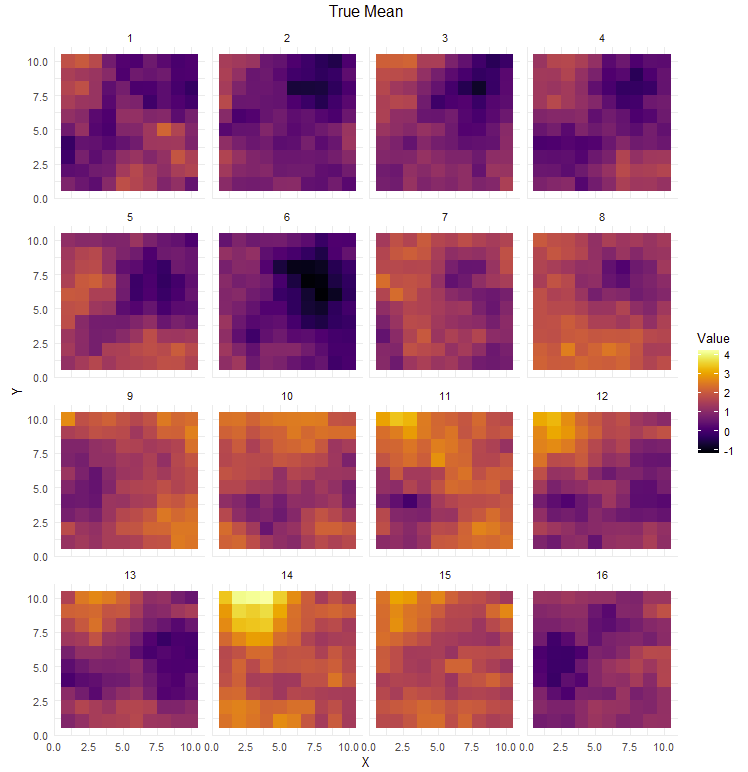}
    \end{subfigure}
    \hfill
    \begin{subfigure}{0.3\textwidth}
        \includegraphics[width=\linewidth]{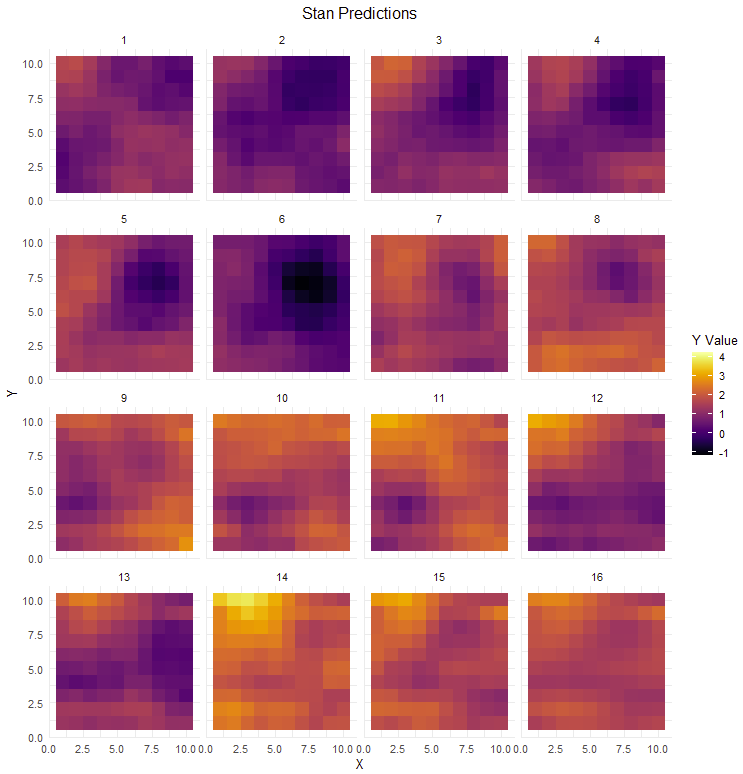}
    \end{subfigure}
    \hfill
    \begin{subfigure}{0.3\textwidth}
        \includegraphics[width=\linewidth]{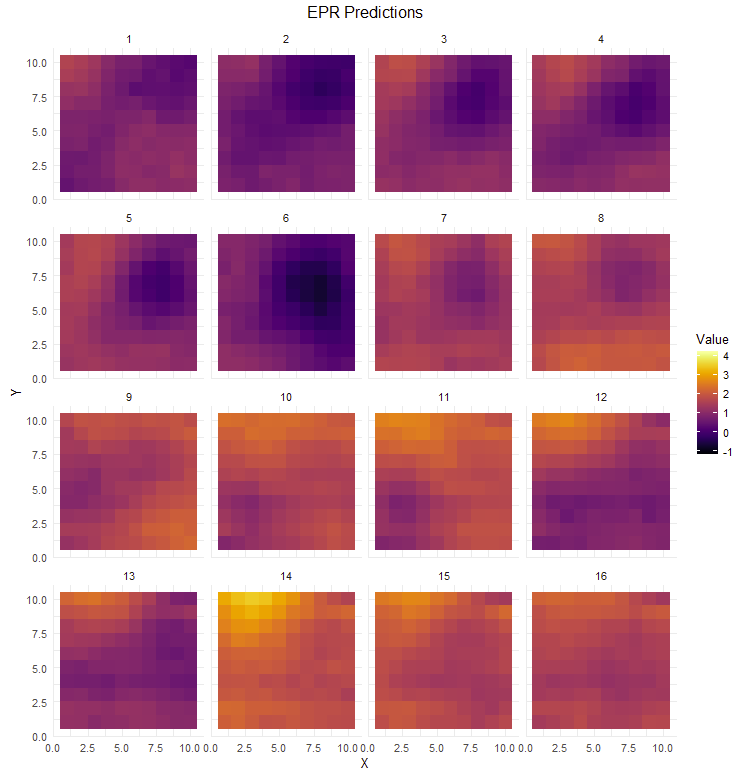}
    \end{subfigure}
    
    \caption{The first column is the true process simulated from the GQN, the second column is the posterior mean from the GQN model implemented with Stan, and the third column is the posterior mean from the linear model calibrated toward a GQN using the Frobenius norm matching strategy implemented with EPR. Each row is the comparison with \(T = 3, 7, 11, 15\) observed time points, respectively. The \(T+1\) time point in each example was foretasted. From top to bottom, the Frobenius norm matching strategy was implemented with 388, 776, 1164, and 1552 spatio-temporal basis functions, respectively.} 
    \label{fig:mcmc_comp}
\end{figure}

The true GQN implemented via MCMC is denoted GQN-MCMC and the linear model calibrated toward a GQN using Frobenius norm matching implemented via EPR is denoted FNM-EPR. Table \ref{fig:mcmc.compare} presents predictive and computational evaluation metrics across various time points using each of the two approaches, and visual comparison of the two approaches is presented in Figure \ref{fig:mcmc_comp}. The FNM-EPR approach yields more accurate forecasts compared to GQN-MCMC. The FNM-EPR approach tends to produce smoother forecasts, whereas the GQN-MCMC forecasts may overfit the previous time point, potentially contributing to larger forecast errors. In most cases, the mean squared estimation error (MSE) of the true regression coefficient \(\beta\) is similar for both approaches. GQN-MCMC achieves better performance in terms of spatial smoothing measured by the mean squared prediction error (MSPE) and also lower values of continuous rank probability score (CRPS) compared to FNM-EPR since it fits the correctly specified GQN model. The FNM-EPR fits a linear model that has been calibrated toward the GQN using Frobenius norm matching, which leads to over smoothing which can be seen in the slightly larger MSPE and CRPS. While GQN-MCMC outperforms FNM-EPR in terms of smoothing, the differences are practically smaller. 

FNM-EPR has significant computational advantages over GQN-MCMC. GQN-MCMC takes days to run whereas the FNM-EPR model has a central processing unit (CPU) time that is under an hour even in largest scenario for \(T\) that was considered. Although this simulation involves a relatively small parameter space with only 10 unknown parameters, the computational burden of GQN-MCMC will become worse with larger datasets or more complex model specifications. FNM-EPR provides a computationally efficient alternative by calibrating towards the true GQN and then fitting a linear mixed effects model.

\subsection{Frobenius Norm Matching for Larger Parameter Space}\label{ss:complex.sim}
Next, we simulate data from a GQN model with a more complicated specification than the one used in Section \ref{ss:simstudy1}. In this simulation, we compare the Frobnius norm matching strategy implemented with EPR and MCMC. That is, we are calibrating the covariance structure of a linear model to that of the GQN and then we fit the linear model with EPR and then with MCMC implemented via R Stan. Note that in this section, GQN-MCMC is not computationally feasible as the parameter space is too large. We simulate data on a two-dimensional spatial grid of \(n_t = 10 \times 10\) locations over \(T = 14\) time points. The observed data \(Z_t(\mathbf{s}_i)\) is simulated from a Gaussian distribution with mean \({Y}_t(\mathbf{s}_i) = \mathbf{x}_t(\mathbf{s}_i)'\boldsymbol{\beta} + {U}_t(\mathbf{s}_i)\) and variance \(\sigma_z^2 = 0.03\) where \(\mathbf{x}_t\) is defined to be the \(n\)-dimensional vector of ones for an intercept only model with a true regression coefficient \(\beta = 1\). where the coefficients \(a_{ij}\) and \(b_{i,kl}\) are defined as follows: 
\begin{align}\label{sim.specification}
a_{ij} &=
\begin{cases}
\delta_1, & \text{if } d(\mathbf{s}_i, \mathbf{s_j}) = 0 \\
\delta_2* \text{Bernoulli}(p_a), & \text{if } 0 <d(\mathbf{s}_i, \mathbf{s_j}) < \rho \\
0, & \text{otherwise}
\end{cases}, \hspace{2mm} 
b_{i,kl} =
\begin{cases}
\nu*\text{Bernoulli}(p_b), & \text{if } d(\mathbf{s}_i, \mathbf{s}_k),\ d(\mathbf{s}_i, \mathbf{s}_\ell) < \rho\\
0, & \text{otherwise}
\end{cases}
\end{align}
where \(\delta_1 = 0.005\), \(\delta_2 = 0.007\), \(\nu = 0.028\), \(p_a = p_b =0.9\), \(\rho = 3\), and \(d(\mathbf{s}_i, \mathbf{s}_j) \equiv \vert \vert \mathbf{s}_i - \mathbf{s}_j \vert \vert\),. The nonlinear function \(g(\cdot)\) is defined in Section \ref{s:gqn}, with parameters \(\gamma_0 = 0.01\) and \(\gamma_1 = 25\). We assume \(\boldsymbol{\eta}_t\sim \text{MVN}(\mathbf{0}, \boldsymbol{\Sigma}_{\eta})\) and \(\mathbf{U}_{0} \sim \text{MVN}(\mathbf{0}, \boldsymbol{\Sigma}_0)\) where \(\boldsymbol{\Sigma}_{\eta} = \sigma_{\eta}^2 \text{exp}\left(-\frac{\mathbf{D}}{\phi_{\eta}} \right)\), \(\sigma^2_{\eta} = 0.2\), \(\boldsymbol{\Sigma}_0 = \sigma_0^2 \text{exp}\left(-\frac{\mathbf{D}}{\phi_{0}} \right)\), \(\sigma^2_0 = 0.4\), \(\phi_{\eta}=15\), \(\phi_0 = 20\) and \(\mathbf{D}\) is the \(n \times n \) matrix of pairwise Euclidean distances between spatial locations.
\begin{table}[H]
\centering
\begin{tabular}{lcccccc}
\toprule
&Approach & Forecast & MSPE & MSE & CRPS & CPU Time (sec) \\ 
\midrule 
\multicolumn{7}{c}{\textbf{Scenario:} $r=225$} \\ \midrule
&FNM-EPR  &  0.359 & 0.079 & 0.016   & 0.170  & 11.8 \\
&\phantom{EPR}  & (0.000, 0.947) & (0.027, 0.131) & (0.000, 0.032) & (0.156, 0.184) & (11.3, 12.3) \\
&FNM-MCMC  & 1.271  & 0.066  & 0.706  & 0.152  & 849.8  \\
&\phantom{MCMC} & (0.919, 1.623) & (0.060, 0.072) & (0.308, 1.104) & (0.144, 0.160) & (798.7, 900.9) \\
\midrule
\multicolumn{7}{c}{\textbf{Scenario:} $r=450$} \\ \midrule
&FNM-EPR  & 0.287 &  0.061 & 0.016  & 0.161  & 44.9  \\
&\phantom{EPR}& (0.177, 0.397) & (0.047, 0.075) & (0.002, 0.030) & (0.149, 0.173) & (44.7, 45.1) \\
&FNM-MCMC  & 1.477  & 0.057 & 0.805  & 0.135  & 1749.0  \\
&\phantom{MCMC}  & (0.963, 1.991) & (0.051, 0.063) & (0.343, 1.267) & (0.127, 0.143) & (1408.9, 2089.1) \\
\midrule
\multicolumn{7}{c}{\textbf{Scenario:} $r=675$} \\ \midrule
&FNM-EPR & 0.212  & 0.056 & 0.013  & 0.159  & 104.5  \\
&\phantom{EPR}  & (0.134, 0.290) & (0.034, 0.078) & (0.000, 0.027) & (0.145, 0.173) & (102.6, 106.4) \\
&FNM-MCMC  & 1.575  & 0.062  & 1.054  & 0.140  & 3216.0 \\
&\phantom{MCMC} & (0.921, 2.229) & (0.052, 0.072) & (0.572, 1.536) & (0.130, 0.150) & (3138.1, 3293.9) \\
\midrule
\multicolumn{7}{c}{\textbf{Scenario:} $r=900$} \\ \midrule
&FNM-EPR  & 0.172 & 0.054  & 0.014  & 0.158  & 185.6 \\
&\phantom{EPR}  & (0.122, 0.222) & (0.048, 0.060) & (0.002, 0.026) & (0.142, 0.174) & (171.3, 199.9) \\
&FNM-MCMC  & 1.623  & 0.096 & 0.896 & 0.173  & 5578.1 \\
&\phantom{MCMC} & (0.811, 2.435) & (0.078, 0.114) & (0.424, 1.368) & (0.155, 0.191) & (4619.6, 6536.6) \\
\bottomrule
\end{tabular}
\caption{Results comparing the Frobenius norm matching strategy implemented with EPR and MCMC. Various number of basis functions were tested for the covariance calibration. Each approach is used to forecast the \(T + 1\) time point. The results presented are averages over 50 independent replications. Below each row is the corresponding interval \((\text{mean} \pm 2 \times \text{sd})\).}
\label{fig:fnm.compare.gauss}
\end{table}

\begin{figure}[H]
    \centering
    
    \begin{subfigure}{0.3\textwidth}
        \includegraphics[width=\linewidth]{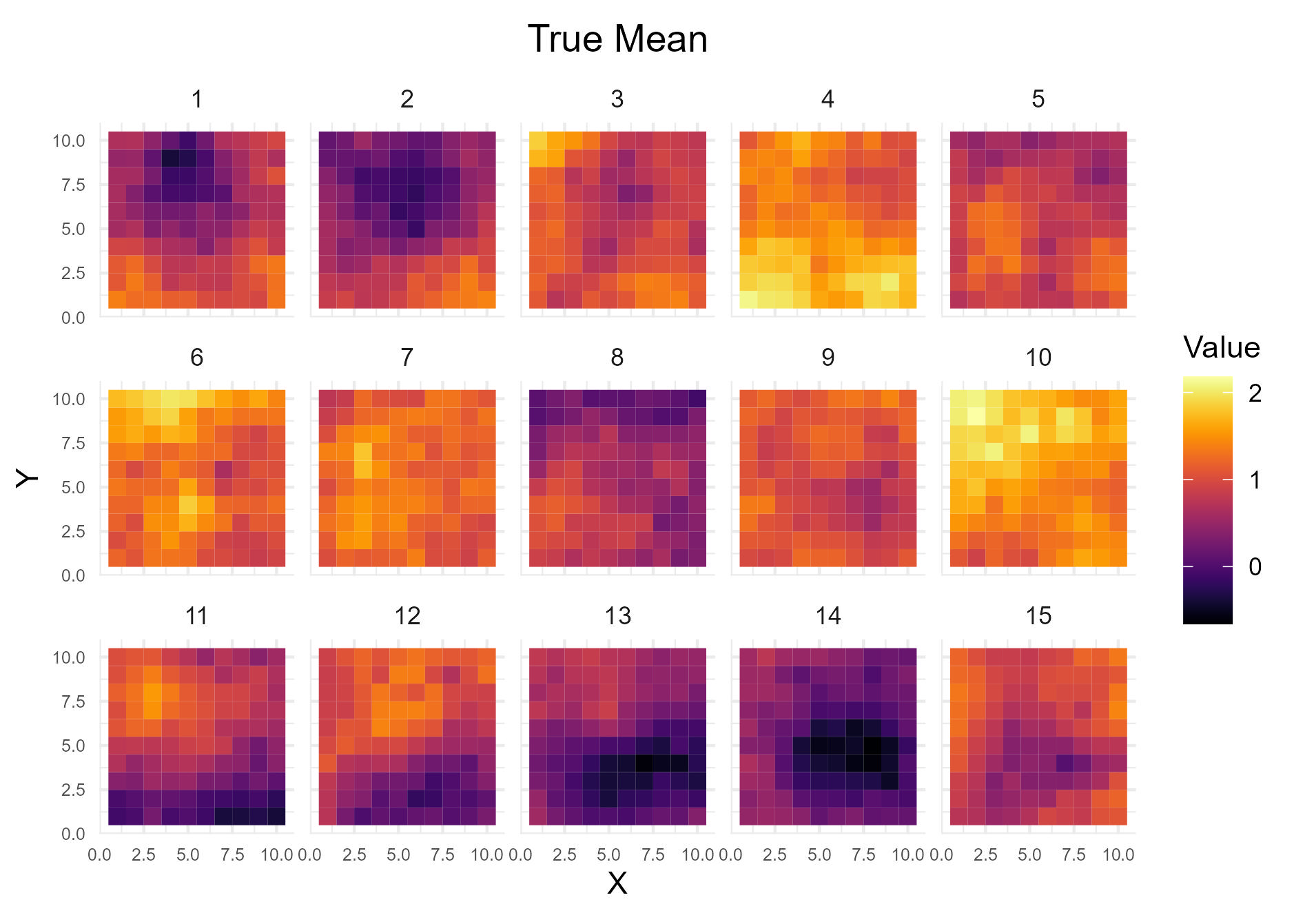}
    \end{subfigure}
    \hfill
    \begin{subfigure}{0.3\textwidth}
        \includegraphics[width=\linewidth]{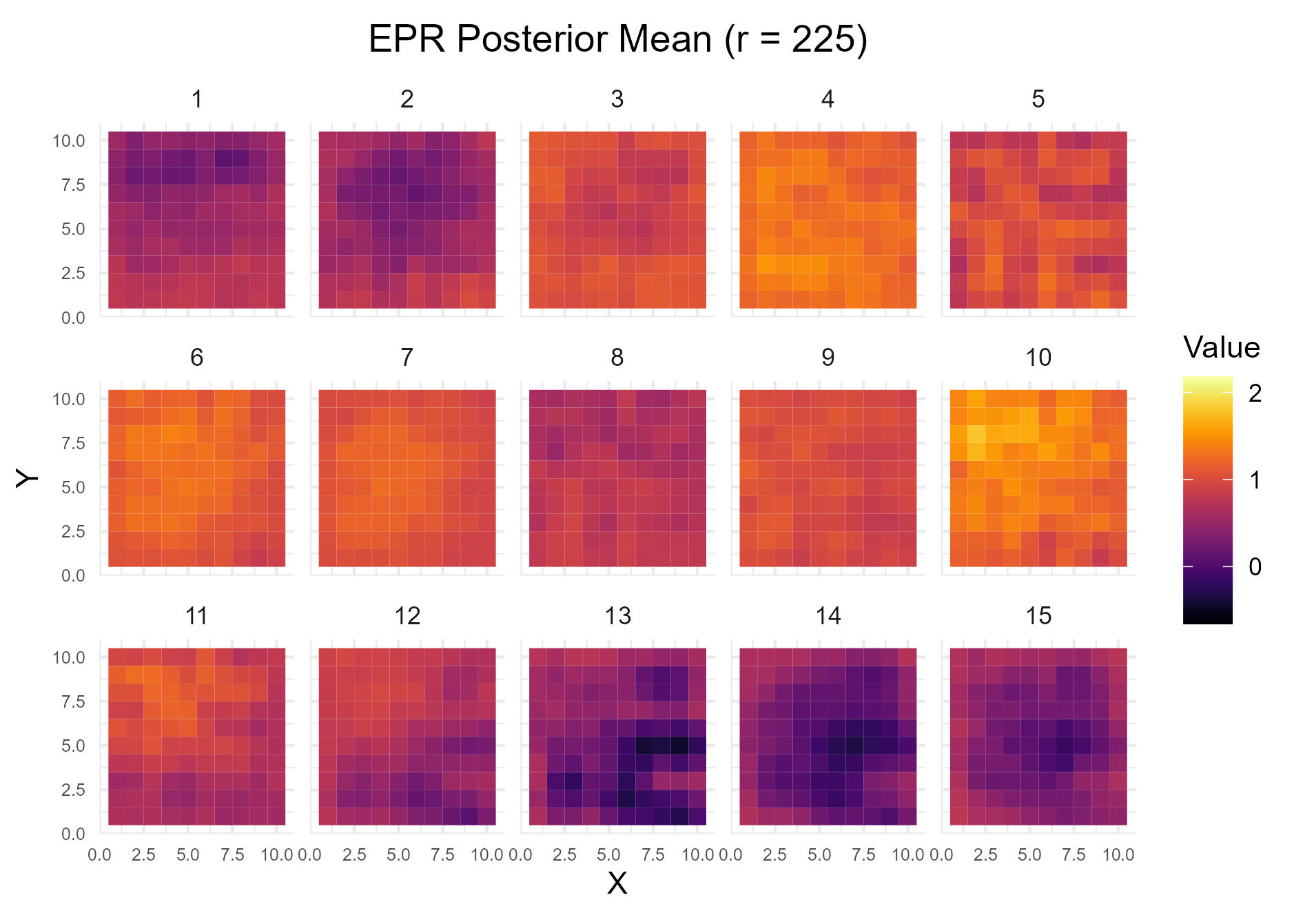}
    \end{subfigure}
    \hfill
    \begin{subfigure}{0.3\textwidth}
        \includegraphics[width=\linewidth]{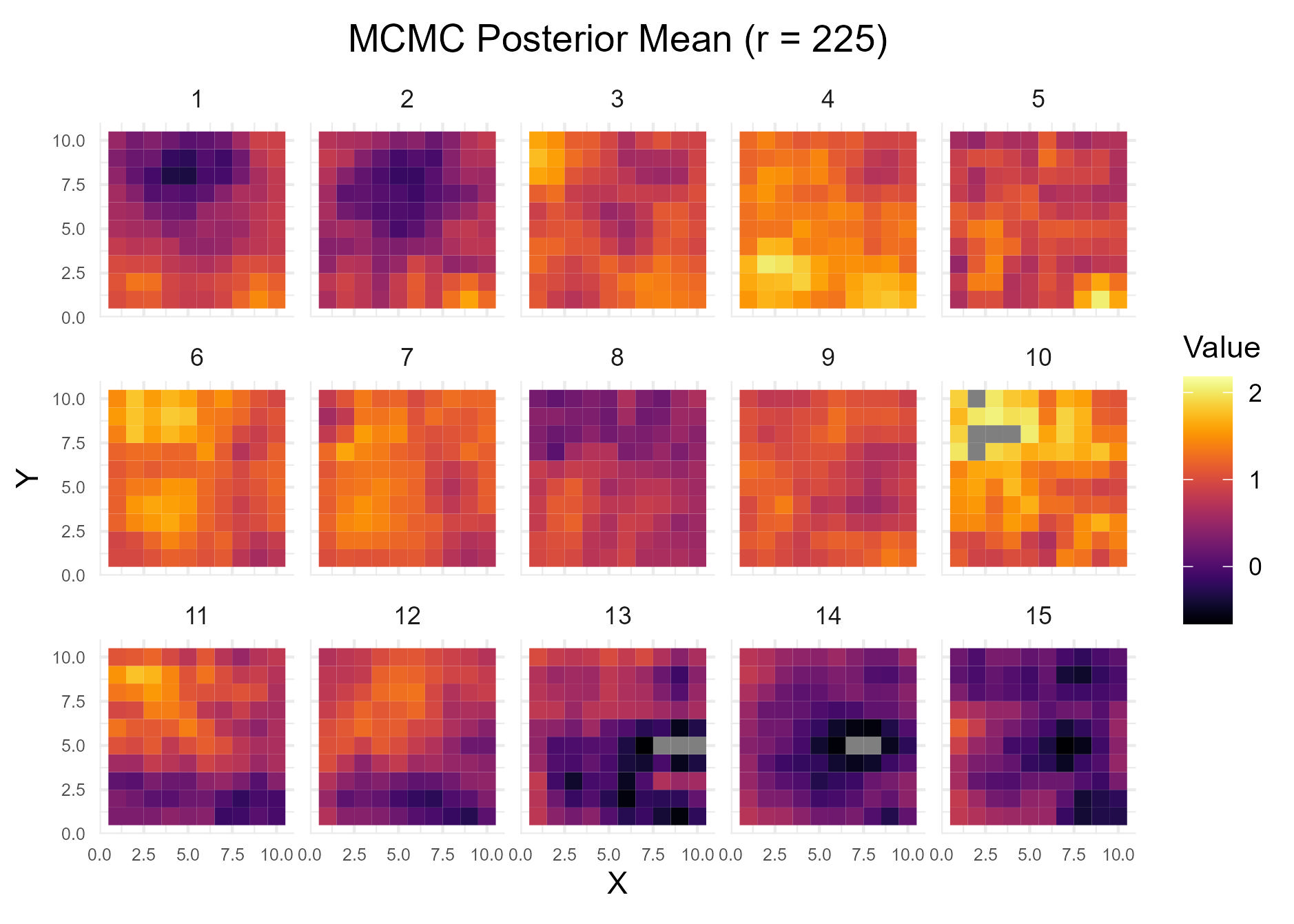}
    \end{subfigure}
    
    \par\bigskip
    \begin{subfigure}{0.3\textwidth}
        \includegraphics[width=\linewidth]{true.data.g.jpeg}
    \end{subfigure}
    \hfill
    \begin{subfigure}{0.3\textwidth}
        \includegraphics[width=\linewidth]{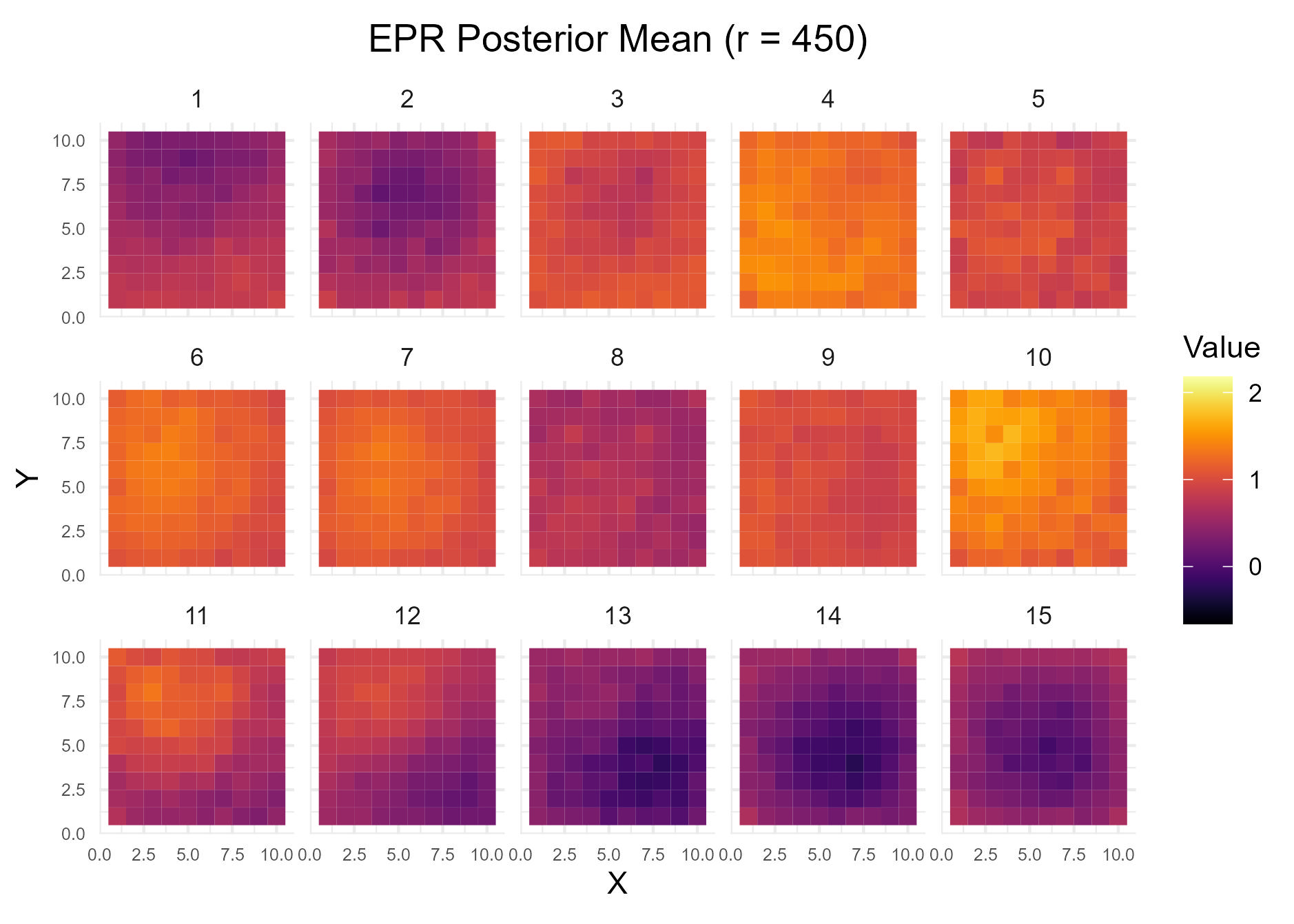}
    \end{subfigure}
    \hfill
    \begin{subfigure}{0.3\textwidth}
        \includegraphics[width=\linewidth]{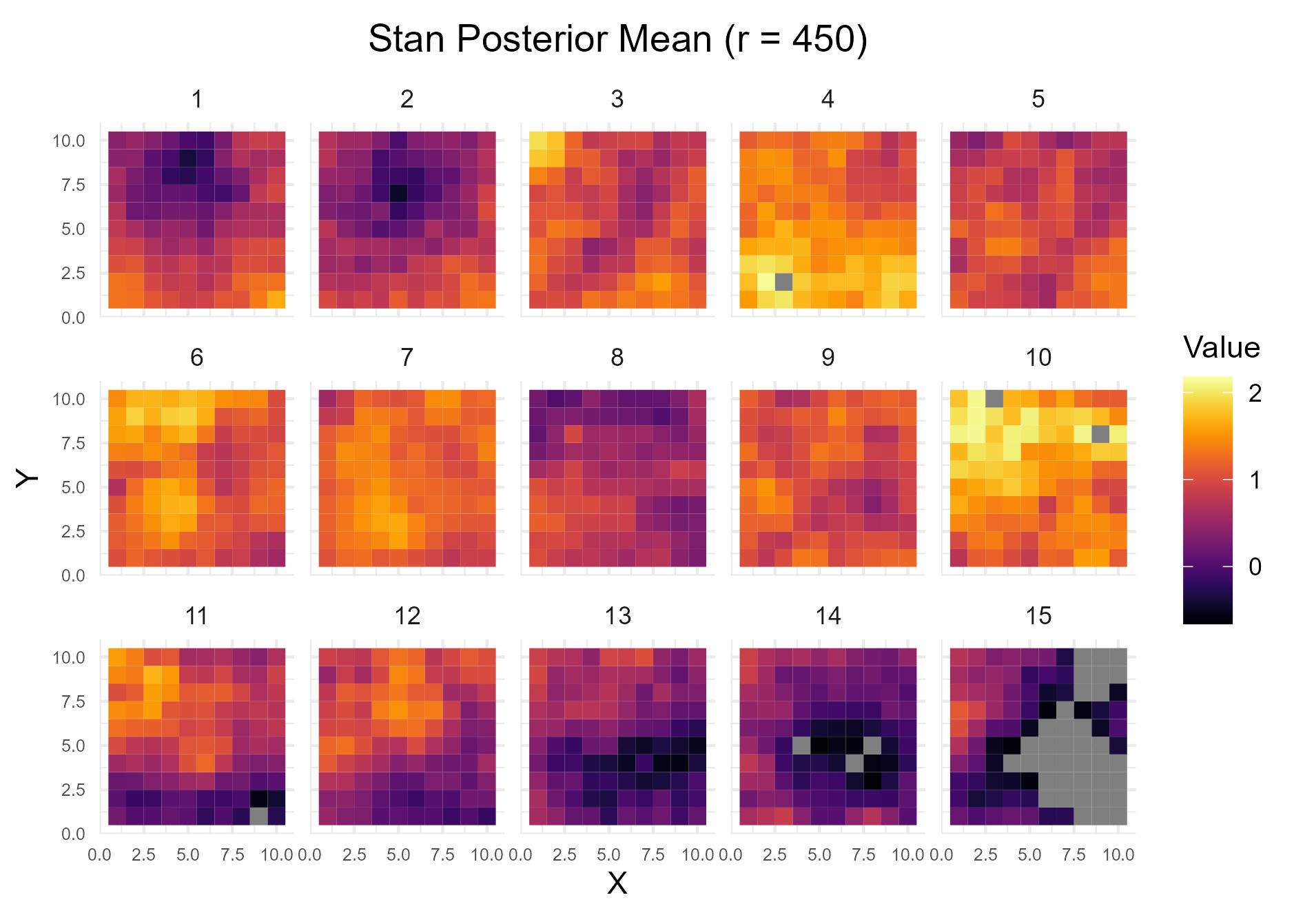}
    \end{subfigure}
    
    \par\bigskip
    \begin{subfigure}{0.3\textwidth}
        \includegraphics[width=\linewidth]{true.data.g.jpeg}
    \end{subfigure}
    \hfill
    \begin{subfigure}{0.3\textwidth}
        \includegraphics[width=\linewidth]{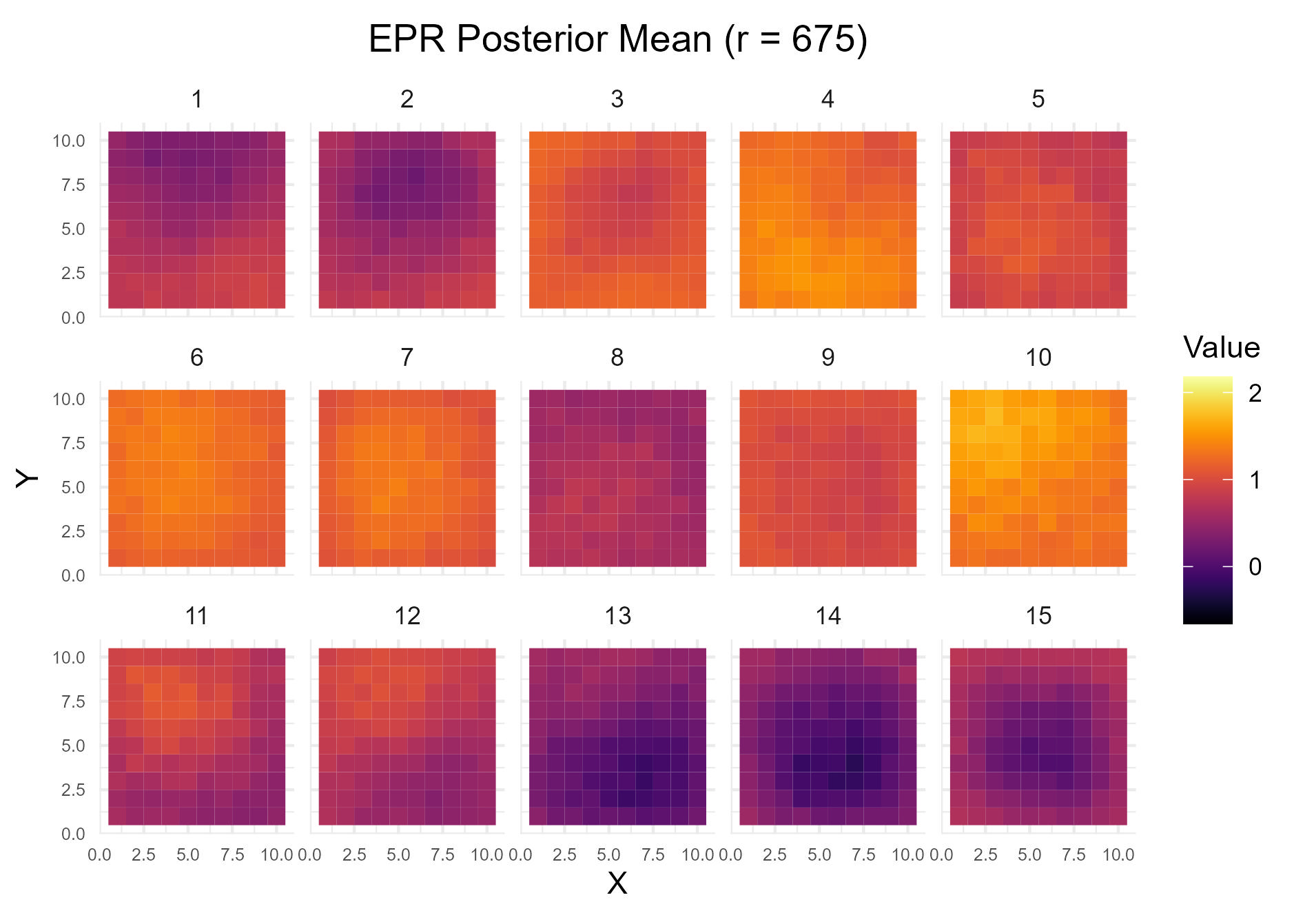}
    \end{subfigure}
    \hfill
    \begin{subfigure}{0.3\textwidth}
        \includegraphics[width=\linewidth]{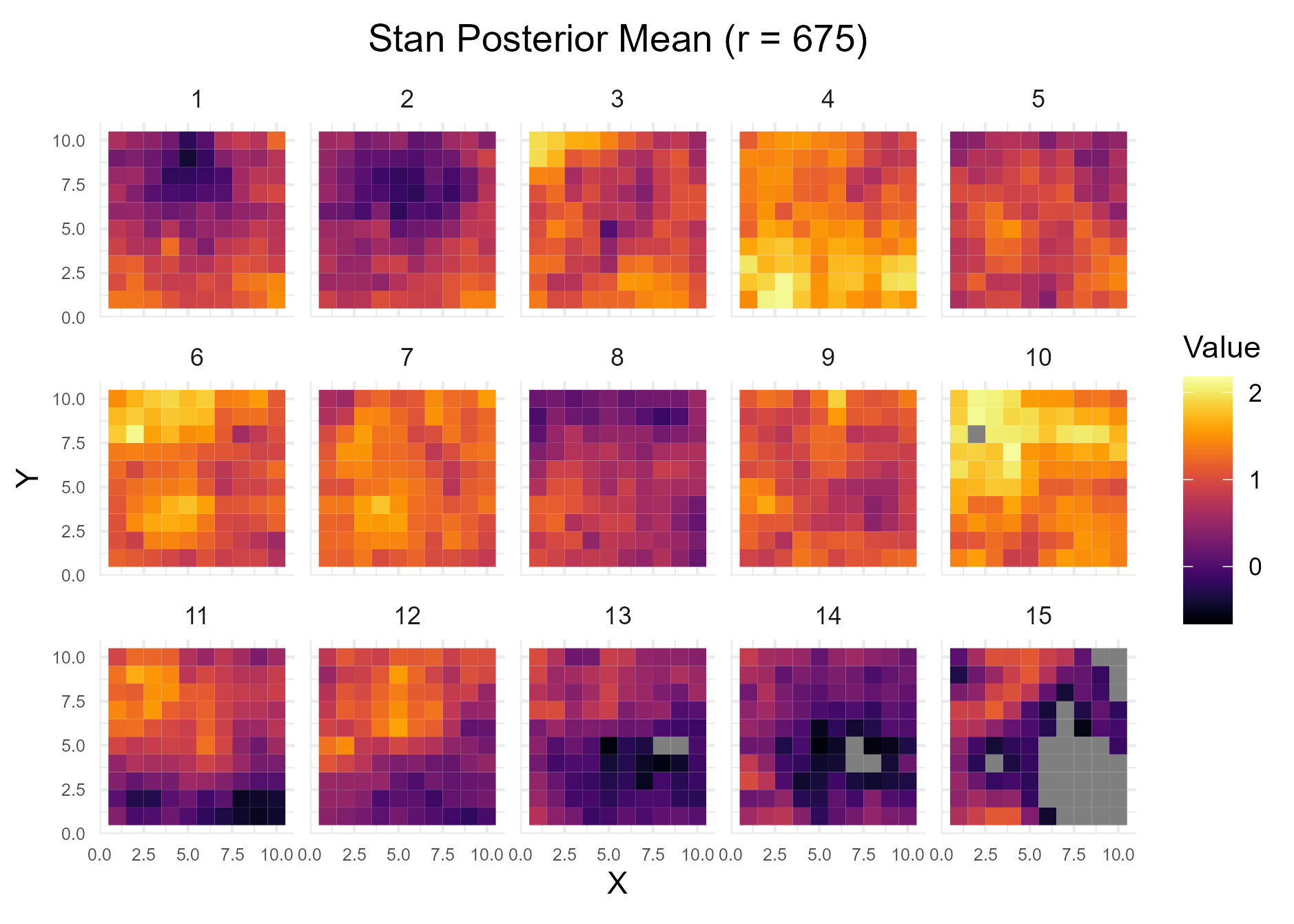}
    \end{subfigure}
    
    \par\bigskip
    \begin{subfigure}{0.3\textwidth}
        \includegraphics[width=\linewidth]{true.data.g.jpeg}
    \end{subfigure}
    \hfill
    \begin{subfigure}{0.3\textwidth}
        \includegraphics[width=\linewidth]{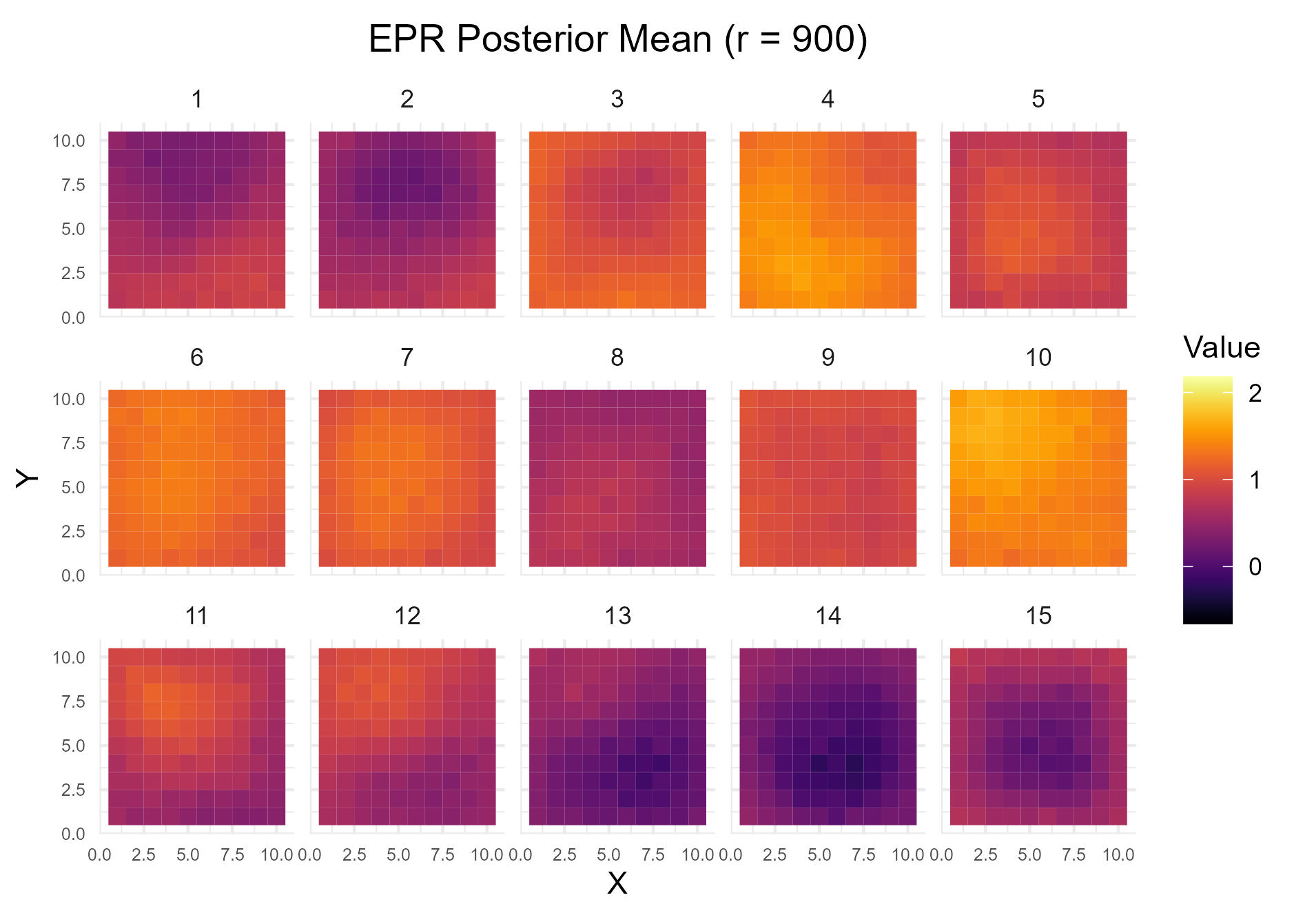}
    \end{subfigure}
    \hfill
    \begin{subfigure}{0.3\textwidth}
        \includegraphics[width=\linewidth]{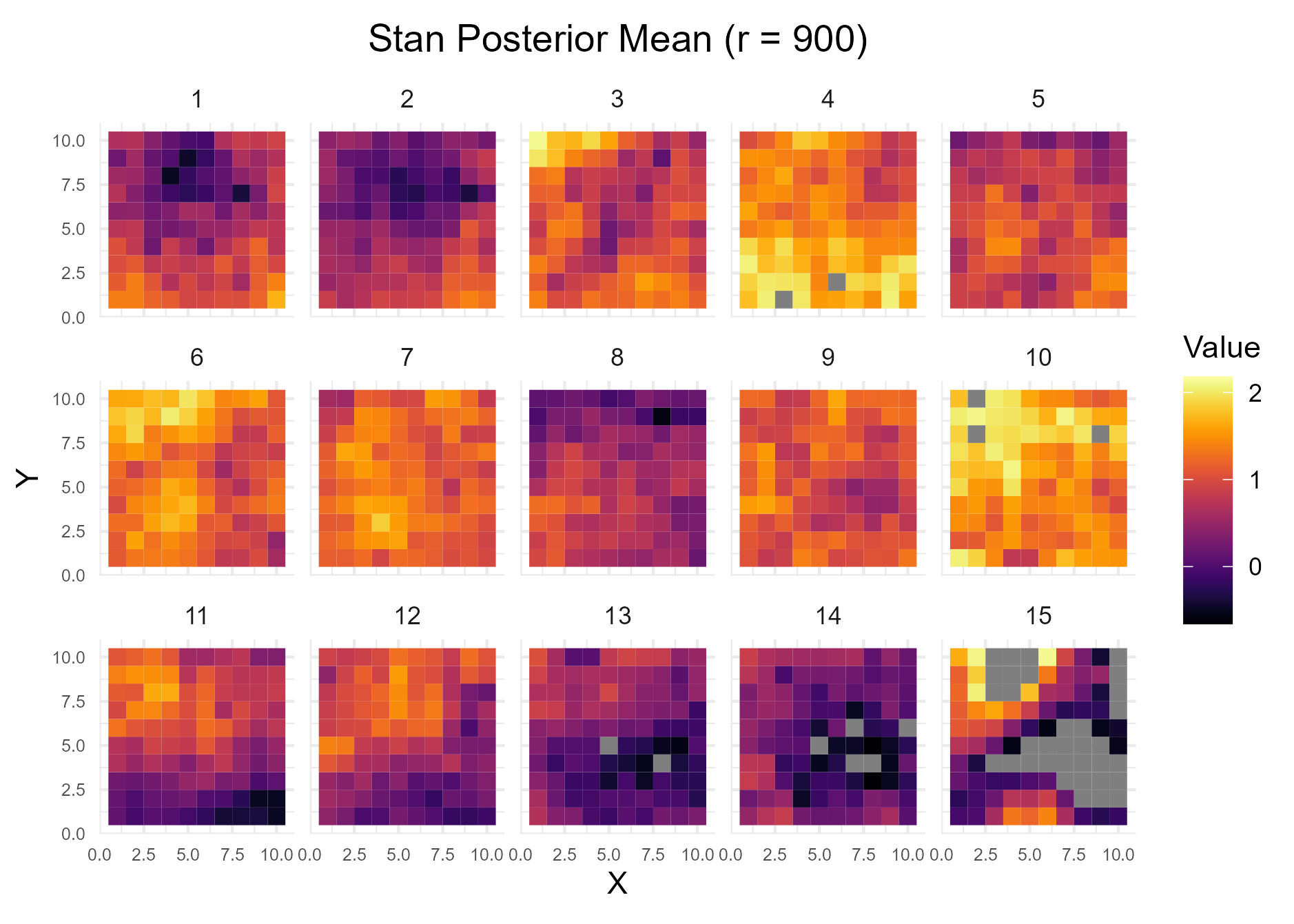}
    \end{subfigure}
    
    \caption{The first column is the true process simulated from the GQN, the second column is the posterior mean from the model implemented with EPR, and the third column is the posterior mean from the model implemented with MCMC. Each row is the comparison with \(r = 225, 450, 675, 900\) spatio-temporal basis functions used in the Frobenius norm matching, respectively. The \(T+1\) time point in each example was forecasted.} 
    \label{fig:mcmc_comp_gauss}
\end{figure}

Table \ref{fig:fnm.compare.gauss} presents predictive and computational evaluation metrics for the calibrated linear model implemented with EPR and MCMC. Results are reported across different values of \(r\), the number of spatio-temporal basis functions used to construct the \(\mathbf{G}\) matrix. Increasing the number of basis functions improves the ability of the calibrated linear model to mimic the covariance structure of the GQN model. As \(r\) increases, the forecast error, MSPE, and CRPS decrease when using FNM-EPR. In contrast, increasing the rank of \(\mathbf{G}\) leads to worse performance with FNM-MCMC due to a higher-dimensional parameter space resulting in longer computation time and potential convergence issues. These simulation results highlight the computational advantages of the Frobenius norm matching strategy implemented with an exact sampler rather than MCMC. The models calibrated with higher rank \(\mathbf{G}\) have improved forecasting and spatial interpolation. Figure \ref{fig:mcmc_comp_gauss} illustrates predictions from each approach across the different number of basis functions. Additional simulation results for Bernoulli and Poisson distributed data are provided in Supplementary Appendix \ref{appen:add.sims}. This involves swapping the normal data model with the Bernoulli with logit link to \(\mathbf{y}\) and Poisson with log link to \(\mathbf{y}\).

\vspace{-25pt}
\section{Data Analysis}\label{ch2:s:analysis}
Our motivating data comprises county-level birth rate data in Florida over a \(34\)-year period from \(1990\) to \(2023\). Annual birth rate data were downloaded from Florida Health Charts \url{https://www.flhealthcharts.gov/}. We incorporate relevant covariates capturing demographic, socioeconomic, and environmental factors that were used in previous studies \citep{wang2025spatio, shoff2012spatially, baghestani2016factors}. The environmental covariates including the average temperature and average precipitation were downloaded from NOAA National Centers for Environmental Information \url{https://www.ncei.noaa.gov/}. Demographic predictors such as proportion of Hispanic mothers, marriage rates, proportion of females age 15-44, proportion of population ages 65 and older, and total population were collected from Florida Health Charts. Socioeconomic variables include per capita personal income and proportion of mothers without a high school diploma and were obtained from the Federal Reserve Economic Data (FRED) \url{https://fred.stlouisfed.org/} and Florida Health Charts, respectively. We include proportion of births with low birth weight (less than 2500 grams) which is not a direct socioeconomic measure but is associated with maternal health and access to care \citep{thorsen2019operational, grbic2024factors}. We also include the behavioral and health-related variable proportion of maternal smoking during pregnancy downloaded from Florida Health Charts.

For our application, we use the Frobenius norm matching strategy to calibrate the covariance structure of a linear mixed effects model to the covariance structure from a GQN model. For comparison, we also use Frobenius norm matching to calibrate the linear model to a Mat\'{e}rn covariance and a VAR(1). This comparison allows us to assess whether the nonlinear dynamics of the GQN has improved predictive performance over the Mat\'{e}rn and VAR(1). The VAR(1) was specified according to Equation (\ref{eq:var1}) in Section \ref{method:var1}. The three calibrated linear models were implemented with EPR as we saw from our simulation study that EPR produced significant computational advantages over MCMC. 

All three models include the set of covariates described above. The temporal domain is the years 1990-2022, with forecasting for year 2023. To mimic a realistic forecasting scenario, we use the covariate values from 2022 to predict outcomes in 2023, assuming that the covariates for the forecast year would not yet be observed in practice. Additionally, to evaluate the model's ability to make accurate predictions at unobserved spatial locations, we randomly hold out \(10\%\) of the spatial locations at each year. The holdout locations are not the same across years. Evaluation metrics for the GQN calibrated model, the Mat\'{e}rn calibrated model, and the VAR(1) calibrated model are presented in Table \ref{ch3:birth.rate.anal}.

\begin{table}[H]
\centering
\caption{Comparison of forecasting, spatial smoothing, spatial interpolating, and computational performance for different covariance structures.}
\begin{tabular}{lcccc}
\toprule
Covariance & Forecast Error & In-sample MSPE & Out-of-sample MSPE & CPU Time \\ 
\midrule
GQN        & 2.214          & 1.355          & 1.567             & 30.51 mins \\
Mat\'{e}rn     & 3.047         & 1.192          & 1.638              & 29.80 mins \\
VAR(1) & 2.493 & 1.387 & 1.583 & 31.14 mins \\
\bottomrule
\end{tabular}\label{ch3:birth.rate.anal}
\end{table}
The model calibrated to the GQN covariance structure demonstrates improved forecast accuracy and out-of-sample mean squared prediction error (MSPE), suggesting that this model is better at making predictions for future time points and at unobserved spatial locations. The model calibrated toward the Mat\'{e}rn covariance has better in-sample MSPE, but does not produce as accurate forecasts or predictions at unobserved locations as the GQN calibrated model. The model calibrated toward the GQN has improved performance compared to the VAR(1) for all metrics. The CPU time is similar for all three models. This is expected because the number of basis functions used for the Frobenius norm matching was the same for all models. These results highlight the benefits of incorporating nonlinear spatio-temporal dynamics through the GQN model without increasing the computation time. 

\begin{table}[H]
\centering
\caption{One-, two-, and three-step ahead forecasting errors for the GQN and VAR(1) calibrated models.}
\begin{tabular}{lcccccc}
\toprule
Covariance & 1-step & 2-step  & 3-step  & In-sample MSPE & Out-of-sample MSPE & CPU Time \\ \midrule
\multicolumn{7}{c}{\textbf{Observe 1990-2022, Forecast 2023}} \\ \midrule
GQN     & 2.214 & - & - & 1.355 & 1.567 & 30.51 mins \\ 
VAR(1)  & 2.493 & - & - & 1.387 & 1.583 & 31.14 mins \\ \midrule
\multicolumn{7}{c}{\textbf{Observe 1990-2021, Forecast 2022-2023}} \\ \midrule
GQN     & 1.821 & 2.270 & - & 1.352 & 1.590 & 29.25 mins \\ 
VAR(1)  & 1.978 & 2.407 & - & 1.393 & 1.596 & 31.28 mins \\ \midrule
\multicolumn{7}{c}{\textbf{Observe 1990-2020, Forecast 2021-2023}} \\ \midrule
GQN    & 2.357 & 3.109 & 2.834 & 1.362 & 1.557 & 30.82 mins \\ 
VAR(1)  & 3.402 & 4.352 & 3.030 & 1.386 & 1.562 & 29.07 mins \\ \bottomrule
\end{tabular}\label{tab:forecast_comparison}
\end{table}
To further demonstrate the superior forecasting performance of the GQN model over the VAR(1) model, we evaluate its accuracy in two-step and three-step ahead forecasting scenarios. For the two-step ahead forecast, we use data from 1990-2021 to predict birth rates for 2022-2023. For the three-step ahead forecast, we use data from 1990-2020 to predict 2021-2023. In both scenarios, covariates from the final observed year are used for forecasting. These results are summarized in Table \ref{tab:forecast_comparison}. Across all scenarios, the GQN calibrated model yields consistently lower forecasting error than the VAR(1), providing further evidence of nonlinear dynamics in the birth rate data.

\begin{table}[H]
\centering
\caption{Posterior mean estimates and 95\% credible intervals for covariate effects under GQN, Mat\'{e}rn, and VAR(1) covariance structures. Statistically significant covariates are shown in red.}
\begin{tabular}{lccc}
\toprule
\textbf{Covariate} & \textbf{GQN} & \textbf{Mat\'{e}rn} & \textbf{VAR(1)} \\
\midrule
Proportion of Births Under 2500g & 0.13(-0.11,0.38) & 0.15(-0.07,0.38) & 0.12(-0.04,0.29)\\
Per Capita Personal Income   & 0.09(-0.42,0.52)  & -0.06(-0.55,0.37) & 0.16(-0.18,0.53)\\
Elderly Proportion  &\textcolor{red}{-0.66}(-1.19,-0.09) & \textcolor{red}{-0.59}(-0.99,-0.14) & \textcolor{red}{-0.76}(-1.05,-0.42)\\
Female Proportion & \textcolor{red}{1.02}(0.60,1.49) & \textcolor{red}{1.00} (0.58,1.36)& \textcolor{red}{0.94}(0.64,1.31)\\
Marriage Rate  & \textcolor{red}{0.37}(0.04,0.77) & \textcolor{red}{0.42}(0.13,0.73) & \textcolor{red}{0.32}(0.12,0.52)\\
Average Precipitation  &0.05(-0.23,0.38)  &0.05(-0.22,0.32) & 0.05 (-0.24,0.31)\\
Average Temperature  & \textcolor{red}{-1.02}(-1.86,-0.32) & -0.05(-1.22,1.12) & \textcolor{red}{-1.02}(-1.53,-0.39)\\
Total Population & \textcolor{red}{0.34}(0.06,0.65) & 0.25(-0.04,0.63)& \textcolor{red}{0.34}(0.05,0.66)\\
Proportion of Mothers who Smoke & 0.08(-0.29,0.48) & 0.01(-0.42,0.45)& 0.10(-0.25,0.56)\\
Proportion of Mothers without HS Diploma & \textcolor{red}{0.97}(0.65,1.32) & \textcolor{red}{1.00}(0.66,1.38)& \textcolor{red}{0.99}(0.67,1.33)\\
Proportion of Hispanic Mothers & \textcolor{red}{0.67}(0.29,1.08) & \textcolor{red}{0.74}(0.17,1.30)& \textcolor{red}{0.64}(0.11,1.22)\\
\bottomrule
\end{tabular}\label{cov.tab}
\end{table}
Table \ref{cov.tab} presents the posterior mean estimates and \(95\%\) credible intervals for each covariate for the GQN, Mat\'{e}rn, and VAR(1) calibrated models. All models have similar inference, identifying the similar covariates as statistically significant, which are determined by the credible intervals not containing zero. Specifically, the proportion of births with low birth weight, elderly population proportion, female population proportion, marriage rate, proportion of mothers without a high school diploma, and proportion of Hispanic mothers are all significant for all three covariance structures. In addition average temperature and total population are significant for the GQN and VAR(1) models. These findings are consistent with previous studies \citep{wang2025spatio, jiang2024analysis, haynes2020analyzing}. We fit the calibrated linear model with static and time varying coefficients and used the WAIC \citep{WAIC} to select which model to use in the analysis. See Supplementary Appendix \ref{appen:static.vs.time.vary} for a comparison of these two models. 

\begin{figure}[H]
    \centering
    \begin{subfigure}[b]{0.47\textwidth}
        \centering
        \includegraphics[width=\textwidth]{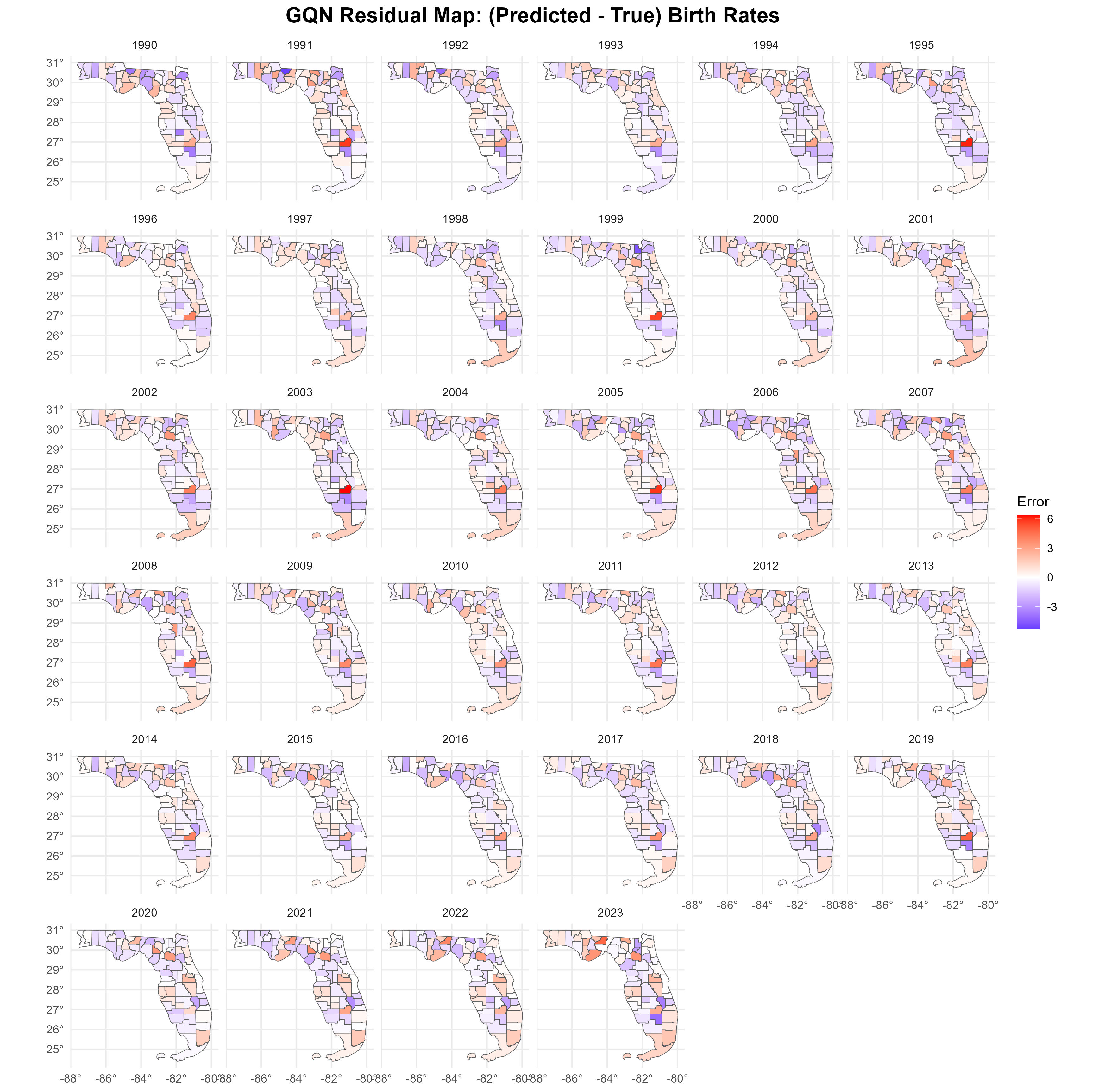}
    \end{subfigure}
    \hfill
    \begin{subfigure}[b]{0.45\textwidth}
        \centering
        \includegraphics[width=\textwidth]{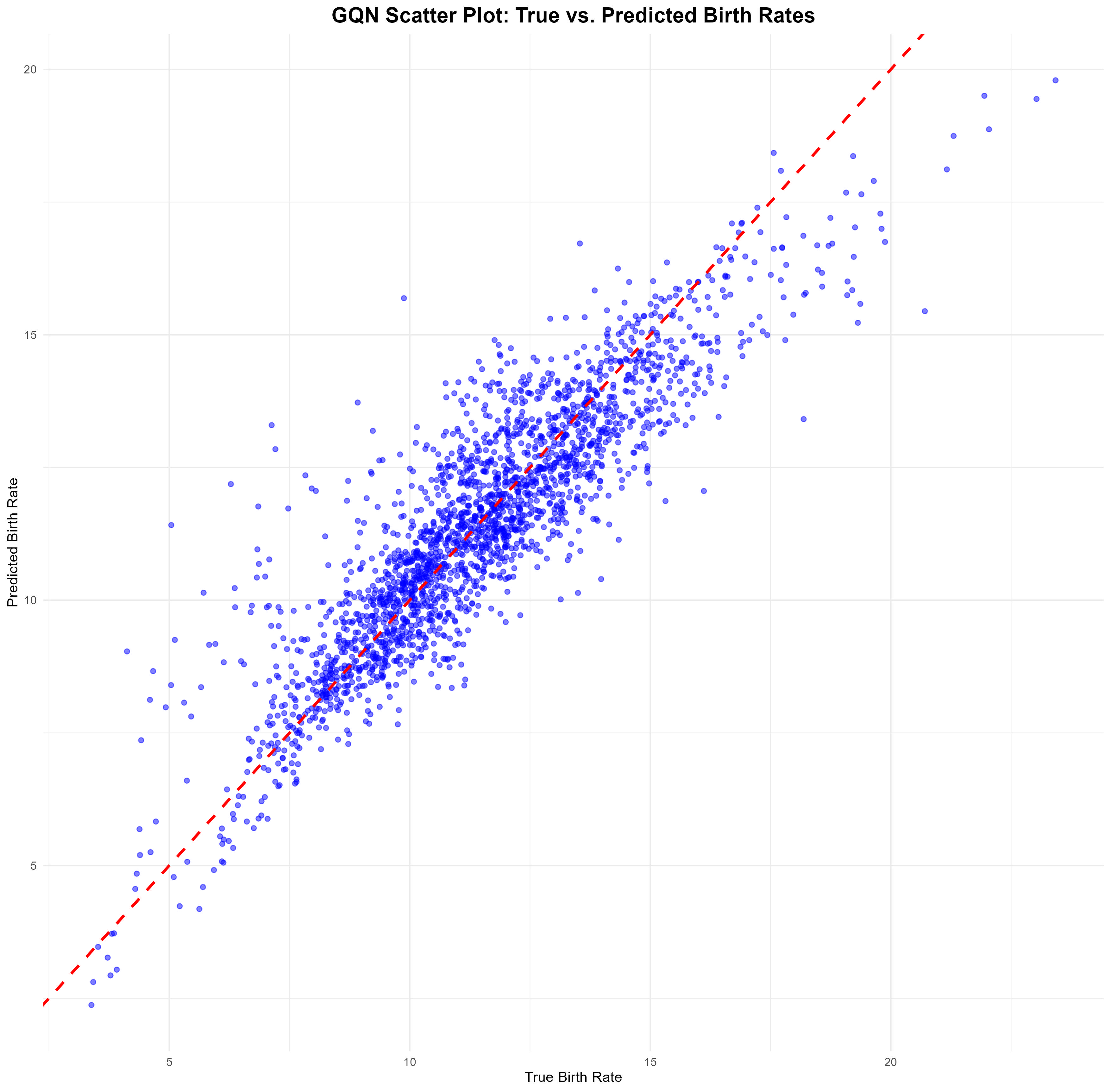}
    \end{subfigure}
    \caption{The left figure is the residual plot by year. The residuals are defined as the (predicted birth rate - true birth rate). The right figure is the scatter plot of the true vs. predicted birth rates. This plot contains the birth rates for all counties over all years.}
    \label{fig:birth.rate.anal}
\end{figure}

Figure \ref{fig:birth.rate.anal} presents residual plots by year and a scatter plot comparing the predicted versus observed birth rates across all counties and years. The residual plots show that most residuals are close to zero, indicating a good model fit. They also help identify specific counties and years where the model overestimated or underestimated the birth rate. Notably, there are a mix of positive and negative residuals which suggests the model is not over-fitting or under-fitting across space and time. As expected, the residuals for the forecasted year 2023 are slightly larger. There is increased uncertainty when making predictions beyond the observed time points. The scatterplot also suggests that we have good model performance as the points are close to the red dashed line, suggesting that the predicted values are close to the true values. These plots show that the model was able to capture the nonlinear spatio-temporal dynamics and provide accurate predictions over the 34-year period. Supplementary Appendix \ref{appen:birth_rate_pred} presents plots of the true birth rate data, the predicted birth rates by year and county, and the posterior standard deviations. 

\section{Discussion}\label{ch2:s:discuss}
In this paper, we introduce a computationally efficient approach for analyzing nonlinear dynamic spatio-temporal data through a novel combination of Frobenius norm matching with exact posterior regression. We evaluate the proposed covariance calibration strategy by comparing it to direct implementation of the GQN model using MCMC. Frobenius norm matching can also be implemented with MCMC and we demonstrate the computational and inferential advantages (i.e., improved predictive metrics) of using EPR instead of MCMC in this scenario. We present an application to population dynamics by modeling birth rate data in Florida. The model calibrated to the GQN covariance had improved forecasting and spatial interpolation compared to a model calibrated to a Mat\'{e}rn covariance and VAR(1). Our application highlights improvements in prediction by incorporating non-linear dynamics when modeling birth rate data and also highlights the computational advantages of using FNM-EPR.  

\vspace{-30pt}
\begin{singlespace}
    \bibliography{citations}

\begin{thebibliography}{}

\bibitem[Aitken, 2022]{aitken2022changing}
Aitken, R.~J. (2022).
\newblock The changing tide of human fertility.

\bibitem[Alessio and Gupta, 2023]{alessio2023reaction}
Alessio, B.~M. and Gupta, A. (2023).
\newblock A reaction-diffusion-chemotaxis model for human population dynamics over fractal terrains.
\newblock {\em arXiv preprint arXiv:2310.07185}.

\bibitem[Baghestani and Malcolm, 2016]{baghestani2016factors}
Baghestani, H. and Malcolm, M. (2016).
\newblock Factors predicting the us birth rate.
\newblock {\em Journal of Economic Studies}, 43(3):432--446.

\bibitem[Beaton and Tukey, 1974]{beaton1974fitting}
Beaton, A.~E. and Tukey, J.~W. (1974).
\newblock The fitting of power series, meaning polynomials, illustrated on band-spectroscopic data.
\newblock {\em Technometrics}, 16(2):147--185.

\bibitem[Bhattacharjee et~al., 2024]{bhattacharjee2024global}
Bhattacharjee, N.~V., Schumacher, A.~E., Aali, A., Abate, Y.~H., Abbasgholizadeh, R., Abbasian, M., Abbasi-Kangevari, M., Abbastabar, H., Abd~ElHafeez, S., Abd-Elsalam, S., et~al. (2024).
\newblock Global fertility in 204 countries and territories, 1950--2021, with forecasts to 2100: a comprehensive demographic analysis for the global burden of disease study 2021.
\newblock {\em The lancet}, 403(10440):2057--2099.

\bibitem[Bonas et~al., 2024]{bonas2024calibrated}
Bonas, M., Wikle, C.~K., and Castruccio, S. (2024).
\newblock Calibrated forecasts of quasi-periodic climate processes with deep echo state networks and penalized quantile regression.
\newblock {\em Environmetrics}, 35(1):e2833.

\bibitem[Bradley and Clinch, 2024]{bradley2024generating}
Bradley, J.~R. and Clinch, M. (2024).
\newblock Generating independent replicates directly from the posterior distribution for a class of spatial hierarchical models.
\newblock {\em Journal of Computational and Graphical Statistics}, pages 1--17.

\bibitem[Bradley et~al., 2015]{bradley2015multivariate}
Bradley, J.~R., Holan, S.~H., and Wikle, C.~K. (2015).
\newblock Multivariate spatio-temporal models for high-dimensional areal data with application to longitudinal employer-household dynamics.
\newblock {\em The Annals of Applied Statistics}, 9(4).

\bibitem[Bradley et~al., 2019]{bradley2019spatio}
Bradley, J.~R., Wikle, C.~K., and Holan, S.~H. (2019).
\newblock Spatio-temporal models for big multinomial data using the conditional multivariate logit-beta distribution.
\newblock {\em Journal of Time Series Analysis}, 40(3):363--382.

\bibitem[Cameletti et~al., 2013]{cameletti2013spatio}
Cameletti, M., Lindgren, F., Simpson, D., and Rue, H. (2013).
\newblock Spatio-temporal modeling of particulate matter concentration through the spde approach.
\newblock {\em AStA Advances in Statistical Analysis}, 97:109--131.

\bibitem[Cheng et~al., 2022]{cheng2022global}
Cheng, H., Luo, W., Si, S., Xin, X., Peng, Z., Zhou, H., Liu, H., and Yu, Y. (2022).
\newblock Global trends in total fertility rate and its relation to national wealth, life expectancy and female education.
\newblock {\em BMC Public Health}, 22(1):1346.

\bibitem[Clinch and Bradley, 2024]{clinch2024exact}
Clinch, M. and Bradley, J.~R. (2024).
\newblock Exact bayesian inference for multivariate spatial data of any size with application to air pollution monitoring.
\newblock {\em arXiv preprint arXiv:2410.02655}.

\bibitem[Comolli and Vignoli, 2021]{comolli2021spreading}
Comolli, C.~L. and Vignoli, D. (2021).
\newblock Spreading uncertainty, shrinking birth rates: A natural experiment for italy.
\newblock {\em European Sociological Review}, 37(4):555--570.

\bibitem[Cressie and Wikle, 2011]{cressie2011statistics}
Cressie, N. and Wikle, C.~K. (2011).
\newblock {\em Statistics for spatio-temporal data}.
\newblock John Wiley \& Sons.

\bibitem[De~Iaco and Maggio, 2016]{de2016dynamic}
De~Iaco, S. and Maggio, S. (2016).
\newblock A dynamic model for age-specific fertility rates in italy.
\newblock {\em Spatial Statistics}, 17:105--120.

\bibitem[Diaconis and Ylvisaker, 1979]{DY}
Diaconis, P. and Ylvisaker, D. (1979).
\newblock Conjugate priors for exponential families.
\newblock {\em The Annals of Statistics}, 7(2):269--281.

\bibitem[Drepper et~al., 1994]{drepper1994nonlinear}
Drepper, F., Engbert, R., and Stollenwerk, N. (1994).
\newblock Nonlinear time series analysis of empirical population dynamics.
\newblock {\em Ecological Modelling}, 75:171--181.

\bibitem[Gilks and Roberts, 1996]{gilks1996strategies}
Gilks, W.~R. and Roberts, G.~O. (1996).
\newblock Strategies for improving mcmc.
\newblock {\em Markov chain Monte Carlo in practice}, 6:89--114.

\bibitem[Grbic et~al., 2024]{grbic2024factors}
Grbic, D., Supic, Z.~T., Todorovic, J., Nesic, D., Karic, S., Jurisic, A., Kocic, S., Bukumiric, Z., Cirkovic, A., and Jankovic, S. (2024).
\newblock Factors associated with low birth weight in low-income populations in the western balkans: insights from the multiple indicator cluster survey.
\newblock {\em Frontiers in Public Health}, 12:1394060.

\bibitem[Haynes and Ogneva-Himmelberger, 2020]{haynes2020analyzing}
Haynes, M. and Ogneva-Himmelberger, Y. (2020).
\newblock Analyzing temporal trends and spatial patterns in adverse birth outcomes in massachusetts from 2000--2014.
\newblock {\em Applied Geography}, 125:102331.

\bibitem[Heckman and Walker, 1989]{heckman1989forecasting}
Heckman, J.~J. and Walker, J.~R. (1989).
\newblock Forecasting aggregate period-specific birth rates: the time series properties of a microdynamic neoclassical model of fertility.
\newblock {\em Journal of the American Statistical Association}, 84(408):958--965.

\bibitem[Henderson and Loreau, 2019]{henderson2019ecological}
Henderson, K. and Loreau, M. (2019).
\newblock An ecological theory of changing human population dynamics.
\newblock {\em People and Nature}, 1(1):31--43.

\bibitem[Hooten and Wikle, 2008]{hooten2008hierarchical}
Hooten, M.~B. and Wikle, C.~K. (2008).
\newblock A hierarchical bayesian non-linear spatio-temporal model for the spread of invasive species with application to the eurasian collared-dove.
\newblock {\em Environmental and Ecological Statistics}, 15:59--70.

\bibitem[Horn and Johnson, 2012]{horn2012matrix}
Horn, R.~A. and Johnson, C.~R. (2012).
\newblock {\em Matrix analysis}.
\newblock Cambridge university press.

\bibitem[Jarzebski et~al., 2021]{jarzebski2021ageing}
Jarzebski, M.~P., Elmqvist, T., Gasparatos, A., Fukushi, K., Eckersten, S., Haase, D., Goodness, J., Khoshkar, S., Saito, O., Takeuchi, K., et~al. (2021).
\newblock Ageing and population shrinking: implications for sustainability in the urban century.
\newblock {\em Npj Urban Sustainability}, 1(1):17.

\bibitem[Jiang et~al., 2024]{jiang2024analysis}
Jiang, X., Pan, Q., Xu, W., Sun, J., and Chen, S. (2024).
\newblock Analysis of the spatiotemporal evolution of birth rates and influencing factors in the yangtze river basin.
\newblock {\em PloS one}, 19(12):e0316139.

\bibitem[Kearney and Levine, 2022]{kearney2022causes}
Kearney, M.~S. and Levine, P.~B. (2022).
\newblock The causes and consequences of declining us fertility.
\newblock {\em Aspen Institute}.

\bibitem[Kearney et~al., 2022]{kearney2022puzzle}
Kearney, M.~S., Levine, P.~B., and Pardue, L. (2022).
\newblock The puzzle of falling us birth rates since the great recession.
\newblock {\em Journal of Economic Perspectives}, 36(1):151--176.

\bibitem[Liu and Raftery, 2020]{liu2020education}
Liu, D.~H. and Raftery, A.~E. (2020).
\newblock How do education and family planning accelerate fertility decline?
\newblock {\em Population and development review}, 46(3):409--441.

\bibitem[McCollough et~al., 2023]{mccollough2023impact}
McCollough, J.~D., Sargsyan, G., and Luo, Z. (2023).
\newblock The impact of declining birth rates on future infrastructure maintenance costs per capita.
\newblock {\em Journal of Economic Studies}, 50(6):1121--1129.

\bibitem[McDermott and Wikle, 2016]{mcdermott2016model}
McDermott, P.~L. and Wikle, C.~K. (2016).
\newblock A model-based approach for analog spatio-temporal dynamic forecasting.
\newblock {\em Environmetrics}, 27(2):70--82.

\bibitem[Neverova et~al., 2016]{neverova2016dynamics}
Neverova, G., Yarovenko, I., and Frisman, E.~Y. (2016).
\newblock Dynamics of populations with delayed density dependent birth rate regulation.
\newblock {\em Ecological Modelling}, 340:64--73.

\bibitem[Pan et~al., 2024]{pan2024bayesian}
Pan, S., Zhang, L., Bradley, J.~R., and Banerjee, S. (2024).
\newblock Bayesian inference for spatial-temporal non-{Gaussian} data using predictive stacking.
\newblock {\em arXiv preprint arXiv:2406.04655}.

\bibitem[Raim et~al., 2021]{raim2021spatio}
Raim, A.~M., Holan, S.~H., Bradley, J.~R., and Wikle, C.~K. (2021).
\newblock Spatio-temporal change of support modeling with r.
\newblock {\em Computational Statistics}, 36:749--780.

\bibitem[Robert et~al., 2018]{robert2018accelerating}
Robert, C.~P., Elvira, V., Tawn, N., and Wu, C. (2018).
\newblock Accelerating mcmc algorithms.
\newblock {\em Wiley Interdisciplinary Reviews: Computational Statistics}, 10(5):e1435.

\bibitem[Roques and Bonnefon, 2016]{roques2016modelling}
Roques, L. and Bonnefon, O. (2016).
\newblock Modelling population dynamics in realistic landscapes with linear elements: A mechanistic-statistical reaction-diffusion approach.
\newblock {\em PloS one}, 11(3):e0151217.

\bibitem[Rue et~al., 2009]{inla}
Rue, H., Martino, S., and Chopin, N. (2009).
\newblock Approximate {Bayesian} inference for latent {Gaussian} models by using integrated nested {Laplace} approximations.
\newblock {\em Journal of the Royal Statistical Society Series B: Statistical Methodology}, 71(2):319--392.

\bibitem[Septier and Peters, 2015]{septier2015overview}
Septier, F. and Peters, G.~W. (2015).
\newblock An overview of recent advances in monte-carlo methods for bayesian filtering in high-dimensional spaces.
\newblock {\em Theoretical aspects of spatial-temporal modeling}, pages 31--61.

\bibitem[Shoff and Yang, 2012]{shoff2012spatially}
Shoff, C. and Yang, T.-C. (2012).
\newblock Spatially varying predictors of teenage birth rates among counties in the united states.
\newblock {\em Demographic research}, 27(14):377.

\bibitem[Sigrist et~al., 2015]{sigrist2015stochastic}
Sigrist, F., K{\"u}nsch, H.~R., and Stahel, W.~A. (2015).
\newblock Stochastic partial differential equation based modelling of large space--time data sets.
\newblock {\em Journal of the Royal Statistical Society Series B: Statistical Methodology}, 77(1):3--33.

\bibitem[Tang and Chen, 2002]{tang2002density}
Tang, S. and Chen, L. (2002).
\newblock Density-dependent birth rate, birth pulses and their population dynamic consequences.
\newblock {\em Journal of Mathematical Biology}, 44(2):185--199.

\bibitem[Thorsen et~al., 2019]{thorsen2019operational}
Thorsen, M.~L., Thorsen, A., and McGarvey, R. (2019).
\newblock Operational efficiency, patient composition and regional context of us health centers: associations with access to early prenatal care and low birth weight.
\newblock {\em Social Science \& Medicine}, 226:143--152.

\bibitem[Tzitiridou-Chatzopoulou et~al., 2024]{tzitiridou2024predicting}
Tzitiridou-Chatzopoulou, M., Zournatzidou, G., and Kourakos, M. (2024).
\newblock Predicting future birth rates with the use of an adaptive machine learning algorithm: A forecasting experiment for scotland.
\newblock {\em International Journal of Environmental Research and Public Health}, 21(7):841.

\bibitem[Vanella and Hassenstein, 2023]{vanella2023stochastic}
Vanella, P. and Hassenstein, M.~J. (2023).
\newblock Stochastic forecasting of regional age-specific fertility rates: An outlook for german nuts-3 regions.
\newblock {\em Mathematics}, 12(1):25.

\bibitem[Vollset et~al., 2020]{vollset2020fertility}
Vollset, S.~E., Goren, E., Yuan, C.-W., Cao, J., Smith, A.~E., Hsiao, T., Bisignano, C., Azhar, G.~S., Castro, E., Chalek, J., et~al. (2020).
\newblock Fertility, mortality, migration, and population scenarios for 195 countries and territories from 2017 to 2100: a forecasting analysis for the global burden of disease study.
\newblock {\em The Lancet}, 396(10258):1285--1306.

\bibitem[Waller and Gotway, 2004]{waller2004applied}
Waller, L.~A. and Gotway, C.~A. (2004).
\newblock {\em Applied spatial statistics for public health data}.
\newblock John Wiley \& Sons.

\bibitem[Wang et~al., 2025]{wang2025spatio}
Wang, K., Zhang, Y., Bai, L., Chen, Y., and Ling, C. (2025).
\newblock Spatio-temporal modelling of extreme low birth rates in us counties.
\newblock {\em BMC Public Health}, 25(1):1--11.

\bibitem[Watanabe and Opper, 2010]{WAIC}
Watanabe, S. and Opper, M. (2010).
\newblock Asymptotic equivalence of {Bayes} cross validation and widely applicable information criterion in singular learning theory.
\newblock {\em Journal of Machine Learning Research}, 11(12).

\bibitem[Wikle and Hooten, 2006]{wikle2006hierarchical}
Wikle, C.~K. and Hooten, M.~B. (2006).
\newblock Hierarchical bayesian spatio-temporal models for population spread.
\newblock {\em Applications of computational statistics in the environmental sciences: hierarchical Bayes and MCMC methods}, 145169.

\bibitem[Wikle and Hooten, 2010]{wikle2010general}
Wikle, C.~K. and Hooten, M.~B. (2010).
\newblock A general science-based framework for dynamical spatio-temporal models.
\newblock {\em Test}, 19:417--451.

\bibitem[Wolpin, 1984]{wolpin1984estimable}
Wolpin, K.~I. (1984).
\newblock An estimable dynamic stochastic model of fertility and child mortality.
\newblock {\em Journal of Political economy}, 92(5):852--874.

\bibitem[Zhou and Bradley, 2024]{zhou2024multiscale}
Zhou, S. and Bradley, J.~R. (2024).
\newblock Multiscale multi-type spatial bayesian analysis of wildfires and population change that avoids mcmc and approximating the posterior distribution.
\newblock {\em arXiv preprint arXiv:2410.02905}.

\end{thebibliography}
\end{singlespace}

\newpage
\bigskip
\begin{center}
{\large\bf SUPPLEMENTARY MATERIAL}
\end{center}

\appendix
\renewcommand{\thesection}{\Alph{section}}

\titleformat{\section}[block]
  {\normalfont\Large\bfseries}{Appendix \thesection:}{1em}{}
\vspace{-30pt}

\section{Code}\label{s:code}
All R Code used for simulations and data analysis is available at \url{https://fsu-my.sharepoint.com/:u:/g/personal/mcc21k_fsu_edu/EVp0HbamlDBIltXB8HaujKMB887iDapQR6OGTyL_GSY3Pw?e=jIwxIC}.

\section{Discretization of Reaction-Diffusion PDE}\label{appen:pde}
The reaction-diffusion PDE is often used in population dynamics and is given by the following,
\begin{align*}
    \frac{\partial u}{\partial t} = \frac{\partial}{\partial s_1}\left(\delta(s_1, s_2)\frac{\partial u}{\partial s_1} \right) + \frac{\partial}{\partial s_2}\left(\delta(s_1, s_2)\frac{\partial u}{\partial s_2} \right) + \gamma_0 u \left(1 - \frac{u}{\gamma_1} \right),
\end{align*}
where \(u_t(s_1,s_2)\) represents the value of the process at two-dimensional spatial location \(\mathbf{s} = (s_1,s_2)\) (i.e, latitude and longitude) and time \(t\), \(\delta(s_1,s_2)\) is a spatially varying diffusion coefficient, \(\gamma_0\) denotes the intrinsic population growth rate, and \(\gamma_1\) is the carrying capacity. We will discretize in both time and space where \(\Delta_t\) denotes the temporal step size and \(\Delta_{s_1}\) and \(\Delta_{s_2}\) denote the spatial step sizes in the \(s_1\) and \(s_2\) directions, respectively. The left hand side of the PDE is discretized as follows,
\begin{align*}
    \frac{\partial u}{\partial t} \approx \frac{u_t (s_1, s_2) - u_{t - \Delta_t}(s_1,s_2)}{\Delta_t}.
\end{align*}
Discretizing the first term on the right hand side leads to the following, 
\begin{align*}
    &\frac{\partial}{\partial s_1}\left(\delta(s_1,s_2) \frac{\partial u_{t - \Delta_t}(s_1,s_2)}{\partial s_1} \right) \\
    &\approx \frac{1}{2\Delta_{s_1}^2}\Big\{\big[\delta(s_1+\Delta_{s_1}, s_2) + \delta(s_1,s_2)\big]\big[u_{t - \Delta_t}(s_1 + \Delta_{s_1}, s_2) - u_{t - \Delta_t}(s_1,s_2)\big] - \big[\delta(s_1, s_2) + \delta(s_1-\Delta_{s_1},s_2)\big]\\
    &\times \big[u_{t - \Delta_t}(s_1 , s_2) - u_{t - \Delta_t}(s_1- \Delta_{s_1},s_2)\big] \Big\} 
\end{align*}
Then expanding all the terms leads to 
\begin{align*}
     &= \frac{1}{2\Delta_{s_1}^2} \Big\{\delta(s_1+\Delta_{s_1},s_2)u_{t - \Delta_t}(s_1 + \Delta_{s_1}, s_2) - \delta(s_1+\Delta_{s_1}, s_2)u_{t - \Delta_t}(s_1,s_2) + \delta(s_1,s_2)u_{t - \Delta_t}(s_1 + \Delta_{s_1},s_2) \\ &-\delta(s_1,s_2)u_{t - \Delta_t}(s_1,s_2) - \delta(s_1,s_2)u_{t - \Delta_t}(s_1,s_2) + \delta(s_1,s_2)u_{t - \Delta_t}(s_1 - \Delta_{s_1}, s_2) \\&- \delta(s_1 - \Delta_{s_1}, s_2)u_{t - \Delta_t}(s_1,s_2) + \delta(s_1-\Delta_{s_1}, s_2)u_{t - \Delta_t}(s_1 - \Delta_{s_1}, s_2) \Big\}
\end{align*}
Identifying and grouping terms that share a common factor of \(u_{t-\Delta_t}\) evaluated at the same spatial location, we simplify the expression to: 
\begin{align*}
    &= \frac{1}{2\Delta_{s_1}^2}\Big\{u_{t - \Delta_t}(s_1+\Delta_{s_1}, s_2)\big[\delta(s_1 + \Delta_{s_1},s_2) + \delta(s_1,s_2) \big] - u_{t - \Delta_t}(s_1,s_2) \big[\delta(s_1 + \Delta_{s_1}, s_2) + 2\delta (s_1, s_2)\\ &+ \delta(s_1 - \Delta_{s_1}, s_2) \big] + u_{t - \Delta_t} (s_1 - \Delta_{s_1}, s_2)\big[\delta(s_1 - \Delta_{s_1}, s_2) + \delta(s_1,s_2) \big]\Big\}
\end{align*}
Distributing the \(\frac{1}{2}\) into each of the bracketed terms gives: 
\begin{align*}
   &= \frac{1}{\Delta_{s_1}^2}\Big\{u_{t - \Delta_t}(s_1+\Delta_{s_1}, s_2)\big[\frac{1}{2}\delta(s_1 + \Delta_{s_1},s_2) + \frac{1}{2}\delta(s_1,s_2) \big] - u_{t - \Delta_t}(s_1,s_2) \big[\frac{1}{2}\delta(s_1 + \Delta_{s_1}, s_2) \\& + \delta (s_1, s_2) + \frac{1}{2}\delta(s_1 - \Delta_{s_1}, s_2) \big] + u_{t - \Delta_t} (s_1 - \Delta_{s_1}, s_2)\big[\frac{1}{2}\delta(s_1 - \Delta_{s_1}, s_2) + \frac{1}{2}\delta(s_1,s_2) \big]\Big\}
\end{align*}
Next, we add and subtract equivalent terms within each coefficient, allowing the expression to be written a form that reflects a centered finite difference approximation.
\begin{align*}
&= \frac{1}{\Delta_{s_1}^2}\Big\{u_{t - \Delta_t}(s_1+\Delta_{s_1}, s_2)\big[\frac{1}{2}\delta(s_1 + \Delta_{s_1},s_2) + \frac{1}{2}\delta(s_1,s_2) 
 + \frac{1}{2}\delta(s_1,s_2) - \frac{1}{4}\delta(s_1 + \Delta_{s_1}, s_2) \\&- \frac{1}{4}\delta(s_1 - \Delta_{s_1}, s_2)\big] - u_{t - \Delta_t}(s_1,s_2) \big[\frac{1}{2}\delta(s_1 + \Delta_{s_1}, s_2) + \delta (s_1, s_2) + \frac{1}{2}\delta(s_1 - \Delta_{s_1}, s_2) \big] 
    \\&+ u_{t - \Delta_t} (s_1 - \Delta_{s_1}, s_2)\big[\frac{1}{2}\delta(s_1 - \Delta_{s_1}, s_2) + \frac{1}{2}\delta(s_1,s_2) + \frac{1}{2}\delta(s_1,s_2) - \frac{1}{4}\delta(s_1 + \Delta_{s_1}, s_2) - \frac{1}{4}\delta(s_1 - \Delta_{s_1}, s_2)\big]\Big\}
\end{align*}
Now, combining like terms results in the following: 
\begin{align*}
&= \frac{1}{\Delta_{s_1}^2}\Big\{u_{t - \Delta_t}(s_1+\Delta_{s_1}, s_2)\big[\frac{1}{4}\delta(s_1 + \Delta_{s_1},s_2) + \delta(s_1,s_2)  - \frac{1}{4}\delta(s_1 - \Delta_{s_1}, s_2)\big] \\&- u_{t - \Delta_t}(s_1,s_2) \big[\frac{1}{2}\delta(s_1 + \Delta_{s_1}, s_2) + \delta (s_1, s_2) + \frac{1}{2}\delta(s_1 - \Delta_{s_1}, s_2) \big] \\&
+ u_{t - \Delta_t} (s_1 - \Delta_{s_1}, s_2)\big[\frac{1}{4}\delta(s_1 - \Delta_{s_1}, s_2) + \delta(s_1,s_2) - \frac{1}{4}\delta(s_1 + \Delta_{s_1}, s_2) \big]\Big\}\\
&= \frac{1}{\Delta_{s_1}^2}\Big\{u_{t - \Delta_t}(s_1+\Delta_{s_1}, s_2)\big[\delta(s_1,s_2) + \frac{\big(\delta(s_1 + \Delta_{s_1},s_2) - \delta(s_1 - \Delta_{s_1}, s_2)\big)}{4}\big] \\&- u_{t - \Delta_t}(s_1,s_2) \big[\frac{1}{2}\delta(s_1 + \Delta_{s_1}, s_2) + \delta (s_1, s_2) + \frac{1}{2}\delta(s_1 - \Delta_{s_1}, s_2) \big] 
\\&+ u_{t - \Delta_t} (s_1 - \Delta_{s_1}, s_2)\big[\delta(s_1,s_2) - \frac{\big(\delta(s_1 + \Delta_{s_1}, s_2) - \delta(s_1 - \Delta_{s_1}, s_2)\big)}{4} \big]\Big\}\\
&= \frac{1}{\Delta_{s_1}^2}\Big\{u_{t - \Delta_t}(s_1+\Delta_{s_1}, s_2)\big[\delta(s_1,s_2) + \frac{\big(\delta(s_1 + \Delta_{s_1},s_2) - \delta(s_1 - \Delta_{s_1}, s_2)\big)}{4}\big] - u_{t - \Delta_t}(s_1,s_2) \big[2 \delta (s_1, s_2) \big] \\&
+ u_{t - \Delta_t} (s_1 - \Delta_{s_1}, s_2)\big[\delta(s_1,s_2) - \frac{\big(\delta(s_1 + \Delta_{s_1}, s_2) - \delta(s_1 - \Delta_{s_1}, s_2)\big)}{4} \big]\Big\}
\end{align*} 
Similarly, for the partial derivative with respect to \(s_2\), applying the same algebraic steps used in the approximation of the derivative with respect to \(s_1\) yields the following expression:
\begin{align*}
    &\frac{\partial}{\partial s_2}\left(\delta(s_1,s_2) \frac{\partial u_{t - \Delta_t}(s_1,s_2)}{\partial s_2} \right) \\
    &\approx_1 \frac{1}{\Delta_{s_2}^2}\Big\{u_{t - \Delta_t}(s_1, s_2 + \Delta_{s_2})\big[\delta(s_1,s_2) + \frac{\big(\delta(s_1,s_2 + \Delta_{s_2}) - \delta(s_1, s_2 - \Delta_{s_2})\big)}{4}\big] - u_{t - \Delta_t}(s_1,s_2) \big[2 \delta (s_1, s_2) \big] \\&
    + u_{t - \Delta_t} (s_1, s_2 - \Delta_{s_2})\big[\delta(s_1,s_2) - \frac{\big(\delta(s_1, s_2 + \Delta_{s_2}) - \delta(s_1, s_2 - \Delta_{s_2})\big)}{4} \big]\Big\}
\end{align*}
Now putting it all together with the nonlinear reaction term we have, 
\begin{align*}
 &\frac{u_t (s_1, s_2) - u_{t - \Delta_t}(s_1,s_2)}{\Delta_t} \\ &= \frac{1}{\Delta_{s_1}^2}\Big\{u_{t - \Delta_t}(s_1+\Delta_{s_1}, s_2)\big[\delta(s_1,s_2) + \frac{\big(\delta(s_1 + \Delta_{s_1},s_2) - \delta(s_1 - \Delta_{s_1}, s_2)\big)}{4}\big] \\&- u_{t - \Delta_t}(s_1,s_2) \big[2 \delta (s_1, s_2) \big] 
    + u_{t - \Delta_t} (s_1 - \Delta_{s_1}, s_2)\big[\delta(s_1,s_2) - \frac{\big(\delta(s_1 + \Delta_{s_1}, s_2) - \delta(s_1 - \Delta_{s_1}, s_2)\big)}{4} \big]\Big\}\\& + \frac{1}{\Delta_{s_2}^2}\Big\{u_{t - \Delta_t}(s_1, s_2 + \Delta_{s_2})\big[\delta(s_1,s_2) + \frac{\big(\delta(s_1,s_2 + \Delta_{s_2}) - \delta(s_1, s_2 - \Delta_{s_2})\big)}{4}\big] \\&- u_{t - \Delta_t}(s_1,s_2) \big[2 \delta (s_1, s_2) \big]
    + u_{t - \Delta_t} (s_1, s_2 - \Delta_{s_2})\big[\delta(s_1,s_2) - \frac{\big(\delta(s_1, s_2 + \Delta_{s_2}) - \delta(s_1, s_2 - \Delta_{s_2})\big)}{4} \big]\Big\} \\&+ \gamma_0 u_{t-\Delta_t}(s_1,s_2)\Big(1 - \frac{u_{t - \Delta_t}(s_1,s_2)}{\gamma_1} \Big) 
\end{align*}
Multiplying both sides by \(\Delta_t\), adding \(u_{t-
\Delta_t}(s_1, s_2)\), and distributing the nonlinear reaction term results in the following expression:
\begin{align*}
 u_t (s_1, s_2) &=  u_{t - \Delta_t}(s_1,s_2)\Bigg(1 - 2\delta(s_1,s_2)\Big\{\frac{\Delta_t}{\Delta_{s_1}^2}+\frac{\Delta_t}{\Delta_{s_2}^2}\Big\} \Bigg) \\
 &+  u_{t - \Delta_t}(s_1+\Delta_{s_1}, s_2)\Big\{\frac{\Delta_t}{\Delta_{s_1}^2}\big[\delta(s_1,s_2) + \frac{\big(\delta(s_1 + \Delta_{s_1},s_2) - \delta(s_1 - \Delta_{s_1}, s_2)\big)}{4}\big]\Big\} \\
     &+ u_{t - \Delta_t} (s_1 - \Delta_{s_1}, s_2)\Big\{\frac{\Delta_t}{\Delta_{s_1}^2}\big[\delta(s_1,s_2) - \frac{\big(\delta(s_1 + \Delta_{s_1}, s_2) - \delta(s_1 - \Delta_{s_1}, s_2)\big)}{4} \big]\Big\} \\
 & + u_{t - \Delta_t}(s_1, s_2 + \Delta_{s_2})\Big\{\frac{\Delta_t}{\Delta_{s_2}^2}\big[\delta(s_1,s_2) + \frac{\big(\delta(s_1,s_2 + \Delta_{s_2}) - \delta(s_1, s_2 - \Delta_{s_2})\big)}{4}\big]\Big\}
    \\&+ u_{t - \Delta_t} (s_1, s_2 - \Delta_{s_2})\Big\{\frac{\Delta_t}{\Delta_{s_2}^2}\big[\delta(s_1,s_2) - \frac{\big(\delta(s_1, s_2 + \Delta_{s_2}) - \delta(s_1, s_2 - \Delta_{s_2})\big)}{4} \big]\Big\} \\&+ u_{t-\Delta_t}(s_1,s_2)\Big\{\Delta_t\gamma_0\Big\} - u^{2}_{t - \Delta_t}(s_1,s_2)\Big\{\frac{\Delta_t \gamma_0}{\gamma_1} \Big\}
\end{align*}
Assuming \(\Delta_t = 1\), this is the discretized PDE that is found in \citet{hooten2008hierarchical,wikle2010general, wikle2006hierarchical} and can be written in the following matrix notation. 
\begin{align*}
    \mathbf{u}_t = \mathbf{A}(\delta)\mathbf{u}_{t-1} +\mathbf{B}(\mathbf{u}_{t-1}\otimes g(\mathbf{u}_{t-1}; \gamma_0, \gamma_1))
\end{align*}
where \(\mathbf{A}(\delta)\) is the \(n \times n\) matrix with five non-zero elements of \(\boldsymbol{\delta}\), the \(n^2 \times n\) matrix \(\mathbf{B} = (\mathbf{B}_1^{\prime}, \dots \mathbf{B}_n^{\prime})^{\prime}\). The elements of matrices \(\mathbf{B}_i, i = 1,\dots n\) are all zero except for one in the \((i,i)\)-th location. The function \(g(\cdot) = \gamma_0\mathbf{u}_{t-1}(1 - \frac{\mathbf{u}_{t-1}}{\gamma_1})\).

\section{Proof of Equation (\ref{eq.fnm})}\label{appen:fnm.proof}
In this appendix, we present the proof of Equation (\ref{eq.fnm}).
\begin{align*}
    \vert\vert \mathbf{G}\boldsymbol{\Sigma}_{\eta}\mathbf{G}^{\prime} - \boldsymbol{\Sigma}_{GQN}\vert\vert^2_F &= 
    \text{tr}\left[(\mathbf{G}\boldsymbol{\Sigma}_{\eta}\mathbf{G}^{\prime} - \boldsymbol{\Sigma}_{GQN})^\prime(\mathbf{G}\boldsymbol{\Sigma}_{\eta}\mathbf{G}^{\prime} - \boldsymbol{\Sigma}_{GQN}) \right] \\
     &= \text{tr}\left[\mathbf{G}\boldsymbol{\Sigma}_{\eta}\mathbf{G}^{\prime}\mathbf{G}\boldsymbol{\Sigma}_{\eta}^{\prime}\mathbf{G}^{\prime}\right]- \text{tr}\left[\boldsymbol{\Sigma}_{GQN}^{\prime}(\mathbf{G}\boldsymbol{\Sigma}_{\eta}\mathbf{G}^{\prime})\right] \\&- \text{tr}\left[(\mathbf{G}\boldsymbol{\Sigma}_{\eta}\mathbf{G}^{\prime})^{\prime}\boldsymbol{\Sigma}_{GQN}\right]+ \text{tr}\left[\boldsymbol{\Sigma}_{GQN}^{\prime}\boldsymbol{\Sigma}_{GQN} \right] \\
    &= \text{tr}\left[\mathbf{G}\boldsymbol{\Sigma}_{\eta}\mathbf{G}^{\prime}\mathbf{G}\boldsymbol{\Sigma}_{\eta}^{\prime}\mathbf{G}^{\prime}\right] - 2\text{tr}\left[\boldsymbol{\Sigma}_{GQN}^{\prime}\mathbf{G}\boldsymbol{\Sigma}_{\eta}\mathbf{G}^{\prime}\right] + \text{tr}\left[\boldsymbol{\Sigma}_{GQN}^{\prime}\boldsymbol{\Sigma}_{GQN} \right]
\end{align*}
Now differentiating with respect to \(\boldsymbol{\Sigma}_{\eta}\) and using the following identities: \(\dfrac{\partial}{\partial\boldsymbol{\Sigma}_{\eta}}\text{trace}(\mathbf{D}\boldsymbol{\Sigma}_{\eta}\mathbf{E}) = \mathbf{D}^{\prime}\mathbf{E}^{\prime}\) and \(\dfrac{\partial}{\partial\boldsymbol{\Sigma}_{\eta}}\text{trace}(\boldsymbol{\Sigma}_{\eta}\mathbf{D}\boldsymbol{\Sigma}_{\eta}^{\prime}\mathbf{E}) = 2\mathbf{E}\boldsymbol{\Sigma}_{\eta}\mathbf{D}\). 
\begin{align*}
    \dfrac{\partial}{\partial\boldsymbol{\Sigma}_{\eta}}\vert\vert \mathbf{G}\boldsymbol{\Sigma}_{\eta}\mathbf{G}^{\prime} - \boldsymbol{\Sigma}_{GQN}\vert\vert^2_F &= \dfrac{\partial}{\partial\boldsymbol{\Sigma}_{\eta}}\text{tr}\left[\mathbf{G}\boldsymbol{\Sigma}_{\eta}\mathbf{G}^{\prime}\mathbf{G}\boldsymbol{\Sigma}_{\eta}^{\prime}\mathbf{G}^{\prime}\right] - 2\dfrac{\partial}{\partial\boldsymbol{\Sigma}_{\eta}}\text{tr}\left[\boldsymbol{\Sigma}_{GQN}^{\prime}\mathbf{G}\boldsymbol{\Sigma}_{\eta}\mathbf{G}^{\prime}\right] + \dfrac{\partial}{\partial\boldsymbol{\Sigma}_{\eta}}\text{tr}\left[\boldsymbol{\Sigma}_{GQN}^{\prime}\boldsymbol{\Sigma}_{GQN} \right] \\
    &= 2\mathbf{G}^{\prime}\mathbf{G}\boldsymbol{\Sigma}_{\eta}\mathbf{G}^{\prime}\mathbf{G} - 2\mathbf{G}^{\prime}\boldsymbol{\Sigma}_{GQN}\mathbf{G}
\end{align*}
Setting the derivative equal to zero and solving gives the solution to our minimization problem. 
\begin{align*}
2\mathbf{G}^{\prime}\mathbf{G}\boldsymbol{\Sigma}_{\eta}\mathbf{G}^{\prime}\mathbf{G} - 2\mathbf{G}^{\prime}\boldsymbol{\Sigma}_{GQN}\mathbf{G} &= 0\\
\mathbf{G}^{\prime}\mathbf{G}\boldsymbol{\Sigma}_{\eta}\mathbf{G}^{\prime}\mathbf{G} &= \mathbf{G}^{\prime}\boldsymbol{\Sigma}_{GQN}\mathbf{G} \\
(\mathbf{G}^{\prime}\mathbf{G})^{-1}\mathbf{G}^{\prime}\mathbf{G}\boldsymbol{\Sigma}_{\eta}\mathbf{G}^{\prime}\mathbf{G}(\mathbf{G}^{\prime}\mathbf{G})^{-1} &= (\mathbf{G}^{\prime}\mathbf{G})^{-1}\mathbf{G}^{\prime}\boldsymbol{\Sigma}_{GQN}\mathbf{G}(\mathbf{G}^{\prime}\mathbf{G})^{-1} \\
\boldsymbol{\Sigma}_{\eta} &= (\mathbf{G}^{\prime}\mathbf{G})^{-1}\mathbf{G}^{\prime}\boldsymbol{\Sigma}_{GQN}\mathbf{G}(\mathbf{G}^{\prime}\mathbf{G})^{-1}
\end{align*}
Now to show the second derivative is positive semi-definite, 
\begin{align*}
 \dfrac{\partial}{\partial\, \text{vec}(\boldsymbol{\Sigma}_{\eta})} \left[\,  \text{tr}\left(\mathbf{G}^{\prime}\mathbf{G}\boldsymbol{\Sigma}_{\eta}\mathbf{G}^{\prime}\mathbf{G}\right) \, \right]  
    = 2\left( \mathbf{G}^{\prime} \mathbf{G} \otimes \mathbf{G}^{\prime} \mathbf{G} \right)
\end{align*}
Since \(\mathbf{G}^{\prime} \mathbf{G}\) is symmetric and positive semidefinite, the Kronecker product \(\mathbf{G}^{\prime} \mathbf{G} \otimes \mathbf{G}^{\prime} \mathbf{G}\) is also positive semidefinite.

\section{Proof of Equations (\ref{ch3:gcm.post}) and (\ref{equation.comp})}\label{appen:epr.proof}
In this Appendix, we will prove Equations (\ref{ch3:gcm.post}) and (\ref{equation.comp}) from the main text. That is, we want to show that the posterior distribution of \((\boldsymbol{\xi}^{\prime}, \boldsymbol{\beta}^{\prime}, \boldsymbol{\eta}^{\prime}, \mathbf{q}^{\prime})^\prime\) is distributed according to the generalized conjugate multivariate distribution. Recall from the main text that we assume our data \(\mathbf{z} \sim N(\mathbf{z} \vert \textbf{X}\boldsymbol{\beta}+\mathbf{L}\boldsymbol{\eta}+\boldsymbol{\xi}-\boldsymbol{\tau}_{y}, \boldsymbol{\Sigma}_z)\). In Supplementary Appendix \ref{appen:add.sims}, we provide additional simulations for Poisson (\(\mathbf{z}\sim \text{Poisson}(\mathbf{z} \vert \text{exp}\{\textbf{X}\boldsymbol{\beta}+\mathbf{L}\boldsymbol{\eta}+\boldsymbol{\xi}-\boldsymbol{\tau}_{y}\})\)) and Bernoulli (\(\mathbf{z}\sim \text{Bernoulli}(\mathbf{z} \vert \text{exp}\{\textbf{X}\boldsymbol{\beta}+\mathbf{L}\boldsymbol{\eta}+\boldsymbol{\xi}-\boldsymbol{\tau}_{y}\}/(1 + \text{exp}\{\textbf{X}\boldsymbol{\beta}+\mathbf{L}\boldsymbol{\eta}+\boldsymbol{\xi}-\boldsymbol{\tau}_{y}\}))\)) distributed data. So, we will prove the the posterior distribution is GCM when the data is Gaussian, Poisson, or Bernoulli distributed. The strategy is to show the following is of the form of a GCM
\begin{align*}
    f(\boldsymbol{\xi}, \boldsymbol{\beta}, \boldsymbol{\eta}, 
\mathbf{q} |\mathbf{z}) &\propto f(\mathbf{z}|\boldsymbol{\xi}, \boldsymbol{\beta}, \boldsymbol{\eta}, \boldsymbol{\theta}, \mathbf{q}) f(\boldsymbol{\xi} | \boldsymbol{\beta}, \boldsymbol{\eta}, \boldsymbol{\theta}, \mathbf{q}) f(\boldsymbol{\beta}| \boldsymbol{\theta}, \mathbf{q}) f(\boldsymbol{\eta}| \boldsymbol{\theta}, \mathbf{q}) f(\mathbf{q}) f(\boldsymbol{\theta}) \\ 
&\propto \int_{\Omega} f(\boldsymbol{\theta})f(\boldsymbol{\xi}, \boldsymbol{\beta}, \boldsymbol{\eta}, \mathbf{q}, \mathbf{z}| \boldsymbol{\theta}) d \boldsymbol{\theta}.
\end{align*}
First, we have the following data model
\begin{align*}
    f(\mathbf{z}|\boldsymbol{\xi}, \boldsymbol{\beta}, \boldsymbol{\eta}, \boldsymbol{\theta}, \mathbf{q}) = N \mbox{exp}\left[\mathbf{a}'\left\{\begin{pmatrix}\mathbf{I}_n & \mathbf{X} & \mathbf{L}\end{pmatrix} \begin{pmatrix}
    \boldsymbol{\xi} \\
    \boldsymbol{\beta}\\
    \boldsymbol{\eta}
\end{pmatrix} - \boldsymbol{\tau}_y\right \} - \mathbf{b}'\boldsymbol{\psi}_{D}\left\{\begin{pmatrix}\mathbf{I}_n & \mathbf{X} & \mathbf{L}\end{pmatrix}\begin{pmatrix}
    \boldsymbol{\xi} \\
    \boldsymbol{\beta}\\
    \boldsymbol{\eta}
\end{pmatrix} - \boldsymbol{\tau}_y\right \} \right].
\end{align*}
The parameters \(N\), \(\mathbf{a}\), and \(\mathbf{b}\) are defined in Table \ref{data.mod.param.tab} for each of the three data types of interest. 
\begin{table}[H]
\caption{Parameters in data model for Gaussian, Poisson, and Bernoulli distributions.}
\label{data.mod.param.tab}
\begin{center}
\begin{tabular}{c p{13cm}}
\toprule
\textbf{Distribution} & \textbf{Parameters} \\ \midrule 
Gaussian& When the data is Gaussian distributed, \(N = \left(\frac{1}{2\pi}\right)^{n/2} \prod_{i = 1}^n\frac{\mbox{exp}(-Z_i^2/2\sigma_i^2)}{\sigma_i}\), \(\mathbf{a} = \mathbf{D}_{\sigma}^{\prime}\mathbf{z}\) where \(\mathbf{D}_{\sigma} = \text{diag}(\frac{1}{\sigma_1^2},\dots \frac{1}{\sigma_n^2})\), and \(\mathbf{b} = \frac{1}{2}\mathbf{1}_{1,n}\mathbf{D}_{\sigma}^{\prime}\). The unit log partition function is \(\mathbf{\psi}_D(\cdot) = (\cdot)^2\).  \\ \midrule
Poisson&  When the data is Poisson distributed \(N =\prod_{i =1}^n \frac{1}{Z_i!} \), \(\mathbf{a} = \mathbf{z}\), and \(\mathbf{b} = \mathbf{1}_{n,1}\). The unit log partition function is \(\mathbf{\psi}_D(\cdot) = \text{exp}(\cdot)\).\\ \midrule 
Bernoulli& When the data is Bernoulli distributed \(N = 1 \), \(\mathbf{a} = \mathbf{z}\), and \(\mathbf{b} = \mathbf{1}_{n,1}\). The unit log partition function is \(\mathbf{\psi}_D(\cdot) = \text{log}\{1 + \text{exp}(\cdot)\}\). \\ \bottomrule
\end{tabular}
\end{center}
\end{table}
Next, we assume the following distribution for \(f(\boldsymbol{\xi}|\boldsymbol{\beta}, \boldsymbol{\eta}, \boldsymbol{\theta}, \mathbf{q})\)
\begin{align}
        f(\boldsymbol{\xi}|\boldsymbol{\beta}, \boldsymbol{\eta}, \boldsymbol{\theta}, \mathbf{q}) &\propto \left(\frac{1}{2 \pi \sigma_{\xi}^2} 
 \right)^{n/2} \mbox{exp} \left[\boldsymbol{\alpha}_{\xi}'\left\{ 
 \begin{pmatrix}
     \mathbf{I}_n & \mathbf{X} & \mathbf{L}\\
     \frac{1}{\sigma_{\xi}}\mathbf{I}_n & \mathbf{0}_{n,p} & \mathbf{0}_{n,r}
 \end{pmatrix} \begin{pmatrix}
     \boldsymbol{\xi} \\ \boldsymbol{\beta} \\ \boldsymbol{\eta} 
 \end{pmatrix} - \begin{pmatrix}
     \boldsymbol{\tau}_{y} \\ \boldsymbol{\tau}_{\xi}
 \end{pmatrix} \right\} \right.\notag\\[1em]
 & \left. - \boldsymbol{\kappa}_{\xi}' \boldsymbol{\psi}_{D,\xi} \left\{ 
 \begin{pmatrix}
     \mathbf{I}_n & \mathbf{X} & \mathbf{L}\\
     \frac{1}{\sigma_{\xi}}\mathbf{I}_n & \mathbf{0}_{n,p} & \mathbf{0}_{n,r}
 \end{pmatrix} \begin{pmatrix}
     \boldsymbol{\xi} \\ \boldsymbol{\beta} \\ \boldsymbol{\eta} 
 \end{pmatrix} - \begin{pmatrix}
     \boldsymbol{\tau}_{y} \\ \boldsymbol{\tau}_{\xi}
 \end{pmatrix} \right\} \right].
 \label{xi.eq}
\end{align}
The distribution in Equation (\ref{xi.eq}) is a conditional generalized conjugate multivariate distribution. The term \(\boldsymbol{\psi}_{D, \xi} \equiv (\boldsymbol{\psi}_D(\cdot)^{\prime}, \boldsymbol{\psi}_{\xi}(\cdot)^{\prime})\) is the \(2n\)-dimensional vector-valued function that has the first \(n\)-dimensional block consisting of functions equal to the unit log-partition function of the data and the second \(n\)-dimensional block consisting of elements equal to the unit-log partition function of the Gaussian distribution. That is, \(\boldsymbol{\psi}_D(\cdot)\) is defined in Table \ref{data.mod.param.tab} and \(\boldsymbol{\psi}_{\xi}(\cdot) = (\cdot)^2\). See Table \ref{xi.param.table} for the parameters \(\boldsymbol{\alpha}_{\xi}\) and \(\boldsymbol{\kappa}_{\xi}\) for each of the data types and \(\alpha_{\xi}\) and \(\sigma_{\xi}^2\) are known. 
\begin{table}[H]
\caption{Parameters for the conditional GCM for Gaussian, Poisson, and Bernoulli distributed data.}
\label{xi.param.table}
\begin{center}
\begin{tabular}{c p{13cm}}
\toprule
\textbf{Distribution} & \textbf{Parameters} \\ \midrule 
Gaussian& When the data is Gaussian distributed, \(\boldsymbol{\alpha}_{\xi} = \mathbf{0}_{2n,1}\) and \(\boldsymbol{\kappa}_{\xi} = (\mathbf{0}_{1,n}, \frac{1}{2}\mathbf{1}_{1,n})'\). Note that \( f(\boldsymbol{\xi}|\boldsymbol{\beta}, \boldsymbol{\eta}, \boldsymbol{\theta}, \mathbf{q}) \propto N(\boldsymbol{\xi}\vert \boldsymbol{\tau}_{\xi}, \sigma^2_{\xi}\mathbf{I}_n)\) as specified in the main text. \\ \midrule
Poisson&  When the data is Poisson distributed \(\boldsymbol{\alpha}_{\xi} = (\alpha_{\xi}\mathbf{1}_{1,n}, \mathbf{0}_{1,n})'\) and \(\boldsymbol{\kappa}_{\xi} = (\mathbf{0}_{1,n}, \frac{1}{2}\mathbf{1}_{1,n})'\).\\ \midrule 
Bernoulli& When the data is Binomial distributed \(\boldsymbol{\alpha}_{\xi} = (\alpha_{\xi}\mathbf{1}_{1,n}, \mathbf{0}_{1,n})'\) and \(\boldsymbol{\kappa}_{\xi} = (2 \alpha_{\xi}\mathbf{1}_{1,n}, \frac{1}{2}\mathbf{1}_{1,n})'\). \\ \bottomrule
\end{tabular}
\end{center}
\end{table}
Now, multiplying \(f(\boldsymbol{\xi}, \boldsymbol{\beta}, \boldsymbol{\eta}, 
\mathbf{q} |\mathbf{z})\) and \(f(\boldsymbol{\xi}|\boldsymbol{\beta}, \boldsymbol{\eta}, \boldsymbol{\theta}, \mathbf{q})\) results in  
\begin{align*}
        f(\mathbf{z}|\boldsymbol{\xi}, \boldsymbol{\beta}, \boldsymbol{\eta}, &\boldsymbol{\theta}, \mathbf{q})f(\boldsymbol{\xi}|\boldsymbol{\beta}, \boldsymbol{\eta}, \boldsymbol{\theta}, \mathbf{q}) \propto \\ &N \left(\frac{1}{2 \pi \sigma_{\xi}^2} 
 \right)^{n/2} \mbox{exp} \left[\mathbf{c}'\left\{ 
 \begin{pmatrix}
     \mathbf{I}_n & \mathbf{X} & \mathbf{L}\\
     \frac{1}{\sigma_{\xi}}\mathbf{I}_n & \mathbf{0}_{n,p} & \mathbf{0}_{n,r}
 \end{pmatrix} \begin{pmatrix}
     \boldsymbol{\xi} \\ \boldsymbol{\beta} \\ \boldsymbol{\eta} 
 \end{pmatrix} - \begin{pmatrix}
     \boldsymbol{\tau}_{y} \\ \boldsymbol{\tau}_{\xi}
 \end{pmatrix} \right\} \right.\\[1em]
 & \left. - \mathbf{d}' \boldsymbol{\psi}_{D,\xi} \left\{ 
 \begin{pmatrix}
     \mathbf{I}_n & \mathbf{X} & \mathbf{G}\\
     \frac{1}{\sigma_{\xi}}\mathbf{I}_n & \mathbf{0}_{n,p} & \mathbf{0}_{n,r}
 \end{pmatrix} \begin{pmatrix}
     \boldsymbol{\xi} \\ \boldsymbol{\beta} \\ \boldsymbol{\eta} 
 \end{pmatrix} - \begin{pmatrix}
     \boldsymbol{\tau}_{y} \\ \boldsymbol{\tau}_{\xi}
 \end{pmatrix} \right\} \right].
\end{align*}
Table \ref{multiply.params} defines the parameters \(\mathbf{c}\) and \(\mathbf{d}\) for each of the data types.
\begin{table}[H]
\caption{Parameters for the product \( f(\mathbf{z}|\boldsymbol{\xi}, \boldsymbol{\beta}, \boldsymbol{\eta}, \boldsymbol{\theta}, \mathbf{q})f(\boldsymbol{\xi}|\boldsymbol{\beta}, \boldsymbol{\eta}, \boldsymbol{\theta}, \mathbf{q})\) for Gaussian, Poisson, and Bernoulli distributed data.}
\label{multiply.params}
\begin{center}
\begin{tabular}{c p{13cm}}
\toprule
\textbf{Distribution} & \textbf{Parameters} \\ \midrule 
Gaussian& When the data is Gaussian distributed \(\mathbf{c} = (\mathbf{z}^{\prime}\mathbf{D}_{\sigma},\mathbf{0}_{1,n})^{\prime}\) and \(\mathbf{d} = (\frac{1}{2}\mathbf{1}_{1,n}\mathbf{D}_{\sigma}, \frac{1}{2}\mathbf{1}_{1,n})^{\prime}\). \\ \midrule
Poisson &  When the data is Poisson distributed \(\mathbf{c} = (\mathbf{z}^{\prime} + \alpha_{\xi}\mathbf{1}_{1,n}, \mathbf{0}_{1,n})^{\prime}\) and \(\mathbf{d} = (\mathbf{1}_{1,n}, \frac{1}{2}\mathbf{1}_{1,n})'\). \\ \midrule 
Bernoulli& When the data is Bernoulli distributed \(\mathbf{c} = (\mathbf{z}^{\prime} + \alpha_{\xi}\mathbf{1}_{1,n}, \mathbf{0}_{1,n})'\) and \(\mathbf{d} = (\mathbf{1}_{1,n} + 2 \alpha_{\xi}\mathbf{1}_{1,n}, \frac{1}{2}\mathbf{1}_{1,n})^{\prime}\).  \\ \bottomrule
\end{tabular}
\end{center}
\end{table}
Finally, multiplying \(f(\mathbf{z}|\boldsymbol{\xi}, \boldsymbol{\beta}, \boldsymbol{\eta}, \boldsymbol{\theta}, \mathbf{q}) f(\boldsymbol{\xi} | \boldsymbol{\beta}, \boldsymbol{\eta}, \boldsymbol{\theta}, \mathbf{q})\) by \(f(\boldsymbol{\beta}| \boldsymbol{\theta}, \mathbf{q}) f(\boldsymbol{\eta}| \boldsymbol{\theta}, \mathbf{q}) f(\mathbf{q}) f(\boldsymbol{\theta})\) gives the following distribution, 
\begin{align}
f&(\boldsymbol{\xi}, \boldsymbol{\beta}, \boldsymbol{\eta}, 
\mathbf{q} |\mathbf{z}) \propto \notag\\
& \frac{\pi(\boldsymbol{\theta})N^*}{\mbox{det}\{\mathbf{D}(\boldsymbol{\theta})\}}\mbox{exp} \left[\boldsymbol{\alpha}'\left\{ 
 \begin{pmatrix}
     \mathbf{I}_n & \mathbf{X} & \mathbf{L}\\
     \bm{0}_{p,n} & \mathbf{D}_{\beta}(\theta)^{-1} & \bm{0}_{p,r} \\
     \bm{0}_{r,n} & \bm{0}_{r,p} & \mathbf{D}_{\eta}(\theta)^{-1} \\
     \frac{1}{\sigma_{\xi}}\mathbf{I}_n & \mathbf{0}_{n,p} & \mathbf{0}_{n,r}
 \end{pmatrix} \begin{pmatrix}
     \boldsymbol{\xi} \\ \boldsymbol{\beta} \\ \boldsymbol{\eta} 
 \end{pmatrix} - \begin{pmatrix}
      \boldsymbol{\tau}_{y} \\ \boldsymbol{\tau}_{\beta} \\ \boldsymbol{\tau}_{\eta} \\\boldsymbol{\tau}_{\xi} 
 \end{pmatrix} \right\} \right.\notag\\[1em]
 & \left. - \boldsymbol{\kappa}' \boldsymbol{\psi} \left\{ 
 \begin{pmatrix}
     \mathbf{I}_n & \mathbf{X} & \mathbf{L}\\
     \bm{0}_{p,n} & \mathbf{D}_{\beta}(\theta)^{-1} & \bm{0}_{p,r} \\
     \bm{0}_{r,n} & \bm{0}_{r,p} & \mathbf{D}_{\eta}(\theta)^{-1} \\
     \frac{1}{\sigma_{\xi}}\mathbf{I}_n & \mathbf{0}_{n,p} & \mathbf{0}_{n,r}
 \end{pmatrix} \begin{pmatrix}
     \boldsymbol{\xi} \\ \boldsymbol{\beta} \\ \boldsymbol{\eta} 
 \end{pmatrix} - \begin{pmatrix}
     \boldsymbol{\tau}_{y} \\ \boldsymbol{\tau}_{\beta} \\ \boldsymbol{\tau}_{\eta} \\\boldsymbol{\tau}_{\xi} 
 \end{pmatrix} \right\} \right].
 \label{post.prop}
\end{align}
Recall from the main text,  \(\boldsymbol{\tau} = -\mathbf{D}(\boldsymbol{\theta})^{-1}\mathbf{Q}\mathbf{q}\) and let \(\boldsymbol{\zeta} = (\boldsymbol{\xi}', \boldsymbol{\beta}', \boldsymbol{\eta}')'\). Substituting, \(-\mathbf{D}(\boldsymbol{\theta})^{-1}\mathbf{Q}\mathbf{q}\) and \(\boldsymbol{\zeta}\) into Equation (\ref{post.prop}) and integrating across \(\boldsymbol{\theta}\), gives the following distribution, 
\begin{align*}
    f&(\boldsymbol{\xi}, \boldsymbol{\beta}, \boldsymbol{\eta}, 
\mathbf{q} |\mathbf{z}) \propto \\
&\int_{\Omega} \frac{\pi(\boldsymbol{\theta})N^*}{\mbox{det}\{\mathbf{D}(\boldsymbol{\theta})\}} \mbox{exp}\left[\boldsymbol{\alpha}' \left\{ \mathbf{D}(\boldsymbol{\theta})^{-1}\begin{pmatrix}
    \mathbf{H} & \mathbf{Q}
\end{pmatrix} \begin{pmatrix}
    \boldsymbol{\zeta} \\ \mathbf{q}
\end{pmatrix} \right \} - \boldsymbol{\kappa}'\boldsymbol{\psi} \left\{\mathbf{D}(\boldsymbol{\theta})^{-1}\begin{pmatrix}
    \mathbf{H} & \mathbf{Q}
\end{pmatrix} \begin{pmatrix}
    \boldsymbol{\zeta} \\ \mathbf{q}
\end{pmatrix}  \right\}  \right] d \boldsymbol{\theta} \\
&\propto \mbox{GCM}(\boldsymbol{\alpha}, \boldsymbol{\kappa}, \bm{0}_{2n + p + r, 1}, \mathbf{V}, \pi, \mathbf{D}; \boldsymbol{\psi}).
\end{align*}
This is the GCM stated in Equation (\ref{ch3:gcm.post}) of the main text where \(\mathbf{V}^{-1} = (\mathbf{H}, \mathbf{Q})\). The term \(\boldsymbol{\psi} \equiv (\boldsymbol{\psi}_D(\cdot)^{\prime}, \boldsymbol{\psi}_{\xi}(\cdot)^{\prime}, \boldsymbol{\psi}_{\beta}(\cdot)^{\prime}, \boldsymbol{\psi}_{\eta}(\cdot)^{\prime})\) is the \(2n+p+r\)-dimensional vector-valued function that has the first \(n\)-dimensional block consists of functions equal to the unit log-partition function of the data and the second \(n\)-dimensional block, third \(p\)-dimensional block, and fourth
\(r\)-dimensional block consists of elements equal to the unit-log partition function of the Gaussian distribution. The parameters \(\boldsymbol{\alpha}\) and \(\boldsymbol{\kappa}\) are given in Table \ref{gcm.post.param} for each data type. 

\begin{table}[H]
\caption{Parameters for the GCM posterior for Gaussian, Poisson, and Bernoulli distributed data.}
\label{gcm.post.param}
\begin{center}
\begin{tabular}{c p{13cm}}
\toprule
\textbf{Distribution} & \textbf{Parameters} \\ \midrule 
Gaussian&  When the data is Gaussian distributed, \(\boldsymbol{\alpha} = (\mathbf{z}^{\prime}\mathbf{D}_{\sigma}, \bm{0}_{1, n + p + r})^{\prime}\) and \(\boldsymbol{\kappa} = (\frac{1}{2}\mathbf{1}_{1,n}\mathbf{D}_{\sigma}^{\prime}, \frac{1}{2}\mathbf{1}_{1, n + p + r})^{\prime}\). \\ \midrule
Poisson&  When the data is Poisson distributed, \(\boldsymbol{\alpha} = (\mathbf{z}^{\prime} + \alpha_{\xi}\mathbf{1}_{1, n}, \mathbf{0}_{1, n + p + r})^{\prime}\) and \(\boldsymbol{\kappa} = (\mathbf{1}_{1,n}, \frac{1}{2}\mathbf{1}_{1, n + p + r})^{\prime}\).\\ \midrule 
Bernoulli& When the data is Bernoulli distributed, \(\mathbf{\alpha} = (\mathbf{z}^{\prime} + \alpha_{\xi}\mathbf{1}_{1,n}, \mathbf{0}_{1, n + p + r})^{\prime}\) and \(\mathbf{\kappa} = (\mathbf{1}_{1,n} + 2\alpha_{\xi}\mathbf{1}_{1,n}, \frac{1}{2}\mathbf{1}_{1, n + p + r})^{\prime}\). \\ \bottomrule
\end{tabular}
\end{center}
\end{table}
Now recall from Section \ref{rev:gcm} of the main text that a GCM random vector is sampled by first sampling \(\boldsymbol{\theta}\) from its prior and then computing the following transformation
\begin{align}
    \mathbf{h} = \boldsymbol{\mu} + \mathbf{V}\mathbf{D}(\boldsymbol{\theta})\mathbf{w}
\end{align}
We have \(\mathbf{V}^{-1} = (\mathbf{H}, \mathbf{Q})\) and we can verify that 
\begin{align*}
    \mathbf{V} = (\mathbf{H}, \mathbf{Q})^{-1}  =    \begin{pmatrix}
        (\mathbf{H}'\mathbf{H})^{-1}\mathbf{H}^{\prime} \\ \mathbf{Q}^{\prime}
    \end{pmatrix}.
\end{align*}
Recall from the main text, \(\mathbf{Q}\mathbf{Q}^{\prime} =  \mathbf{I}_{2n+p +r} - \mathbf{H}(\mathbf{H}^{\prime}\mathbf{H})^{-1}\mathbf{H}^{-1}\) and \(\mathbf{H}^{\prime}\mathbf{Q} = \mathbf{0}_{n + p + r, n}\). We have, 
\begin{align*}
    \begin{pmatrix}
        \mathbf{H} & \mathbf{Q}
    \end{pmatrix}\begin{pmatrix}
        \mathbf{H} & \mathbf{Q}
    \end{pmatrix}^{-1} &= \begin{pmatrix}
        \mathbf{H} & \mathbf{Q}
    \end{pmatrix}\begin{pmatrix}
        (\mathbf{H}^{\prime}\mathbf{H})^{-1}\mathbf{H}^{\prime} \\
        \mathbf{Q}^{\prime}
    \end{pmatrix}\\
    &= \mathbf{H}(\mathbf{H}^{\prime}\mathbf{H})^{-1}\mathbf{H}^{\prime}+ \mathbf{Q}\mathbf{Q}^{\prime} \\
    &= \mathbf{H}(\mathbf{H}^{\prime}\mathbf{H})^{-1}\mathbf{H}^{\prime}+ \mathbf{I}_{2n + p + r} - \mathbf{H}(\mathbf{H}^{\prime}\mathbf{H})^{-1}\mathbf{H}^{\prime} \\
    &= \mathbf{I}_{2n + p + r}
\end{align*}
We also have, 
\begin{align*}
    \begin{pmatrix}
        \mathbf{H} & \mathbf{Q}
    \end{pmatrix}^{-1}\begin{pmatrix}
        \mathbf{H} & \mathbf{Q}
    \end{pmatrix} &= \begin{pmatrix}
        (\mathbf{H}^{\prime}\mathbf{H})^{-1}\mathbf{H}^{\prime} \\
        \mathbf{Q}^{\prime}
    \end{pmatrix}\begin{pmatrix}
        \mathbf{H} & \mathbf{Q}
    \end{pmatrix}\\
    &= \begin{pmatrix}
        (\mathbf{H}^{\prime}\mathbf{H})^{-1}\mathbf{H}^{\prime}\mathbf{H} &  (\mathbf{H}^{\prime}\mathbf{H})^{-1}\mathbf{H}^{\prime}\mathbf{Q} \\
        \mathbf{Q}^{\prime}\mathbf{H} & \mathbf{Q}^{\prime}\mathbf{Q}
    \end{pmatrix}\\
    &= \begin{pmatrix}
        \mathbf{I}_{n + p + r} & \mathbf{0}_{n + p + r, n} \\
        \mathbf{0}_{n, n + p + r} & \mathbf{I}_{n}
    \end{pmatrix} \\
    &= \mathbf{I}_{2n + p + r}
\end{align*}
The matrix \(\mathbf{D}(\boldsymbol{\theta})\) is defined in the main text and \(\boldsymbol{\mu} = \mathbf{0}_{2n + p +r}\). From the transformation equation, we have the posterior replicates of \((\boldsymbol{\xi}^{\prime}, \boldsymbol{\beta}^{\prime}, \boldsymbol{\xi}^{\prime}, \mathbf{q}^{\prime})\) have the following form, 
\begin{align*}
    \begin{pmatrix}
        \boldsymbol{\xi}_{rep} \\
        \boldsymbol{\beta}_{rep} \\
        \boldsymbol{\eta}_{rep}
    \end{pmatrix} &= (\mathbf{H}'\mathbf{H})^{-1}\mathbf{H}^{\prime}{\mathbf{w}_{rep}} \\
    \textbf{q}_{rep} &= \textbf{Q}^{\prime}{\textbf{w}_{rep}},
\end{align*}
where \({\mathbf{w}_{rep}} = (\mathbf{y}_{rep}^{\prime}, \mathbf{w}_{\beta}^{\prime}, \mathbf{w}_{\eta}^{\prime}, \mathbf{w}_{\xi}^{\prime})\).
\section{Additional Simulations}\label{appen:add.sims}
\subsection{Additional Simulation Setup}\label{ch3:s:add.sims}
In the main text, we provide a simulation study were the spatio-temporal data is assumed to be Gaussian distributed. In this appendix, we provide simulations for Poisson distributed data and Bernoulli distributed data as these data types are also of interest. Similar to Section \ref{ss:complex.sim} we compare the Frobenius norm matching strategy implemented with EPR and MCMC. We simulate spatio-temporal data on a two-dimensional spatial grid of \(n = 10 \times 10\) locations over \(T =14\) time points. The Poisson distributed data is simulated with mean \(\text{exp}(Y_t(\mathbf{s}_i))\) and the Bernoulli distributed data is simulated with probability \(\frac{\text{exp}(Y_t(\mathbf{s}_i))}{1 + \text{exp}(Y_t(\mathbf{s}_i))}\), where \({Y}_t(\mathbf{s}_i) = \mathbf{x}_t(\mathbf{s}_i)'\boldsymbol{\beta} + {U}_t(\mathbf{s}_i)\). The process \({U}_t(\mathbf{s}_i)\) is simulated from Equation (\ref{eq:gqn}) from the main text with parameters \(a_{ij}\) and \(b_{i,kl}\) as defined in Equation (\ref{sim.specification}) from the main text and all other parameters are defined in Table \ref{ch2:sim.param.table}.

\begin{table}[H]
\caption{Parametes used for GQN simulation setup for additional data types of interest.}
\label{ch2:sim.param.table}
\begin{center}
\begin{tabular}{c p{13cm}}
\toprule
\textbf{Distribution} & \textbf{Parameters for Simulation} \\ \midrule 
Poisson&  \(\delta_1 = 0.001\), \(\delta_2 = 0.0015\), \(\nu = 0.1\), \(p_a = 0.9\), \(p_b = 0.75\), \(\rho = 4\), and \(d(\mathbf{s}_i, \mathbf{s}_j) \equiv \vert \vert \mathbf{s}_i - \mathbf{s}_j \vert \vert\),. The nonlinear function \(g(\cdot)\) is defined in Section \ref{s:gqn}, with parameters \(\gamma_0 = 0.0001\) and \(\gamma_1 = 20\). We assume \(\boldsymbol{\eta}_t\sim \text{MVN}(\mathbf{0}, \boldsymbol{\Sigma}_{\eta})\) and \(\mathbf{U}_{0} \sim \text{MVN}(\mathbf{0}, \boldsymbol{\Sigma}_0)\) where \(\boldsymbol{\Sigma}_{\eta} = \sigma_{\eta}^2 \text{exp}\left(-\frac{\mathbf{D}}{\phi_{\eta}} \right)\), \(\sigma^2_{\eta} = 0.2\), \(\boldsymbol{\Sigma}_0 = \sigma_0^2 \text{exp}\left(-\frac{\mathbf{D}}{\phi_{0}} \right)\), \(\sigma^2_0 = 0.4\), \(\phi_{\eta}=20\), \(\phi_0 = 25\) and \(\mathbf{D}\) is the \(n \times n \) matrix of pairwise Euclidean distances between spatial locations. The true value of \(\beta = 1\).\\ \midrule 
Bernoulli& \(\delta_1 = 0.009\), \(\delta_2 = 0.0017\), \(\nu = 0.03\), \(p_a = 0.9\), \(p_b = 0.95\), \(\rho = 3\), and \(d(\mathbf{s}_i, \mathbf{s}_j) \equiv \vert \vert \mathbf{s}_i - \mathbf{s}_j \vert \vert\),. The nonlinear function \(g(\cdot)\) is defined in Section \ref{s:gqn}, with parameters \(\gamma_0 = 0.01\) and \(\gamma_1 = 20\). We assume \(\boldsymbol{\eta}_t\sim \text{MVN}(\mathbf{0}, \boldsymbol{\Sigma}_{\eta})\) and \(\mathbf{U}_{0} \sim \text{MVN}(\mathbf{0}, \boldsymbol{\Sigma}_0)\) where \(\boldsymbol{\Sigma}_{\eta} = \sigma_{\eta}^2 \text{exp}\left(-\frac{\mathbf{D}}{\phi_{\eta}} \right)\), \(\sigma^2_{\eta} = 0.3\), \(\boldsymbol{\Sigma}_0 = \sigma_0^2 \text{exp}\left(-\frac{\mathbf{D}}{\phi_{0}} \right)\), \(\sigma^2_0 = 0.4\), \(\phi_{\eta}=20\), \(\phi_0 = 25\) and \(\mathbf{D}\) is the \(n \times n \) matrix of pairwise Euclidean distances between spatial locations. The true value of \(\beta = -0.5\). \\ \bottomrule
\end{tabular}
\end{center}
\end{table}

\subsection{Poisson GQN Simulation Study}
\begin{figure}[H]
    \centering
    
    \begin{subfigure}{0.3\textwidth}
        \includegraphics[width=\linewidth]{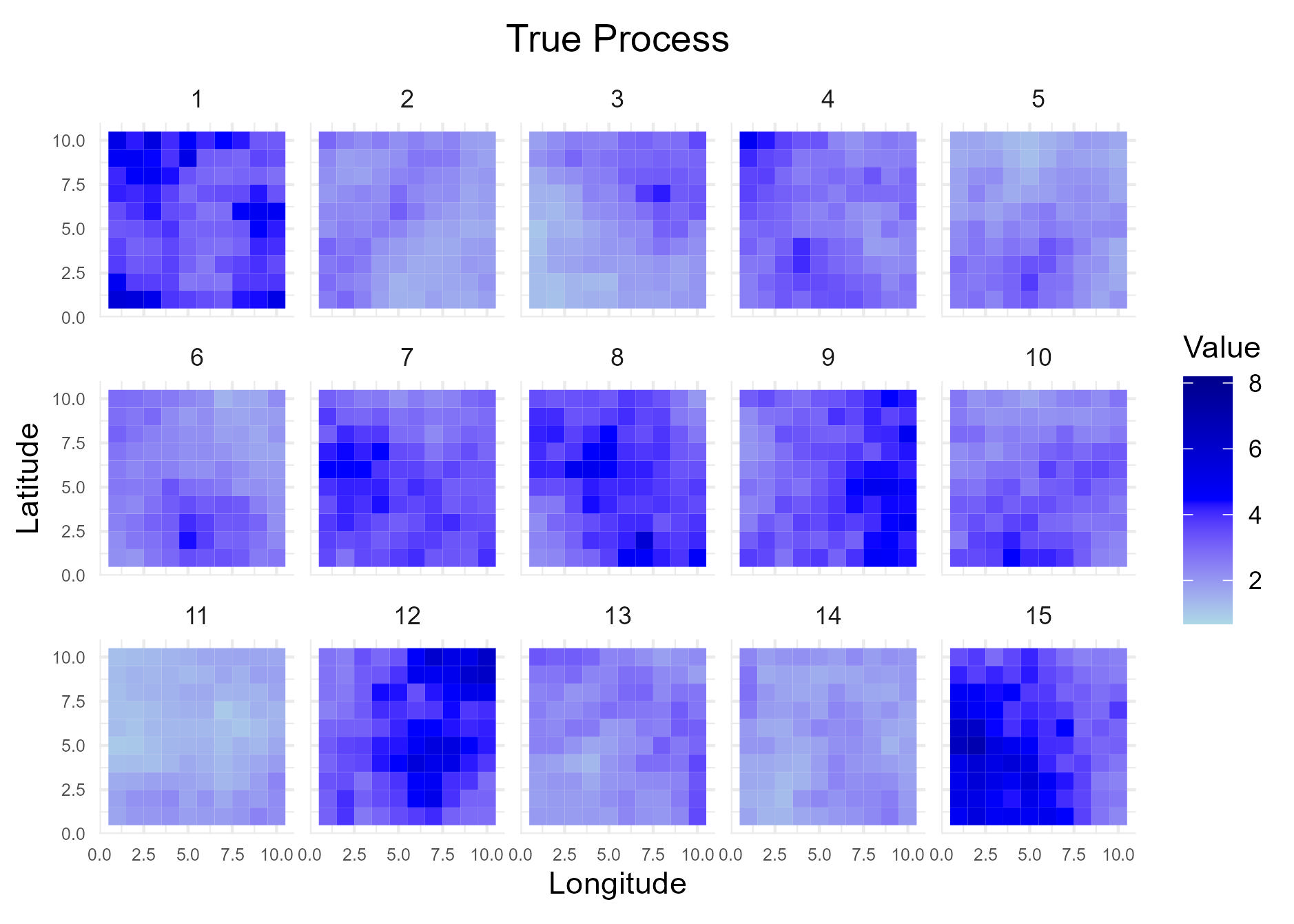}
    \end{subfigure}
    \hfill
    \begin{subfigure}{0.3\textwidth}
        \includegraphics[width=\linewidth]{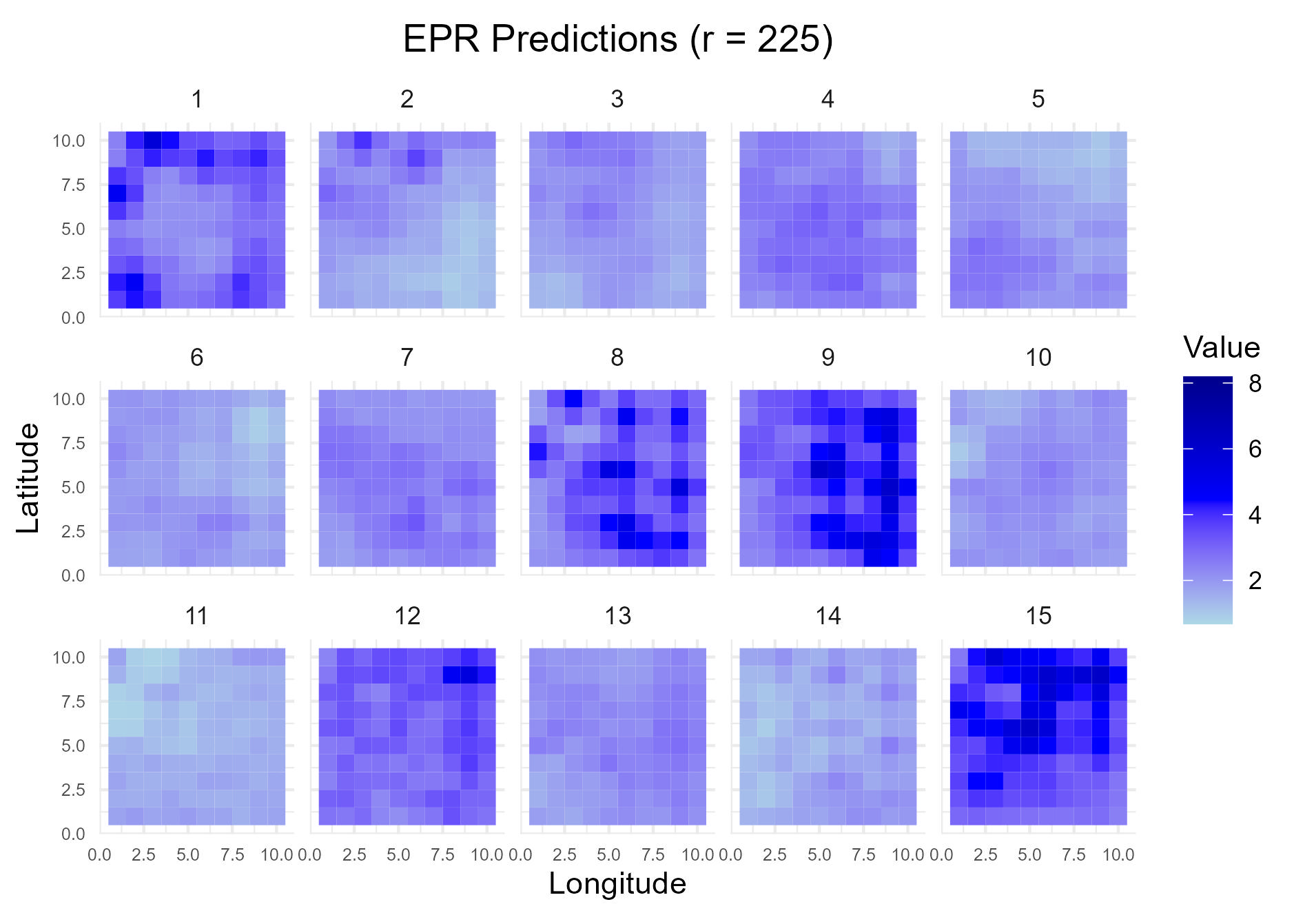}
    \end{subfigure}
    \hfill
    \begin{subfigure}{0.3\textwidth}
        \includegraphics[width=\linewidth]{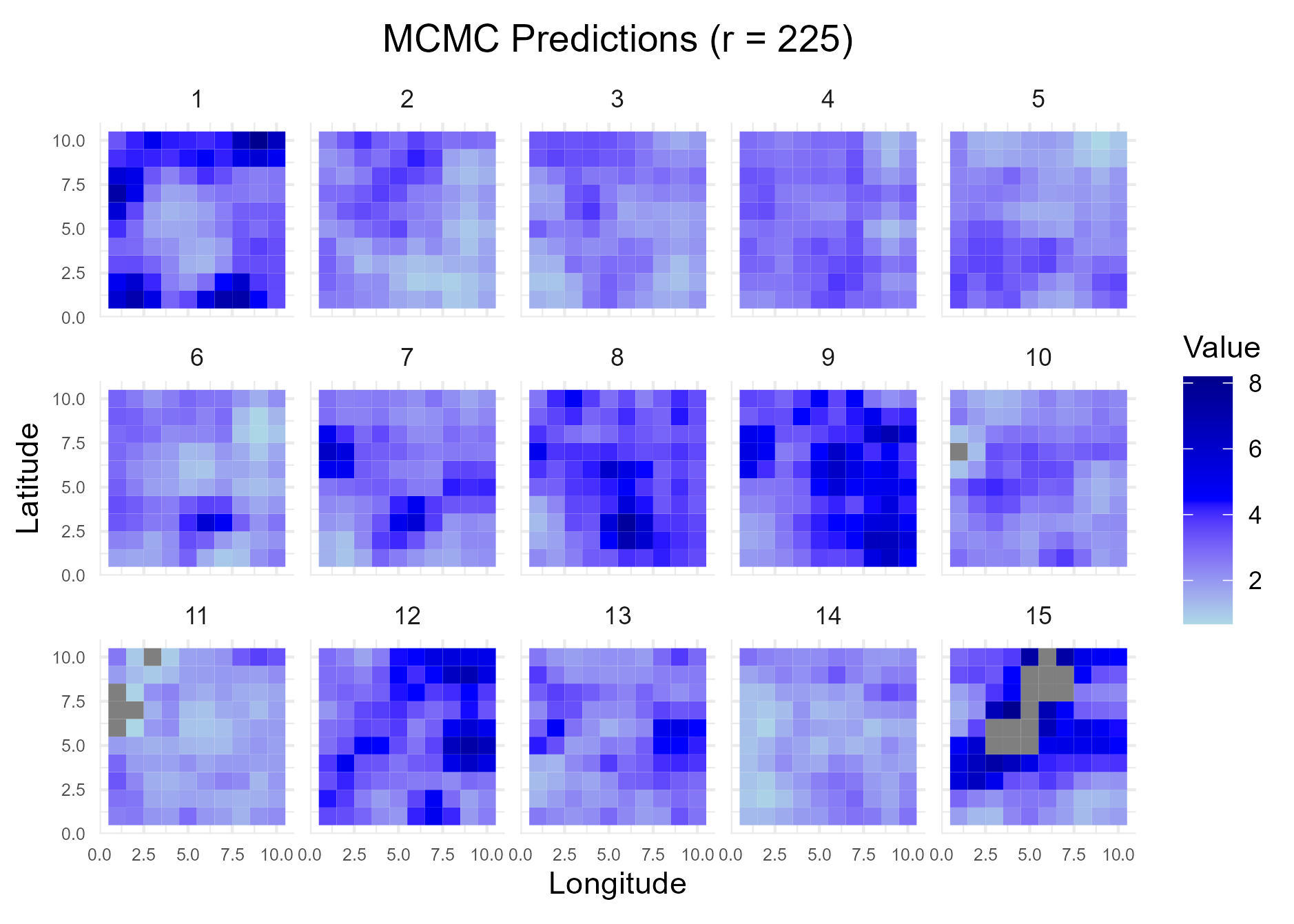}
    \end{subfigure}
    
    \par\bigskip
    \begin{subfigure}{0.3\textwidth}
        \includegraphics[width=\linewidth]{true.data.p.jpeg}
    \end{subfigure}
    \hfill
    \begin{subfigure}{0.3\textwidth}
        \includegraphics[width=\linewidth]{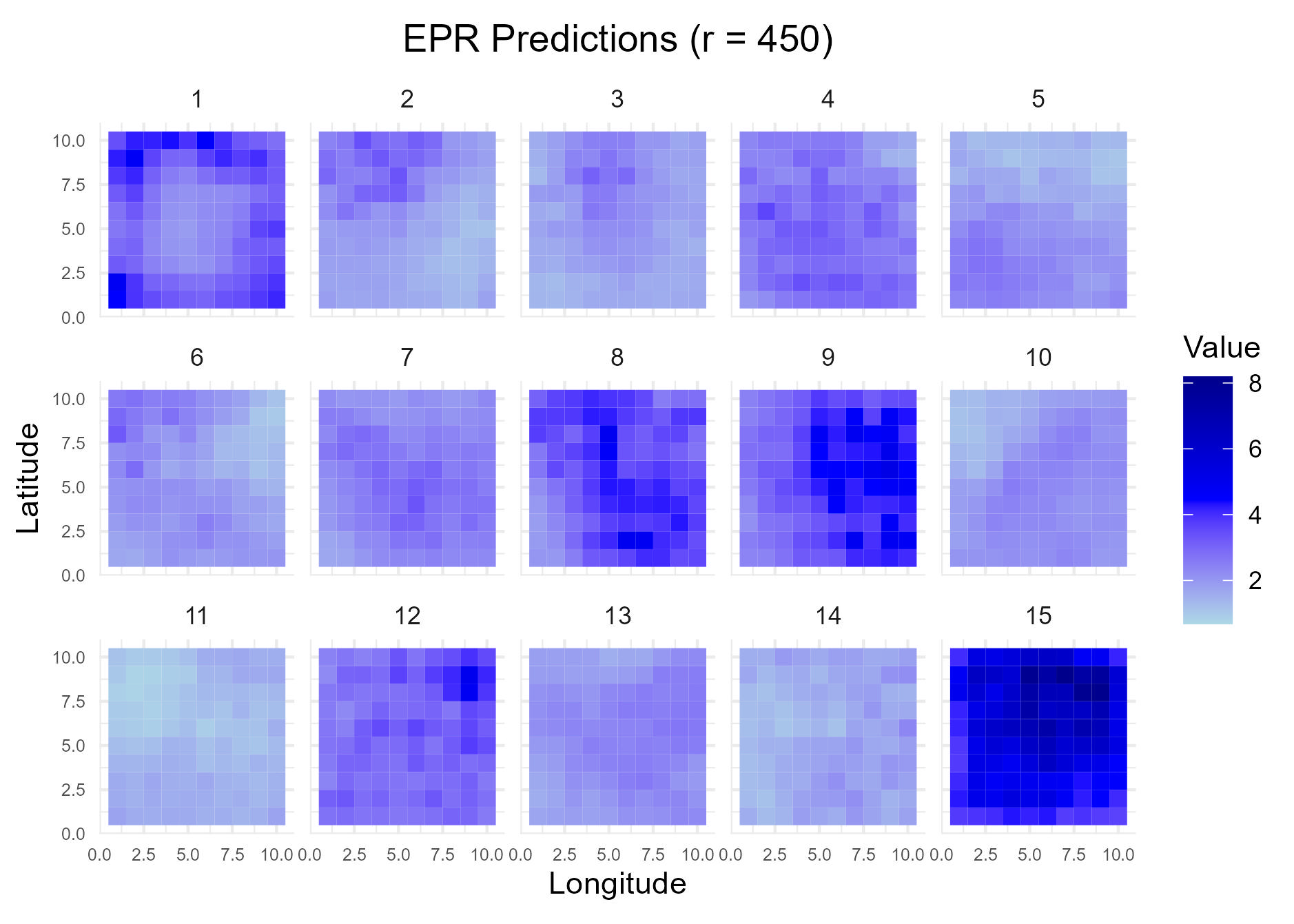}
    \end{subfigure}
    \hfill
    \begin{subfigure}{0.3\textwidth}
        \includegraphics[width=\linewidth]{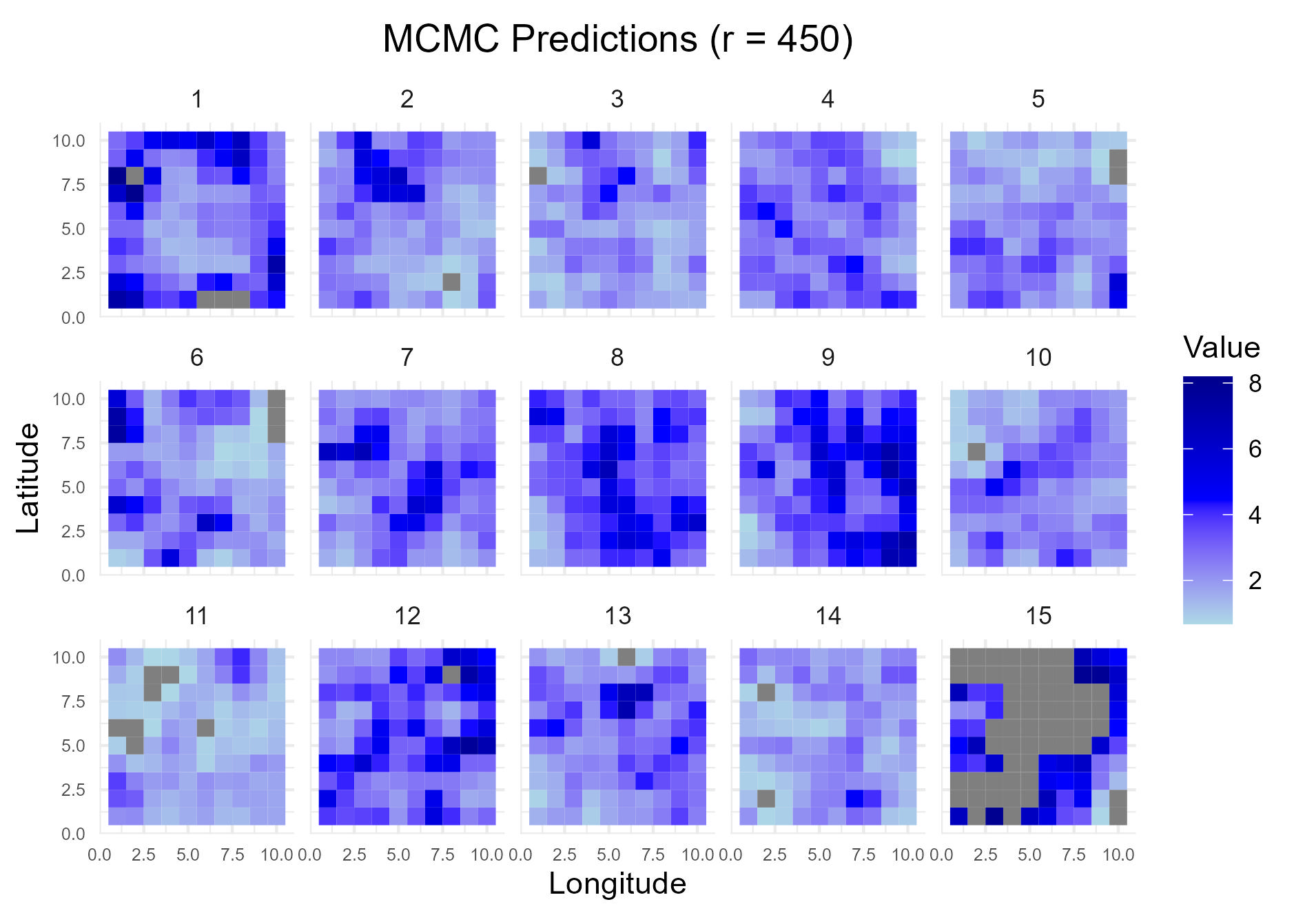}
    \end{subfigure}
    
    \par\bigskip
    \begin{subfigure}{0.3\textwidth}
        \includegraphics[width=\linewidth]{true.data.p.jpeg}
    \end{subfigure}
    \hfill
    \begin{subfigure}{0.3\textwidth}
        \includegraphics[width=\linewidth]{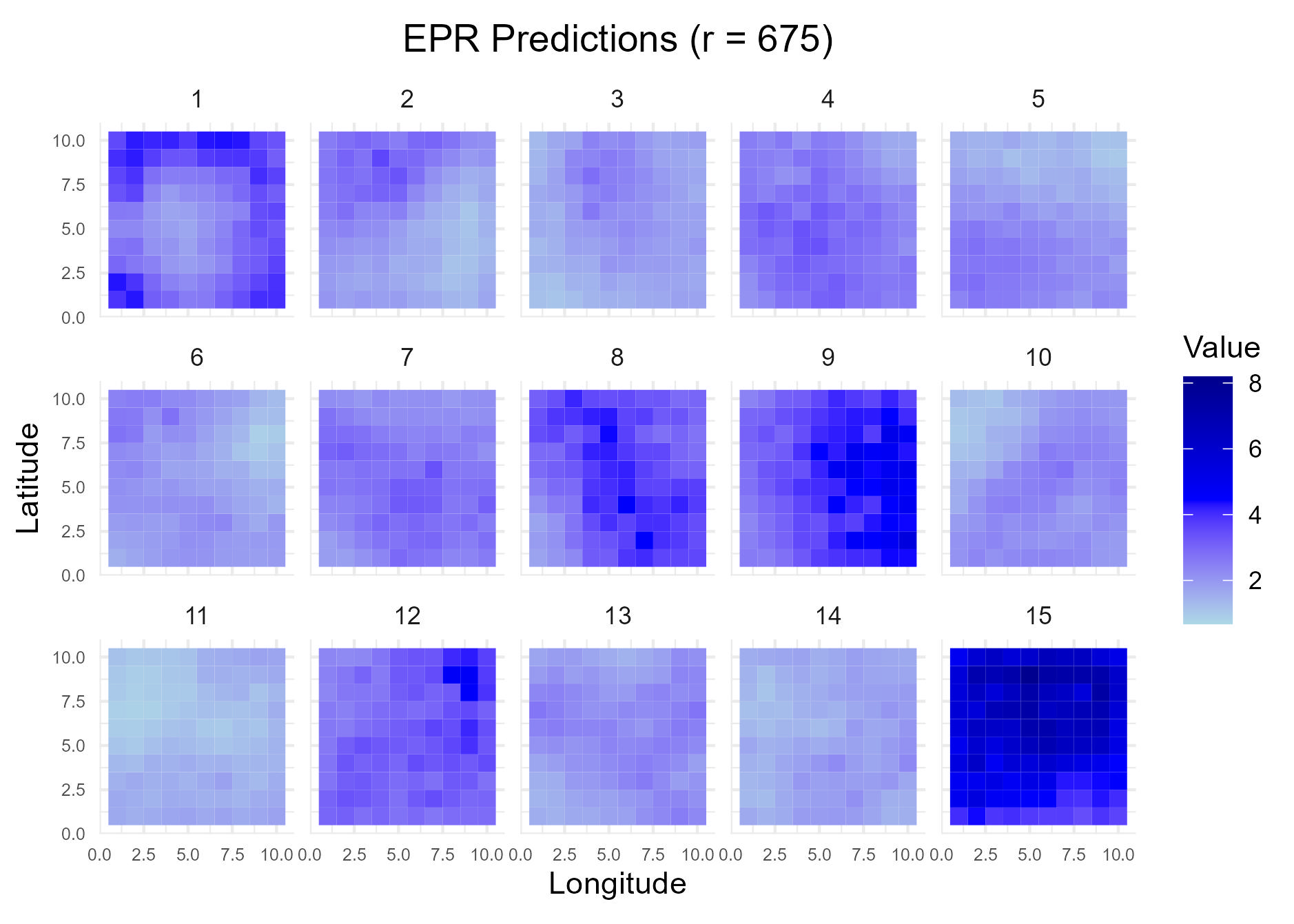}
    \end{subfigure}
    \hfill
    \begin{subfigure}{0.3\textwidth}
        \includegraphics[width=\linewidth]{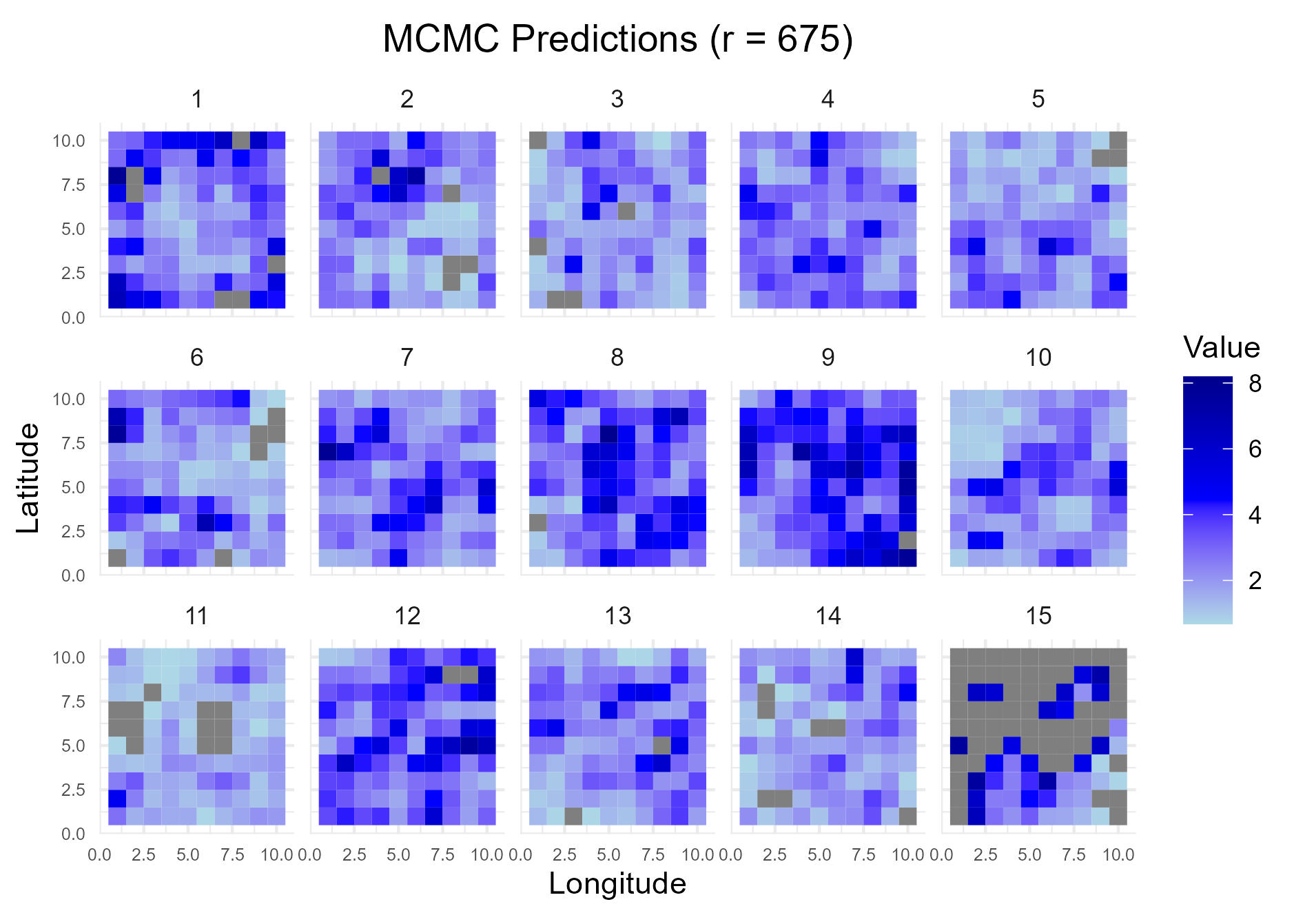}
    \end{subfigure}
    
    \par\bigskip
    \begin{subfigure}{0.3\textwidth}
        \includegraphics[width=\linewidth]{true.data.p.jpeg}
    \end{subfigure}
    \hfill
    \begin{subfigure}{0.3\textwidth}
        \includegraphics[width=\linewidth]{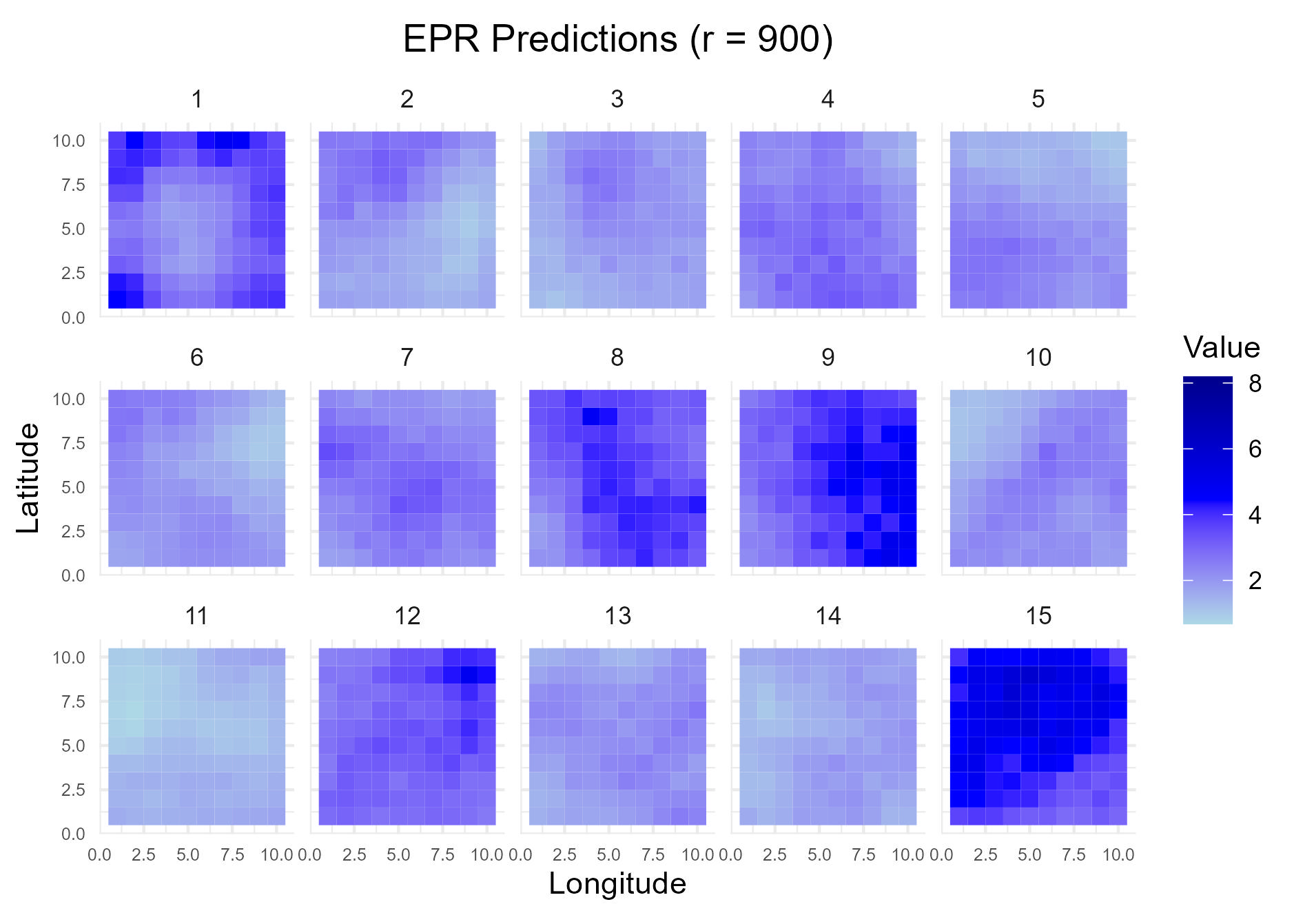}
    \end{subfigure}
    \hfill
    \begin{subfigure}{0.3\textwidth}
        \includegraphics[width=\linewidth]{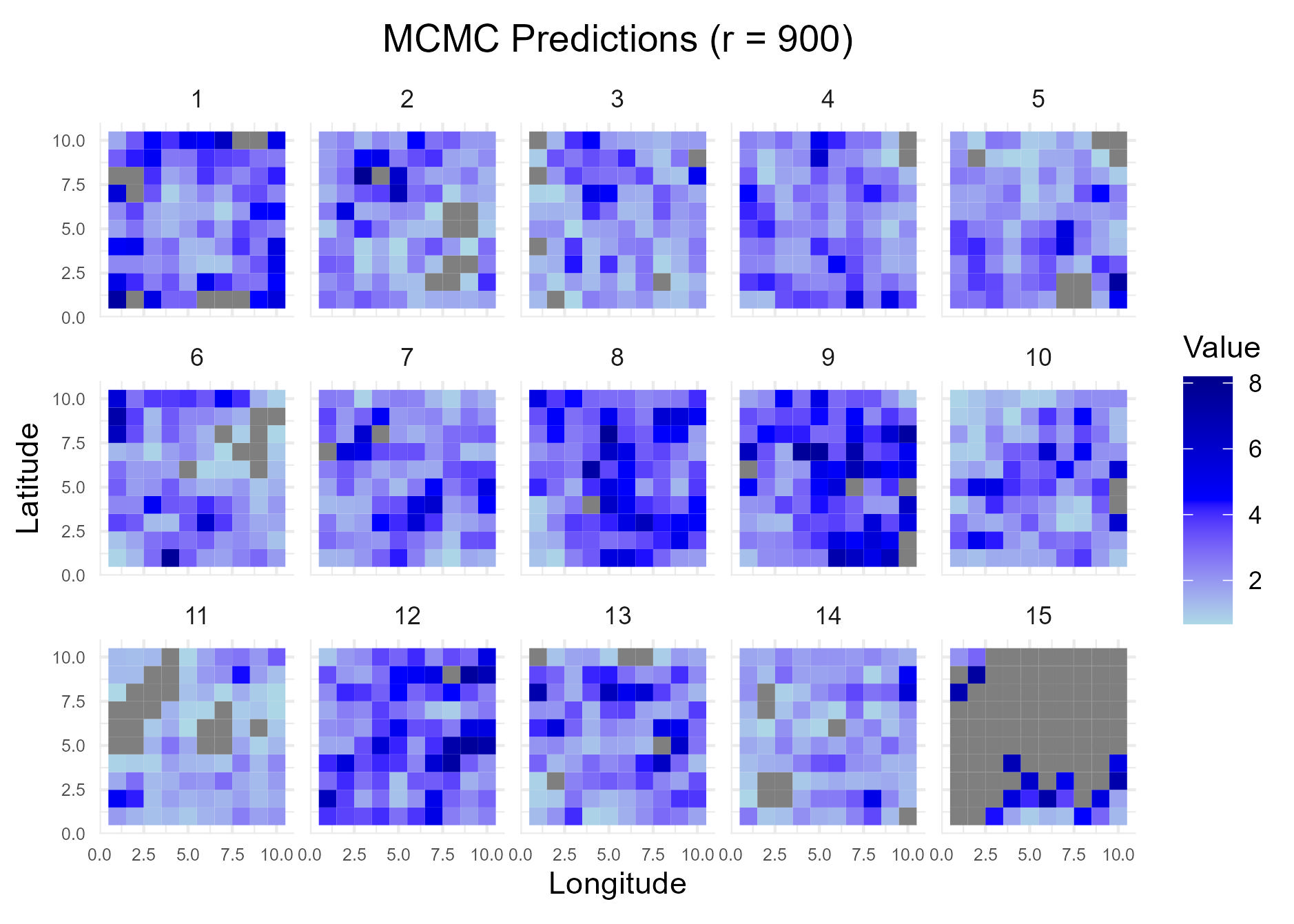}
    \end{subfigure}
    
    \caption{The first column is the true process simulated from the GQN, the second column is the posterior mean from the model implemented with EPR, and the third column is the posterior mean from the model implemented with MCMC. Each row is the comparison with \(r = 225, 450, 675, 900\) spatio-temporal basis functions used in the Frobenius norm matching, respectively. The \(T+1\) time point in each example was forecasted.} 
    \label{fig:mcmc_comp.poiss}
\end{figure}

In this appendix, we present the results for the Poisson distributed simulations. Table \ref{fig:fnm.compare.poiss} presents predictive and computational evaluation metrics for the calibrated linear model implemented with EPR and MCMC. Results are reported across different values of \(r\) where recall from the main text, increase the number of basis functions improves the ability to capture the nonlinear dynamics from the GQN covariance. The MSPE and the CRPS decrease for FNM-EPR as the rank of \(\mathbf{G}\) increases. Similar to the Gaussian simulations, when \(r\) increases, the MSPE and CRPS for the model implemented with FNM-MCMC increases. The forecast error for all scenarios of \(r\) is smaller for FNM-EPR compared to FNM-MCMC. For this simulation setup, the forecast error for FNM-EPR performs similarly for the different values of \(r\) indicated by the overlapping confidence intervals. In all scenarios, we see that FNM-EPR is significantly faster in terms of CPU time compared to FNM-MCMC which is expected since EPR simulates replicates directly from the posterior and does not require a burn in period or large amount of iterations. 
\begin{table}[H]
\centering
\begin{tabular}{lcccccc}
\toprule
&Approach & Forecast & MSPE & MSE & CRPS & CPU Time (sec) \\ 
\midrule 
\multicolumn{7}{c}{\textbf{Scenario:} $r=225$} \\ \midrule
&FNM-EPR &  0.222  & 0.090  &  0.056  & 0.183  &  11.7 \\
&\phantom{EPR} & (0.000, 0.576) & (0.066, 0.114) & (0.000, 0.124) & (0.155, 0.211) & (11.5, 11.9) \\
&FNM-MCMC &  1.130  & 0.134  & 1.012 & 0.201  &  261.5 \\
&\phantom{MCMC} & (0.086, 2.174) & (0.092, 0.176) & (0.608, 1.416) & (0.171, 0.231) & (250.8, 272.2) \\
\midrule
\multicolumn{7}{c}{\textbf{Scenario:} $r=450$} \\ \midrule
&FNM-EPR  & 0.192  & 0.087  & 0.050  & 0.177  & 45.1  \\
&\phantom{EPR}  & (0.000, 0.510) & (0.063, 0.111) & (0.000, 0.122) & (0.149, 0.205) & (44.3, 45.9) \\
&FNM-MCMC & 1.941  & 0.263  & 1.021  & 0.283  & 699.1  \\
&\phantom{MCMC}  & (0.073, 3.809) & (0.199, 0.327) & (0.641, 1.401) & (0.251, 0.315) & (670.9, 727.3) \\
\midrule
\multicolumn{7}{c}{\textbf{Scenario:} $r=675$} \\ \midrule
&FNM-EPR & 0.224  & 0.086  & 0.065  & 0.175 & 100.8  \\
&\phantom{EPR} & (0.000, 0.694) & (0.062, 0.110) & (0.000, 0.153) & (0.145, 0.205) & (99.5, 102.1) \\
&FNM-MCMC & 2.970  & 0.379 & 0.975  & 0.343  & 1134.3  \\
&\phantom{MCMC} & (0.000, 6.228) & (0.281, 0.477) & (0.485, 1.465) & (0.301, 0.385) & (1095.6, 1173.0) \\
\midrule
\multicolumn{7}{c}{\textbf{Scenario:} $r=900$} \\ \midrule
&FNM-EPR & 0.323 & 0.080  & 0.070  & 0.168 & 180.2\\
&\phantom{EPR}  & (0.000, 0.933) & (0.056, 0.104) & (0.000, 0.162) & (0.140, 0.196) & (168.7, 191.7) \\
& FNM-MCMC  & 3.607  & 0.472  & 0.933  & 0.386  & 1718.1  \\
&\phantom{MCMC} & (0.000, 8.607) & (0.366, 0.578) & (0.547, 1.319) & (0.348, 0.424) & (1243.6, 2192.6) \\
\bottomrule
\end{tabular}
\caption{Results comparing the Frobenius norm matching strategy implemented with EPR and MCMC. Various number of basis functions were tested for the covariance calibration. Each approach is used to forecast the \(T + 1\) time point. The results presented are mean (standard deviation) over 50 independent replications. The forecast error and the MSPE are presented on the log scale. Below each row is the corresponding interval \((\text{mean} \pm 2 \times \text{sd})\).}
\label{fig:fnm.compare.poiss}
\end{table}
Figure \ref{fig:mcmc_comp.poiss} displays the true GQN process, the posterior mean from FNM-EPR, and the posterior mean from FNM-MCMC across increasing values of the number of basis functions. As \(r\)  increases, the FNM-MCMC predictions have erratic patterns that deviate from the true process. This suggests that the MCMC chains may not have fully converged, particularly for larger values of \(r\), likely due to the increased dimension of the parameter space. While increasing the number of iterations or burn-in period could improve the convergence, doing so would further increase the already substantial computation required for posterior sampling. In contrast, the FNM-EPR predictions remain stable and visually consistent with the true GQN process across all values of \(r\), highlighting the computational advantages of using EPR. 

\subsection{Bernoulli GQN Simulation Study}
\begin{figure}[H]
    \centering
    
    \begin{subfigure}{0.3\textwidth}
        \includegraphics[width=\linewidth]{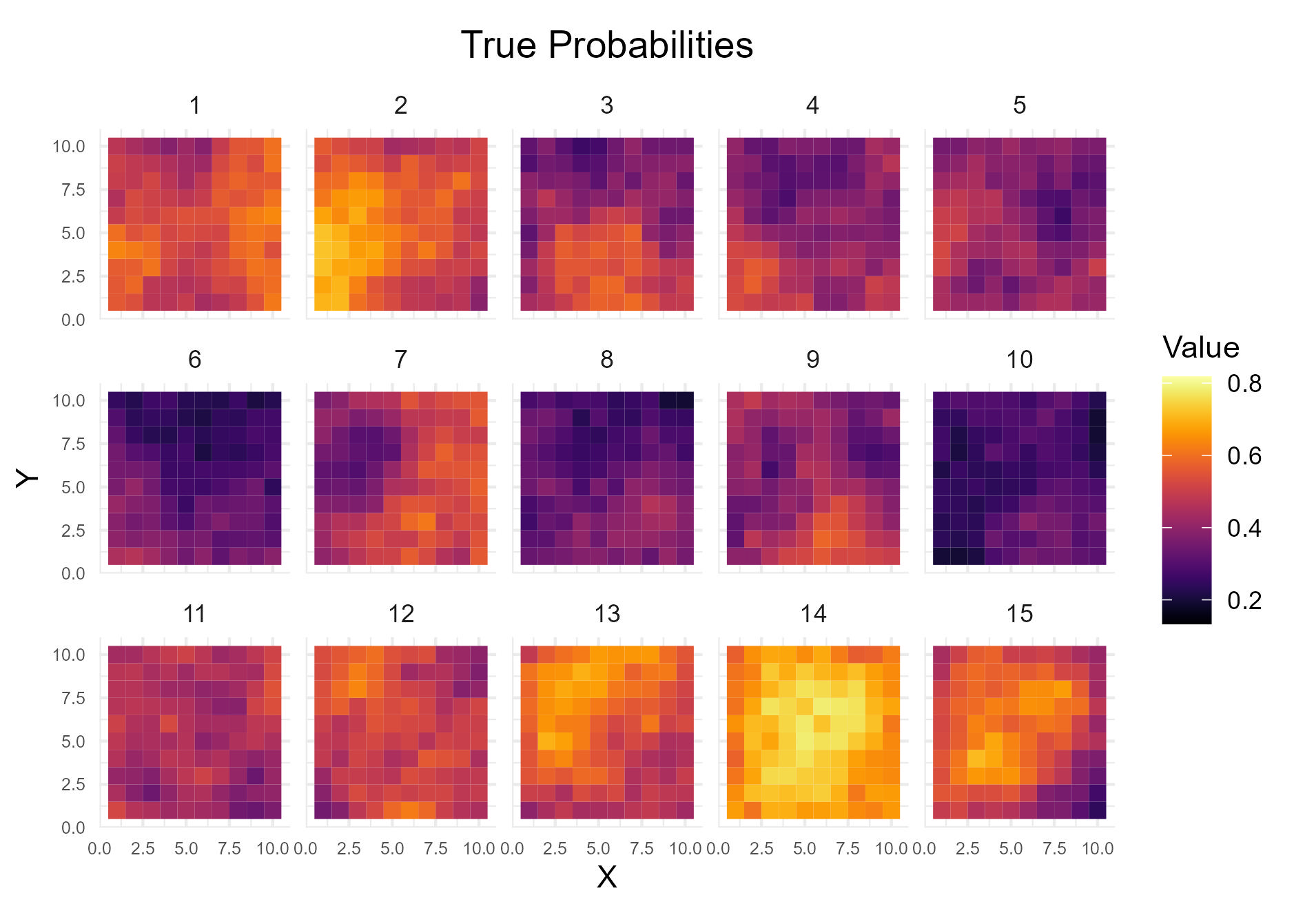}
    \end{subfigure}
    \hfill
    \begin{subfigure}{0.3\textwidth}
        \includegraphics[width=\linewidth]{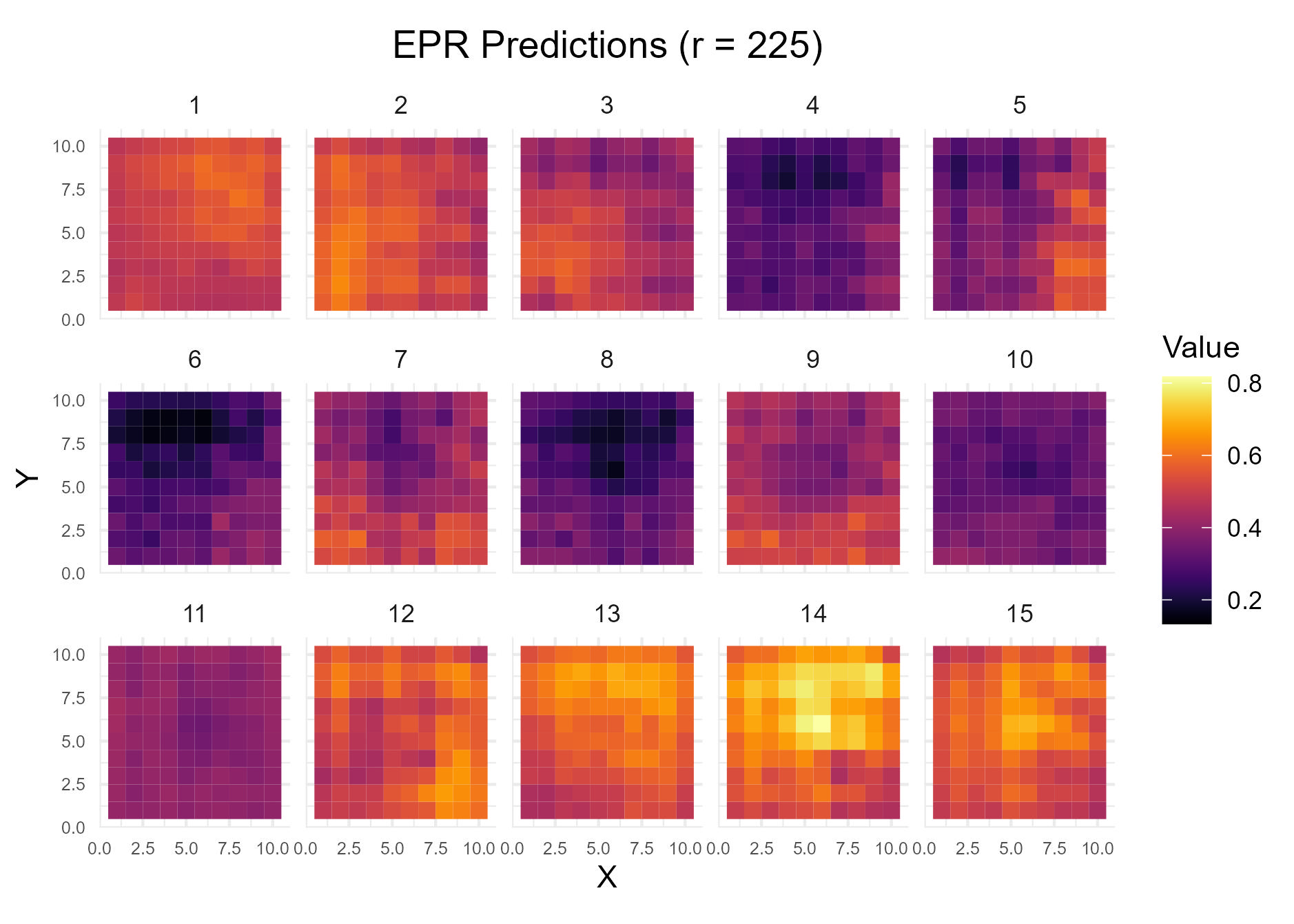}
    \end{subfigure}
    \hfill
    \begin{subfigure}{0.3\textwidth}
        \includegraphics[width=\linewidth]{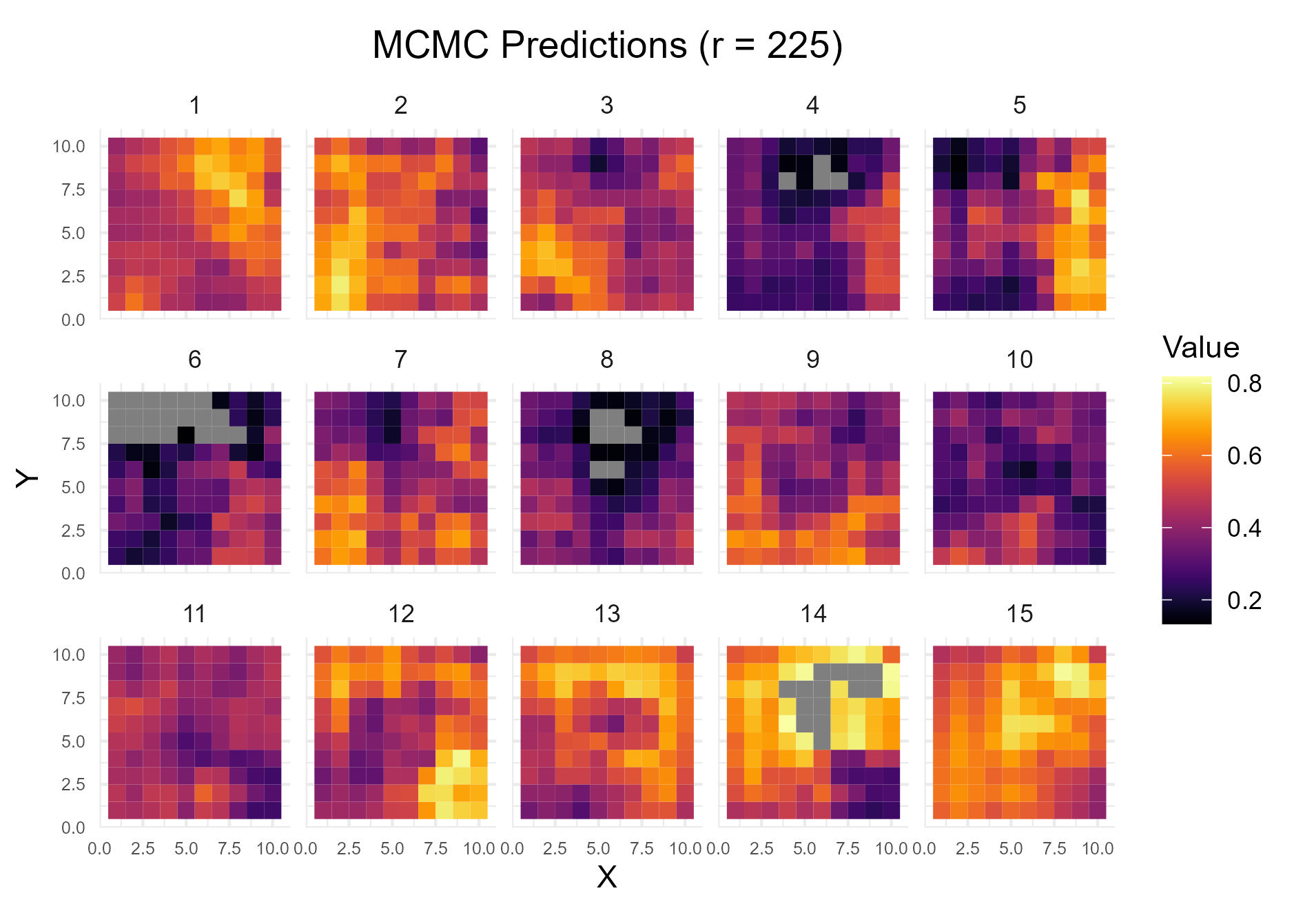}
    \end{subfigure}
    
    \par\bigskip
    \begin{subfigure}{0.3\textwidth}
        \includegraphics[width=\linewidth]{true.data.b.jpeg}
    \end{subfigure}
    \hfill
    \begin{subfigure}{0.3\textwidth}
        \includegraphics[width=\linewidth]{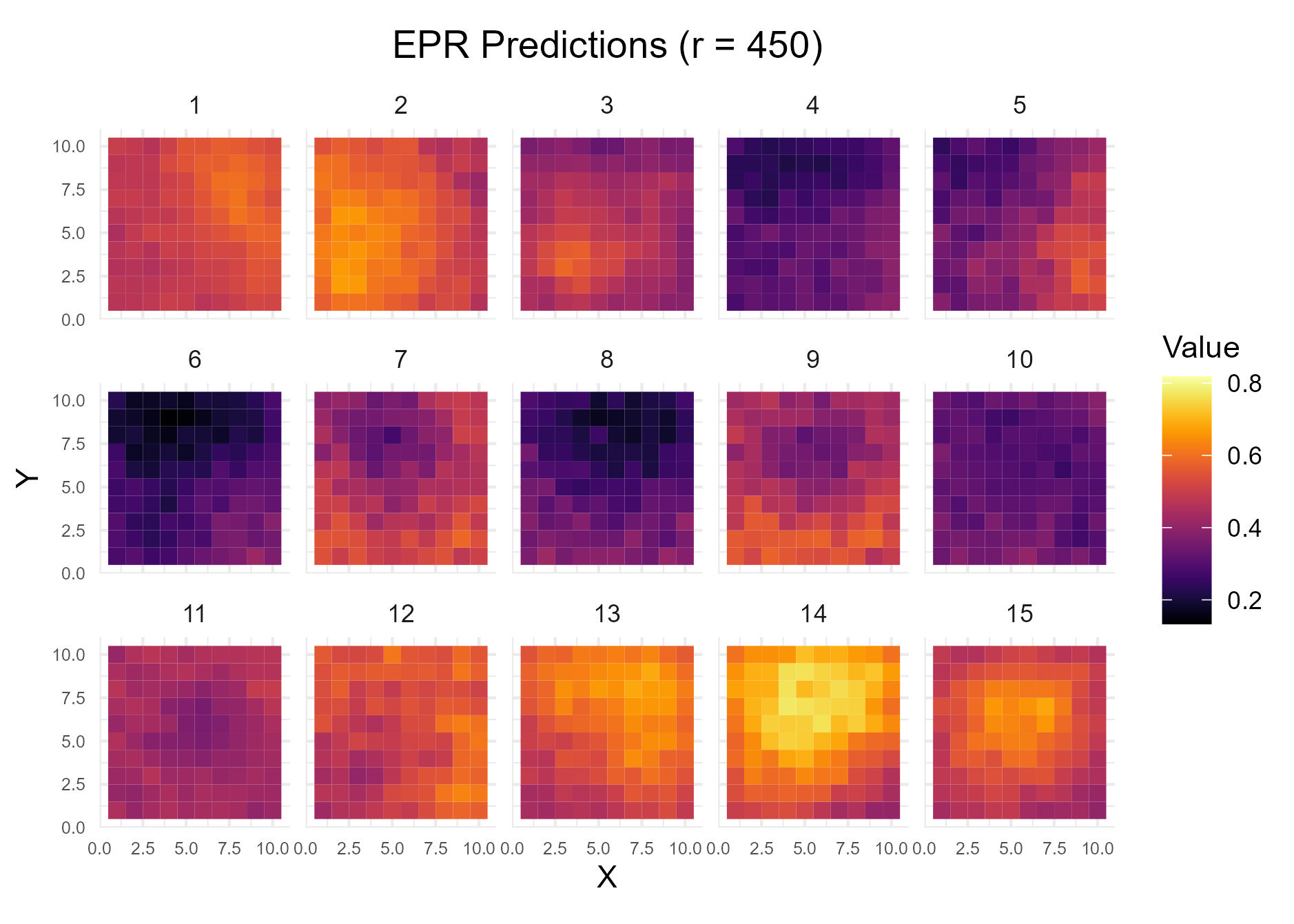}
    \end{subfigure}
    \hfill
    \begin{subfigure}{0.3\textwidth}
        \includegraphics[width=\linewidth]{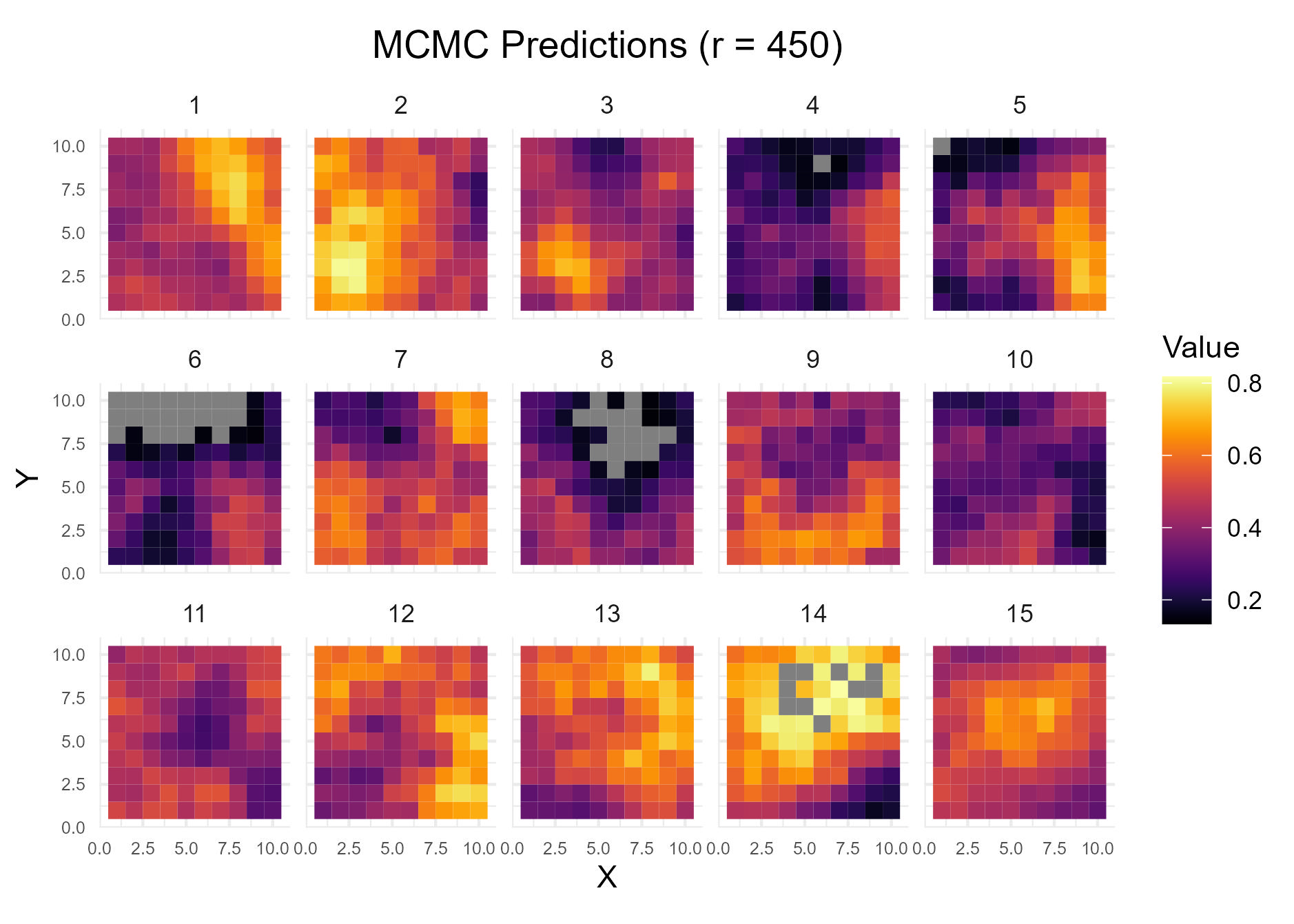}
    \end{subfigure}
    
    \par\bigskip
    \begin{subfigure}{0.3\textwidth}
        \includegraphics[width=\linewidth]{true.data.b.jpeg}
    \end{subfigure}
    \hfill
    \begin{subfigure}{0.3\textwidth}
        \includegraphics[width=\linewidth]{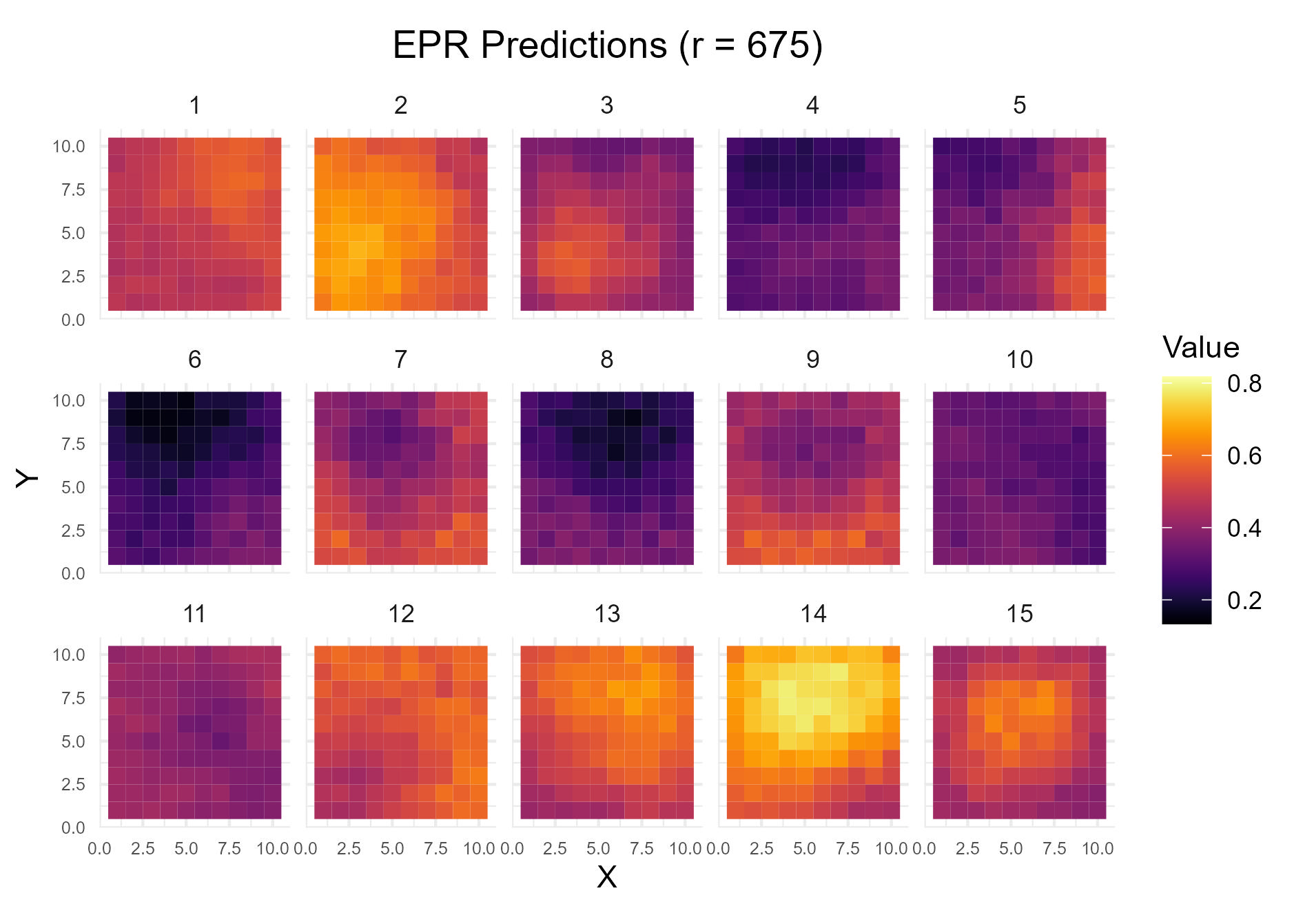}
    \end{subfigure}
    \hfill
    \begin{subfigure}{0.3\textwidth}
        \includegraphics[width=\linewidth]{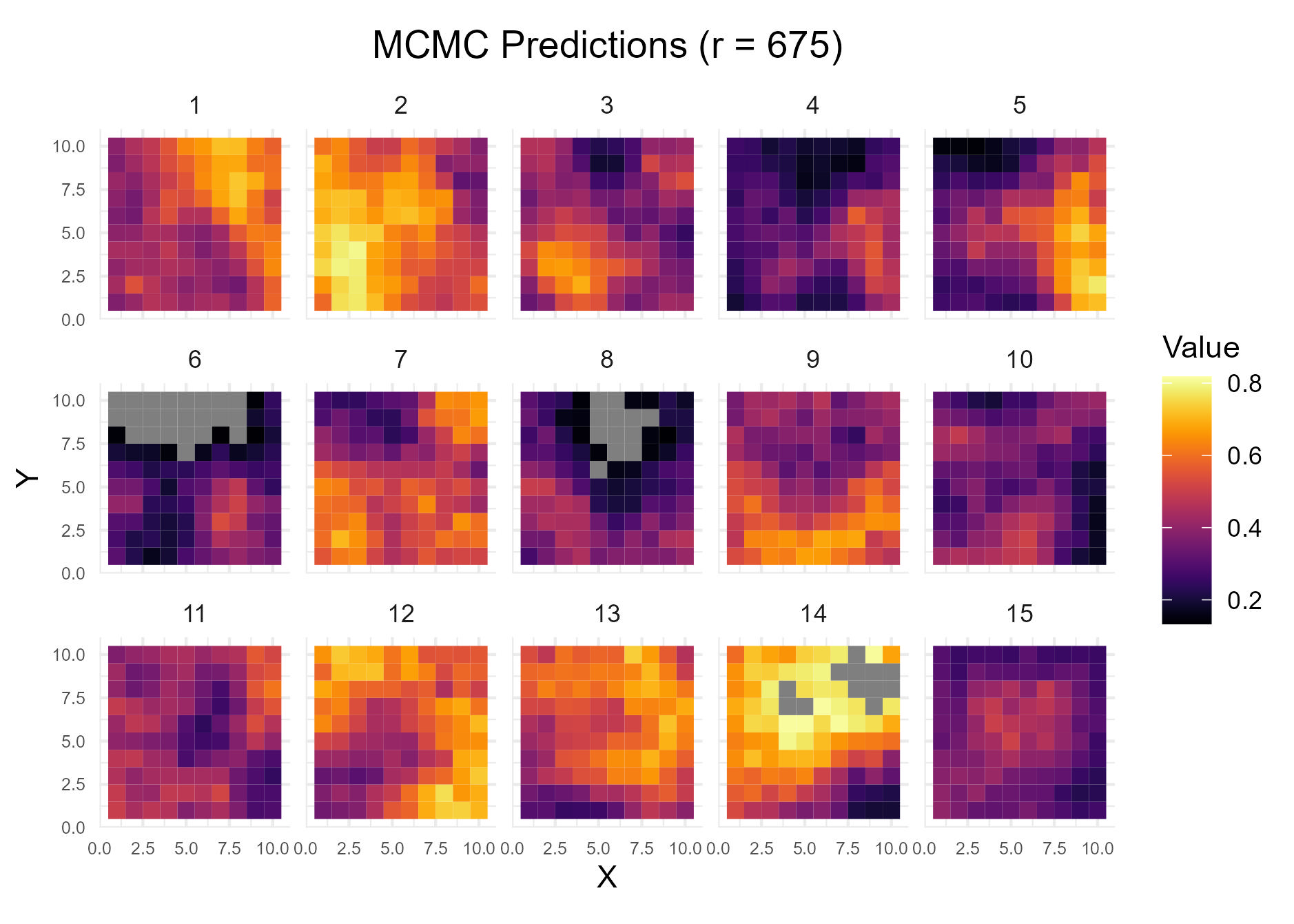}
    \end{subfigure}
    
    \par\bigskip
    \begin{subfigure}{0.3\textwidth}
        \includegraphics[width=\linewidth]{true.data.b.jpeg}
    \end{subfigure}
    \hfill
    \begin{subfigure}{0.3\textwidth}
        \includegraphics[width=\linewidth]{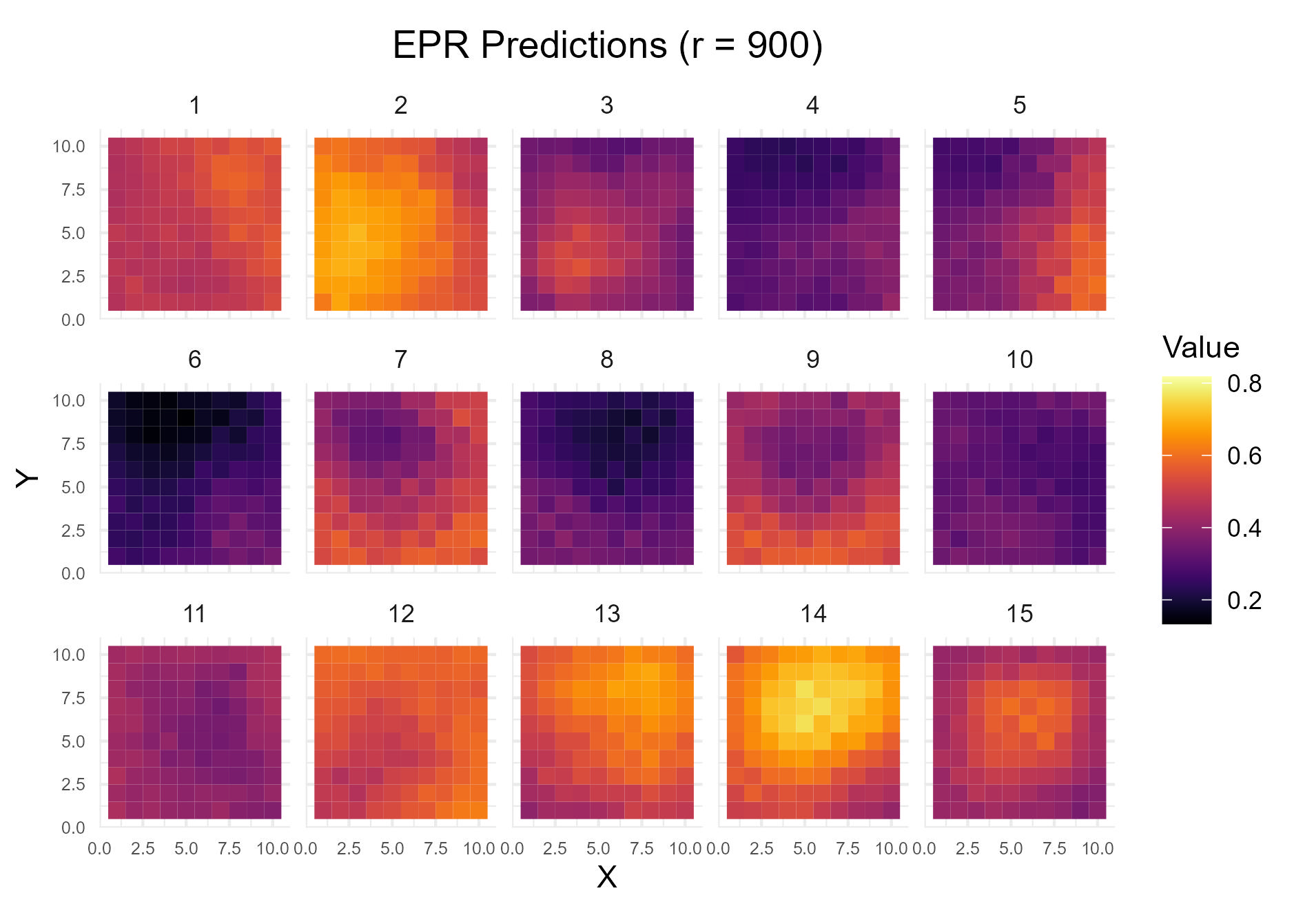}
    \end{subfigure}
    \hfill
    \begin{subfigure}{0.3\textwidth}
        \includegraphics[width=\linewidth]{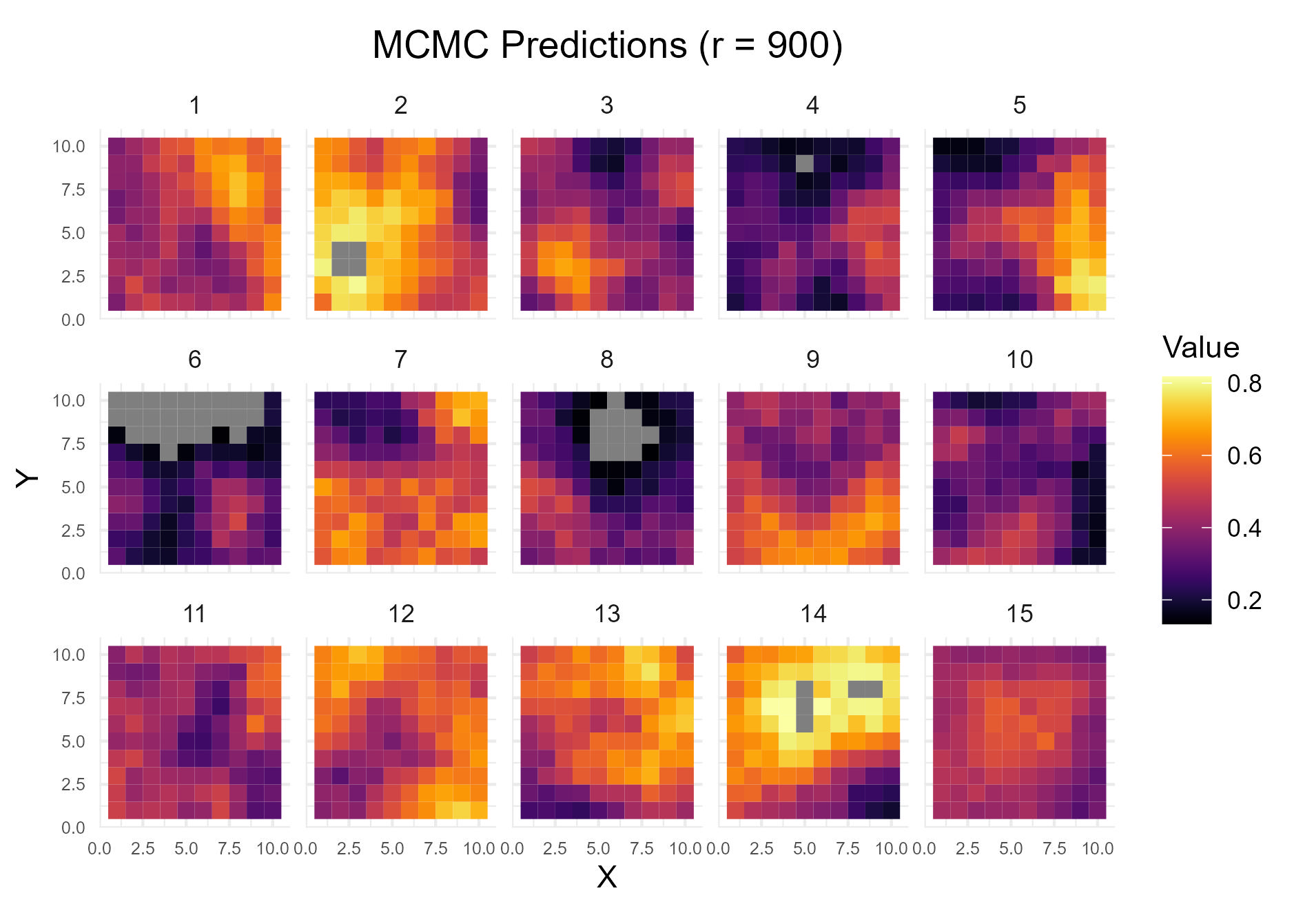}
    \end{subfigure}
    
    \caption{The first column is the true process simulated from the GQN, the second column is the posterior mean from the model implemented with EPR, and the third column is the posterior mean from the model implemented with MCMC. Each row is the comparison with \(r = 225, 450, 675, 900\) spatio-temporal basis functions used in the Frobenius norm matching, respectively. The \(T+1\) time point in each example was forecasted.} 
    \label{fig:mcmc_comp.bern}
\end{figure}

In this appendix, we present the results for the Bernoulli distributed simulation results. Table \ref{fig:mcmc_comp.bern} presents predictive and computational evaluation metric for the calibrated linear model implemented with EPR and MCMC. Results are reported across different values of \(r\) where recall from the main text, increasing the number of basis functions improves the ability to capture the nonlinear dynamics from the GQN covariance.

The linear model implemented with EPR produces smaller forecast error, MSPE, and CRPS compared to the model fitted using MCMC. However, the model implemented with MCMC has superior performance in terms of the area under the receiver operating characteristic curve (AUC). This suggests that EPR produces smoother posterior mean estimates that more closely capture the underlying latent process, whereas MCMC tends to produce predictions that are more closely aligned with the observed data, leading to higher AUC values. There is a trade-off between accurately recovering the true process (should use EPR) and achieving a higher classification accuracy (should use MCMC). 

Consistent with the results observed with the other two data types, EPR offers a significant computational advantage over MCMC. As the number of basis functions increases, the forecast error for EPR decreases similar to the Gaussian data scenario. On the other hand, the forecast error for MCMC does not show a consistent improvement with increasing rank of \(\mathbf{G}\), likely due to convergence difficulties with the increases parameter space.
\begin{table}[H]
\small 
\setlength{\tabcolsep}{3pt}
\centering
\begin{tabular}{llcccccc}
\toprule
 & Approach & Forecast & MSPE & MSE & CRPS & AUC & CPU Time (sec) \\ 
\midrule 
\multicolumn{8}{c}{\textbf{Scenario:} $r=225$} \\ \midrule
&FNM-EPR  &  0.020  & 0.0077  & 0.074 & 0.242  & 0.664  & 11.6  \\
&\phantom{EPR}  & (0.000, 0.052) & (0.0053, 0.0101) & (0.000, 0.230) & (0.212, 0.272) & (0.642, 0.686) & (11.4, 11.8) \\
&FNM-MCMC &  0.045 & 0.0137  & 0.376  & 0.316 & 0.669  & 188.0 \\
&\phantom{MCMC} & (0.000, 0.111) & (0.0099, 0.0175) & (0.000, 1.096) & (0.276, 0.356) & (0.649, 0.689) & (185.2, 190.8) \\
\midrule
\multicolumn{8}{c}{\textbf{Scenario:} $r=450$} \\ \midrule
&FNM-EPR &  0.012 & 0.0071  & 0.069  & 0.242  & 0.672& 46.3 \\
&\phantom{EPR} & (0.000, 0.024) & (0.0045, 0.0097) & (0.000, 0.233) & (0.202, 0.282) & (0.650, 0.694) & (46.0, 46.6) \\
&FNM-MCMC & 0.037  & 0.0137 &  0.349  & 0.327 & 0.692  & 294.2  \\
&\phantom{MCMC} & (0.000, 0.083) & (0.0097, 0.0177) & (0.000, 0.847) & (0.289, 0.365) & (0.674, 0.710) & (232.7, 355.7) \\
\midrule
\multicolumn{8}{c}{\textbf{Scenario:} $r=675$} \\ \midrule
&FNM-EPR  & 0.011  & 0.0067 & 0.064 &  0.236  & 0.676 & 109.2  \\
&\phantom{EPR} & (0.000, 0.025) & (0.0041, 0.0093) & (0.000, 0.210) & (0.200, 0.272) & (0.654, 0.698) & (108.3, 110.1) \\
&FNM-MCMC &  0.043  & 0.0134  & 0.292 & 0.335  & 0.703  & 441.1 \\
&\phantom{MCMC} & (0.000, 0.111) & (0.0096, 0.0172) & (0.000, 0.766) & (0.293, 0.377) & (0.685, 0.721) & (295.9, 586.3) \\
\midrule
\multicolumn{8}{c}{\textbf{Scenario:} $r=900$} \\ \midrule
&FNM-EPR &  0.009  & 0.0066 & 0.065 & 0.236  & 0.677 & 179.5  \\
&\phantom{EPR} & (0.001, 0.017) & (0.0040, 0.0092) & (0.000, 0.223) & (0.194, 0.278) & (0.657, 0.697) & (174.2, 184.8) \\
&FNM-MCMC  & 0.032 & 0.0136  & 0.306  & 0.343  & 0.712  & 495.8 \\
&\phantom{MCMC} & (0.000, 0.082) & (0.0098, 0.0174) & (0.000, 0.620) & (0.301, 0.385) & (0.694, 0.730) & (344.1, 647.5) \\
\bottomrule
\end{tabular}
\caption{Results comparing the Frobenius norm matching strategy implemented with EPR and MCMC. Various number of basis functions were tested for the covariance calibration. Each approach is used to forecast the \(T + 1\) time point. The results presented are mean (standard deviation) over 50 independent replications. Below each row is the corresponding interval \((\text{mean} \pm 2 \times \text{sd})\).}
\label{fig:fnm.compare.bern}
\end{table}
Figure \ref{fig:mcmc_comp.bern} displays the true probabilities simulated from the GQN model alongside the predicted probabilities from EPR and MCMC, across increasing values of the number of basis functions used for Frobenius norm matching. The EPR predictions more closely resemble the true latent probabilities compared to MCMC which supports what we saw from the metrics in Table \ref{fig:fnm.compare.bern}. The MCMC predictions appear noiser, yet achieve better classification performance, as highlighted by thte higher AUC values. This suggests that while MCMC is less effective at recovering the underlying latent process, it may be better for classification tasks compared to EPR. 

\section{Static vs Time-Varying Fixed Effects}\label{appen:static.vs.time.vary}
\begin{figure}[H]
    \centering
    \begin{subfigure}[b]{0.47\textwidth}
        \centering
        \includegraphics[width=\textwidth]{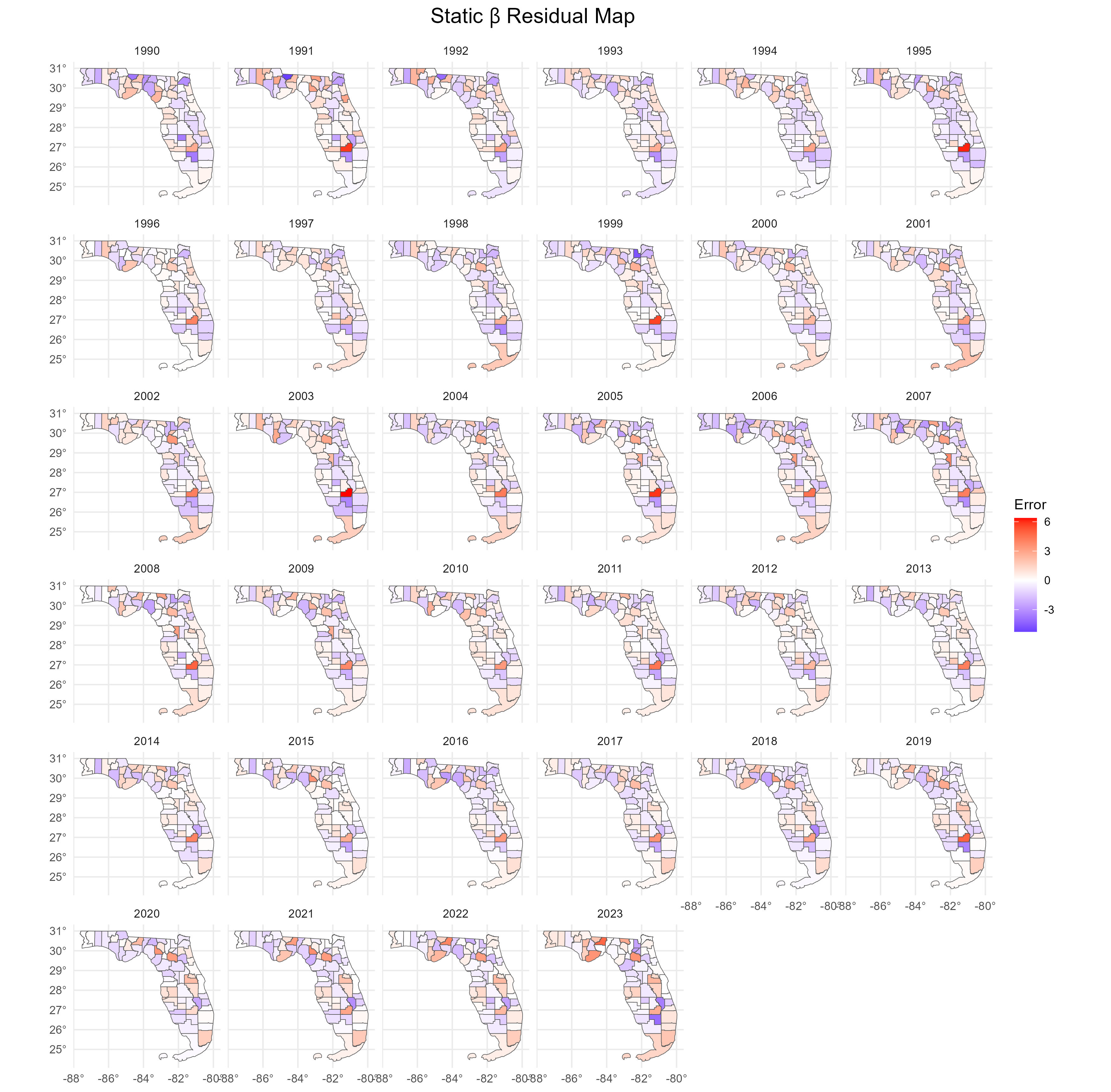}
    \end{subfigure}
    \hfill
    \begin{subfigure}[b]{0.45\textwidth}
        \centering
        \includegraphics[width=\textwidth]{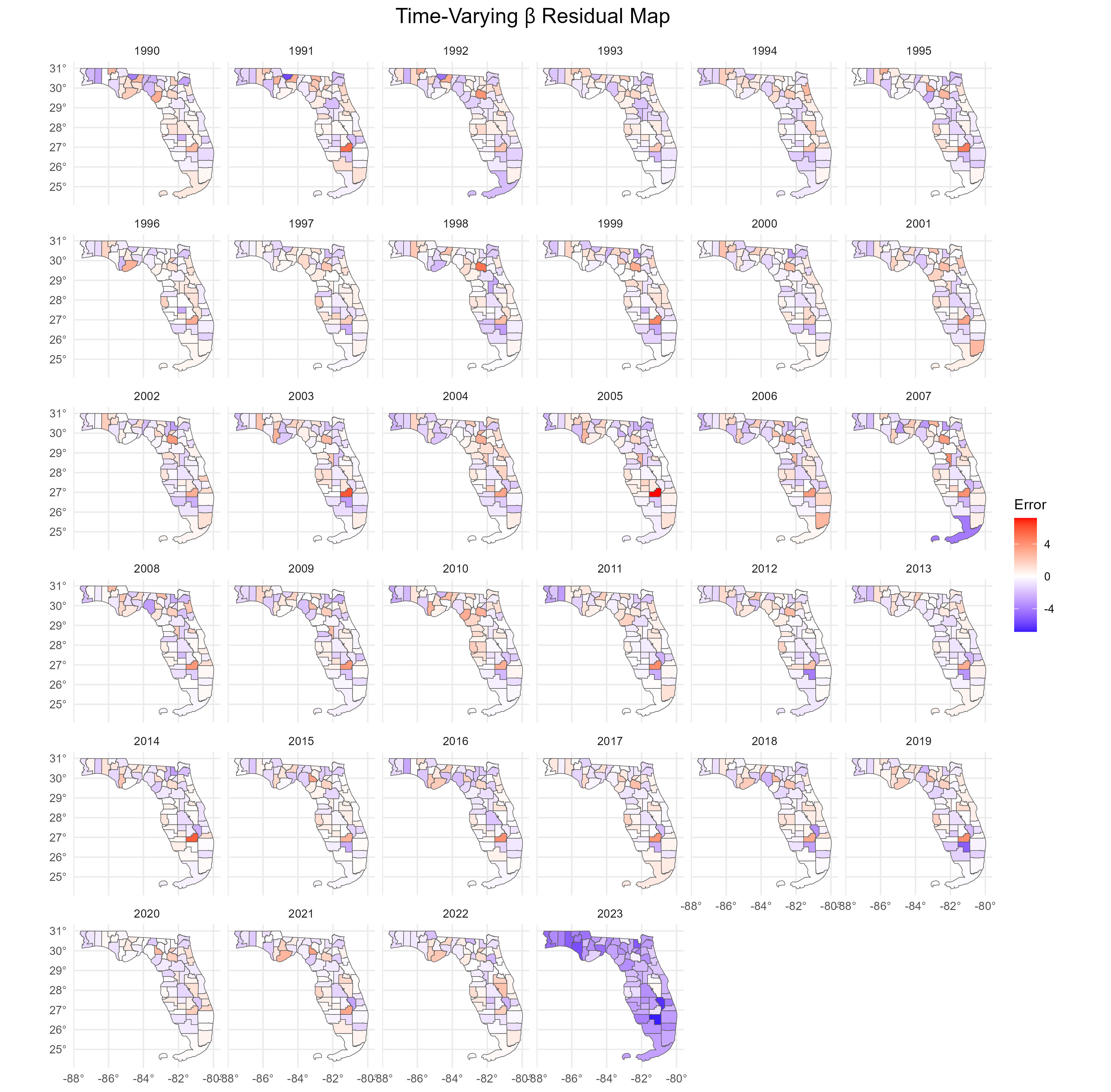}
    \end{subfigure}
    \caption{The left figure is the residual plot by year for the model with static covariate effects over time. The right figure is the residual plot by year for the model with time-varying covariate effects. The residuals are defined as the (predicted birth rate - true birth rate).}
    \label{fig:static.time.vary}
\end{figure}

\begin{table}[H]
\centering
\caption{Comparison of WAIC, forecasting, spatial smoothing, spatial interpolating, and computational performance for static and time-varying covariate effects for FNM-EPR.}
\begin{tabular}{cccccc}
\toprule
Fixed Effects & WAIC & Forecast Error & In-sample MSPE & Out-of-sample MSPE & CPU Time \\ 
\midrule
\(\boldsymbol{\beta}\) &  9.378  & 2.214  & 1.355    & 1.567             & 30.51 mins \\
\(\boldsymbol{\beta}_t\)  & 18.878    & 9.682   & 1.268   & 2.327 & 34.79 mins\\
\bottomrule
\end{tabular}\label{table:time.var}
\end{table}
Table \ref{table:time.var} compares the performance of the FNM-EPR model with static covariate effects and the FNM-EPR model with time-varying covariate effects. Overall, the static FNM-EPR model has better performance compared to the time-varying model with a lower WAIC, smaller forecast error, and smaller out-of-sample MSPE. The WAIC penalizes model complexity. The smaller WAIC for the static model compared to the time-varying model suggests that the time-varying coefficient model introduces addition complexity without significant improvement in the model fit. Allowing covariates to change over time does not improve the model's ability to make predictions at unobserved locations or accurately forecast. This is illustrated in Figure \ref{fig:static.time.vary}, where we see there are larger residuals for the time-varying model compared to the static covariate effects model. 

\section{Birth Rate Predictions}\label{appen:birth_rate_pred}
\begin{figure}[H]
    \centering
    \begin{subfigure}[b]{0.45\textwidth}
        \centering
        \includegraphics[width=\textwidth]{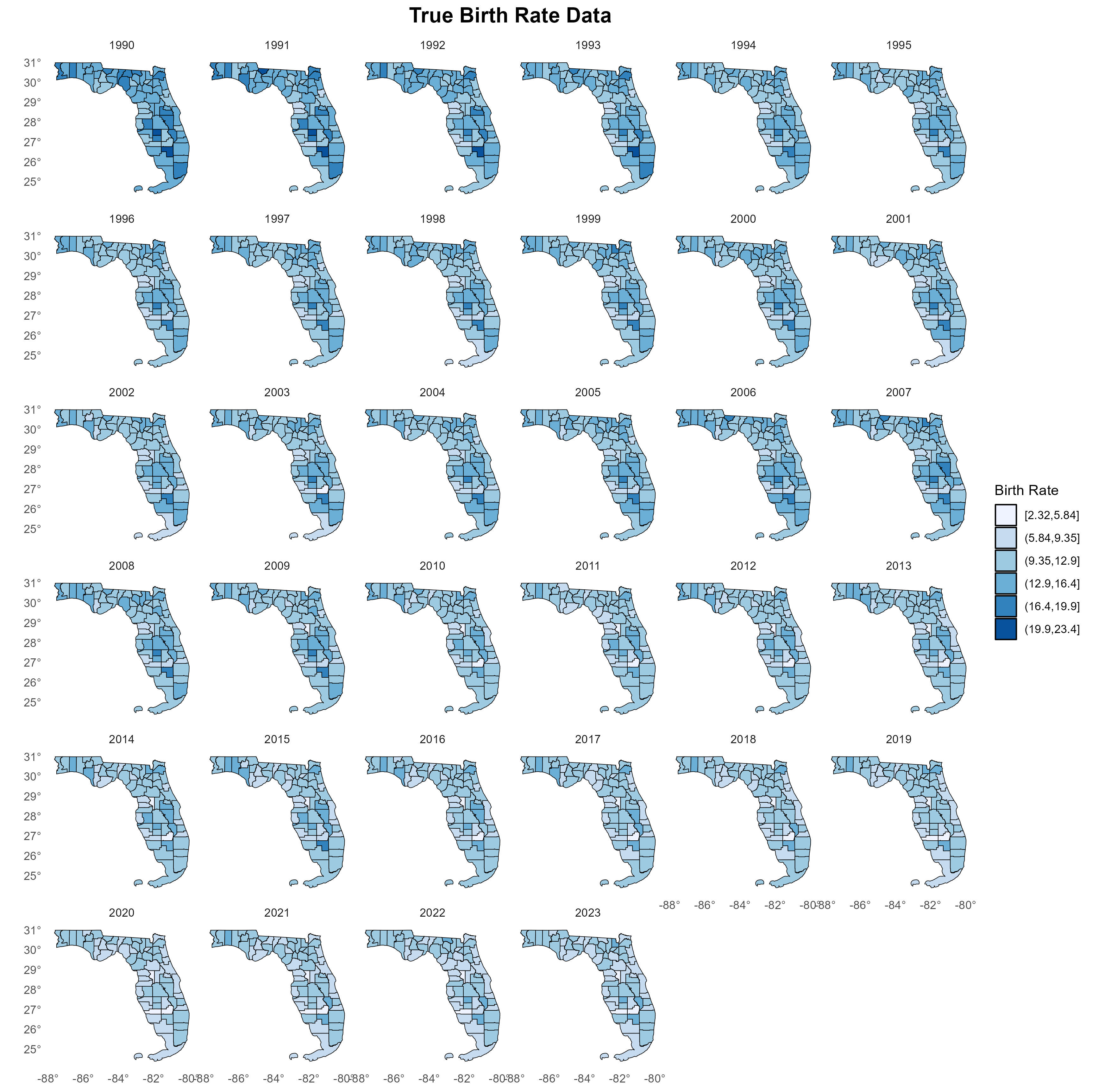}
    \end{subfigure}
    \hfill
    \begin{subfigure}[b]{0.45\textwidth}
        \centering
        \includegraphics[width=\textwidth]{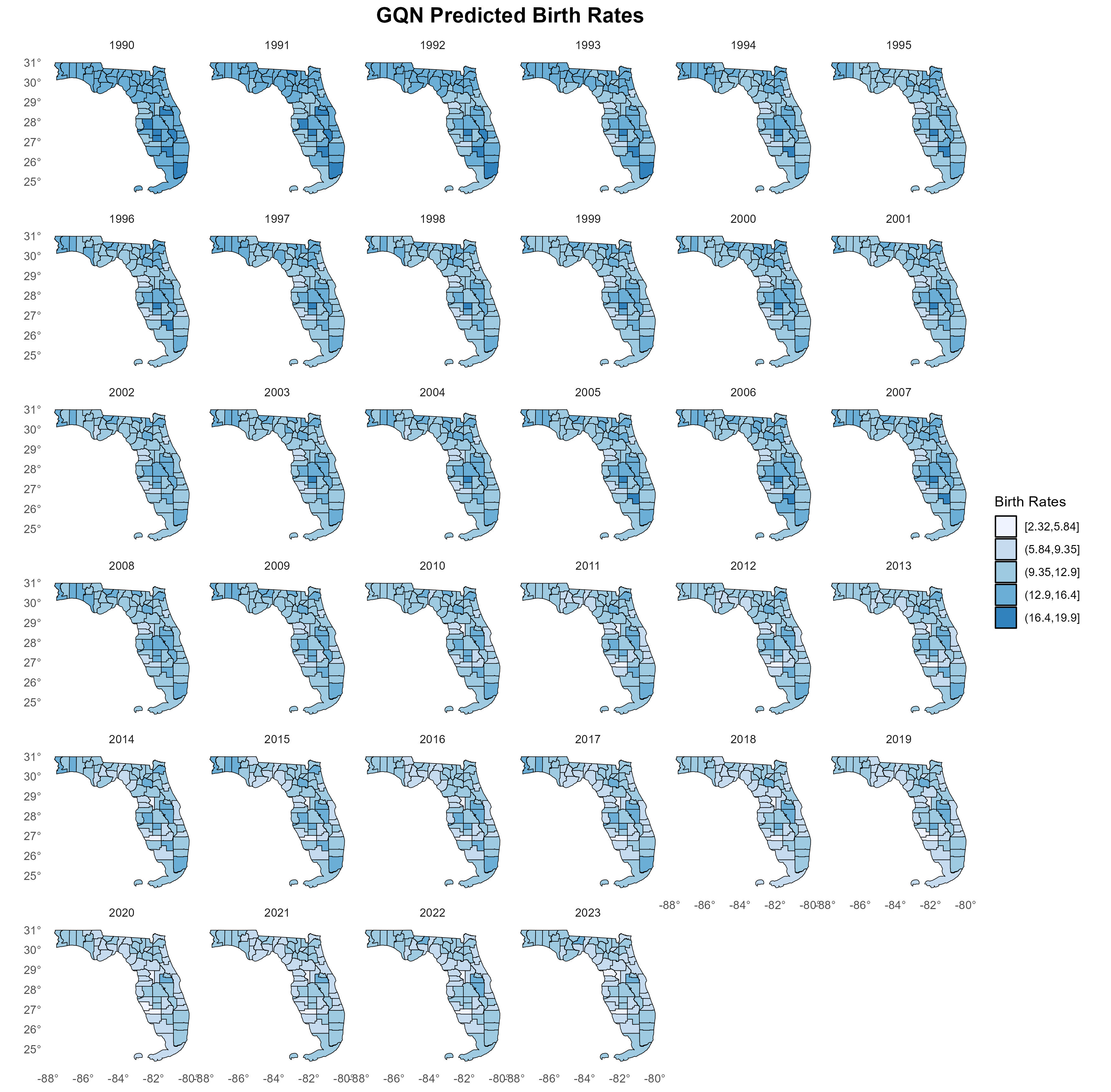}
    \end{subfigure}
    \\[1em]
    \begin{subfigure}[b]{0.45\textwidth}
        \centering
        \includegraphics[width=\textwidth]{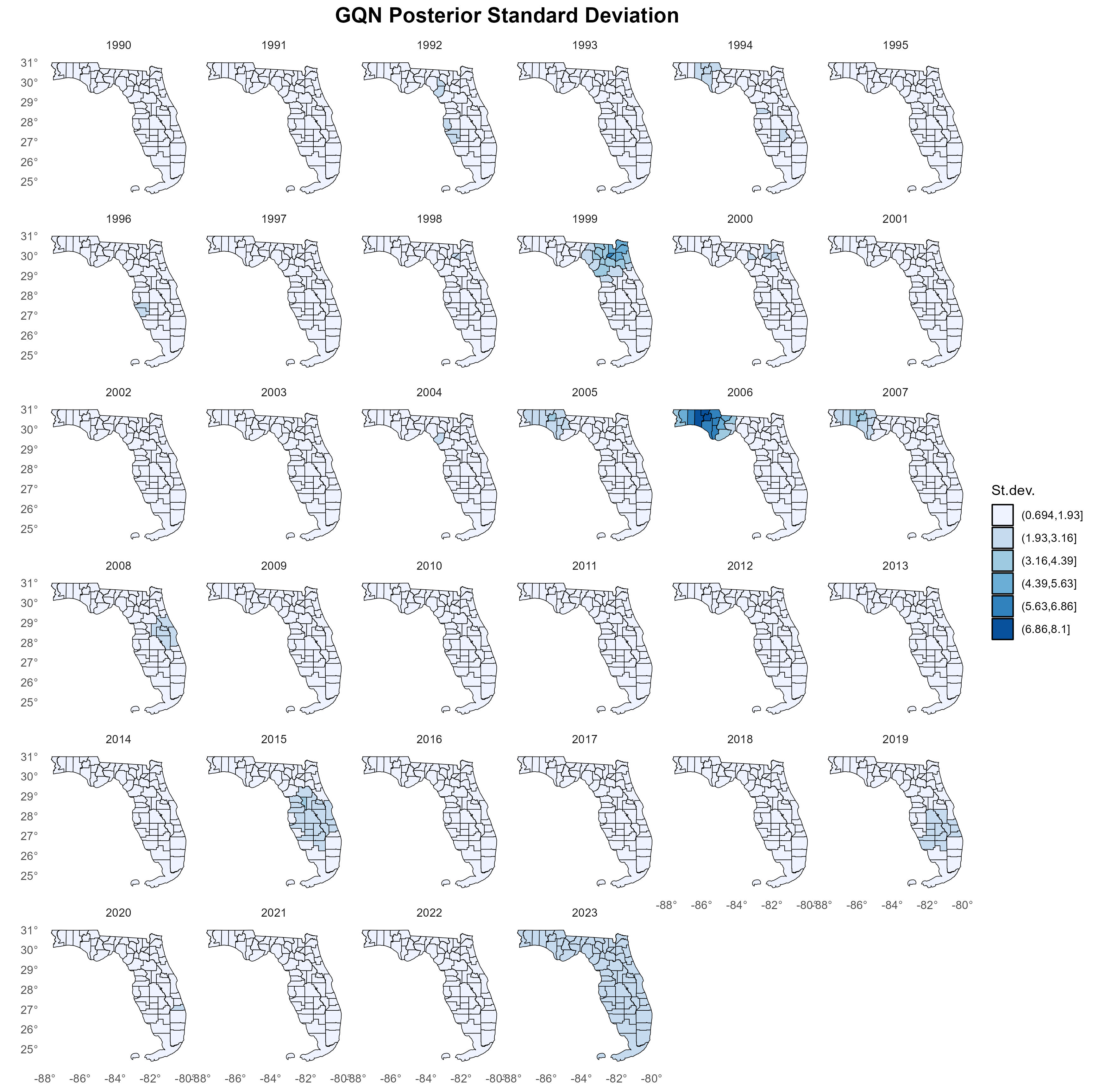}
    \end{subfigure}

    \caption{The top left figure is the true birth rate data and the top right figure is the predicted birth rates. The figure in the second row is the posterior standard deviations.}
    \label{fig:birth.rate.preds}
\end{figure}

\end{document}